\documentclass[reprint,showpacs,superscriptaddress,english,twocolumn,notitlepage,longbibliography]{revtex4-1}
\usepackage[utf8]{inputenc}
\pdfoutput=1
\usepackage{times}
\usepackage{color}
\usepackage{array}
\usepackage{verbatim}
\usepackage{multirow}
\usepackage{amsmath}
\usepackage{amssymb}
\usepackage{dsfont}
\usepackage{graphicx}
\usepackage{esint}
\usepackage[unicode=true,
 bookmarks=true,bookmarksnumbered=true,bookmarksopen=false,
 breaklinks=true,pdfborder={0 0 0},backref=false,colorlinks=true]
 {hyperref}
\usepackage{enumitem}
\usepackage{titlesec}
\usepackage[normalem]{ulem}

\usepackage{footmisc}

\hypersetup{
    linkcolor=red,          
    citecolor=blue,        
    filecolor=magenta,      
    urlcolor=magenta           
}

\usepackage{cleveref}
\crefname{appendix}{Appendix}{Appendices}
\crefname{equation}{Eq.}{Eqs.}
\crefname{figure}{Fig.}{Figs.}
\crefname{table}{Table}{Tables}
\crefname{section}{Sec.}{Secs.}

\renewcommand{\paragraph}[1]{\vspace{0.2cm}{\bf \textit{#1}}}

\usepackage{simpler-wick}
\allowdisplaybreaks[1] 
\newcommand{\mbf}{\mathbf}

\newcommand{\mcl}{\mathcal}

\newcommand{\mrm}{\mathrm}

\newcommand{\td}{\widetilde}
\newcommand{\ovl}{\overline}

\newcommand{\kk}{\mathbf{k}}
\newcommand{\RR}{\mathbf{R}}

\def\pare#1{\left( #1 \right)}
\def\brak#1{\left[#1\right]}

\def\ket#1{| #1 \rangle}

\def\inn#1{\langle #1 \rangle}
\def\Inn#1{\left\langle #1 \right\rangle}

\def\Im{\mathrm{Im}}

\def\sgn{\mathrm{sgn}}

\def\up{\uparrow}
\def\down{\downarrow}

\def\tr{\mathrm{tr}}

\def\mJ{\mathcal{J}}
\def\mV{\mathcal{V}}
\def\spin{\varsigma}
\def\ee{\epsilon}
\def\mG{\mathcal{G}}

\def\rr{\mathbf{r}}

\begin{document}
\title{Molecular Pairing in Twisted Bilayer Graphene Superconductivity}

\author{Yi-Jie Wang}
\thanks{These authors contribute equally to this work.}
\affiliation{International Center for Quantum Materials, School of Physics, Peking University, Beijing 100871, China}

\author{Geng-Dong Zhou}
\thanks{These authors contribute equally to this work.}
\affiliation{International Center for Quantum Materials, School of Physics, Peking University, Beijing 100871, China}

\author{Shi-Yu Peng}
\affiliation{Applied Physics \& Materials Science, California Institute of Technology, Pasadena, California 91125, USA}

\author{Biao Lian}
\affiliation{Department of Physics, Princeton University, Princeton, New Jersey 08544, USA}

\author{Zhi-Da Song}
\email{songzd@pku.edu.cn}
\affiliation{International Center for Quantum Materials, School of Physics, Peking University, Beijing 100871, China}
\affiliation{Hefei National Laboratory, Hefei 230088, China}
\affiliation{Collaborative Innovation Center of Quantum Matter, Beijing 100871, China}

\date{\today}
\begin{abstract}
We propose a theory for how the weak phonon-mediated interaction ($J_{\rm A}\!=\!1\!\sim\!4$meV) wins over the prohibitive Coulomb repulsion ($U\!=\!30\!\sim\!60$meV) and leads to a superconductor in  magic-angle twisted bilayer graphene (MATBG).  
We find the pairing mechanism akin to that in the A$_3$C$_{60}$ family of molecular superconductors: Each AA stacking region of MATBG resembles a C$_{60}$ molecule, in that optical phonons can dynamically lift the degeneracy of the moir\'e orbitals, in analogy to the dynamical Jahn-Teller effect. 
Such induced $J_{\rm A}$ has the form of an inter-valley anti-Hund's coupling and is less suppressed than $U$ by the Kondo screening near a Mott insulator.  
Additionally, we also considered an intra-orbital Hund's coupling $J_{\rm H}$ that originates from the on-site repulsion of a carbon atom. 
Under a reasonable approximation of the realistic model, we prove that the renormalized local interaction between quasi-particles have a pairing (negative) channel in a doped correlated insulator at $\nu=\pm(2+\delta\nu)$, albeit the bare interaction is positive definite. 
The proof is non-perturbative and based on {\it exact} asymptotic behaviors of the vertex function imposed by Ward identities. 
Existence of an optimal $U$ for superconductivity is predicted.
In a large area of the parameter space of $J_{\rm A}$, $J_{\rm H}$, the ground state is found to have a nematic $d$-wave singlet pairing, which, however, can lead to a $p$-wave-like nodal structure due to the Berry's phase on Fermi surfaces (or Euler obstruction). 
\end{abstract}

\maketitle

\paragraph{Introduction.}
A striking feature of magic-angle twisted bilayer graphene (MATBG) \cite{bistritzer_moire_2011} is that superconductivity (SC) emerges at small doping upon the correlated insulator (CI) \cite{cao_unconventional_2018,cao_correlated_2018,lu_superconductors_2019,yankowitz_tuning_2019}. The SC exhibits unconventional properties, such as a small coherence length \cite{cao_unconventional_2018,lu_superconductors_2019}, V-shaped tunneling spectrum \cite{oh_evidence_2021}, nematicity \cite{cao_nematicity_2021}, and $T$-linear resistance \cite{cao_strange_2020,polshyn_large_2019,jaoui_quantum_2022}. 
Despite extensive research on various pairing mechanisms \cite{wu_theory_2018,lian_twisted_2019,liu_electron-k-phonon_2023,yu_euler_2022,you_superconductivity_2019,khalaf_charged_2021,blason_local_2022,lothman_nematic_2022,christos_nodal_2023,islam_unconventional_2023}, understanding the coexistence of CI \cite{po_origin_2018,bultinck_ground_2020,kang_strong_2019,TBG4,TBG6,liu_theories_2021,kennes_strong_2018,koshino_maximally_2018,xu_kekule_2018,venderbos_correlations_2018,ochi_possible_2018,seo_ferromagnetic_2019,classen_competing_2019,kang_non-abelian_2020,xie_nature_2020,cea_band_2020,zhang_correlated_2020,soejima_efficient_2020,liu_nematic_2021,da_liao_correlation-induced_2021,hejazi_hybrid_2021} and SC, and their unconventional behaviors, remains challenging. 
Nevertheless, experimental studies have provided some constraints on the pairing. 
Suppressing the CI gap by screening the Coulomb interaction may enhance SC \cite{stepanov_untying_2020,saito_independent_2020,liu_tuning_2021}. Proximity-induced spin-orbit coupling enhances SC, while spontaneous ferromagnetism suppresses it, implying pairing of time-reversal partners \cite{arora_superconductivity_2020,lin_spin-orbitdriven_2022}. 
These observations are consistent with a phonon-based singlet pairing mechanism, but weak coupling BCS theory cannot explain the unconventional behaviors or how the strong Coulomb repulsion \cite{wong_cascade_2020,choi_interaction-driven_2021} is overcome by a small attractive interaction \cite{wu_theory_2018,lian_twisted_2019,liu_electron-k-phonon_2023}.

Inspired by the recent experimental evidence of significant coupling between flat band electrons and $A_1$, $B_1$ phonons at $\omega_{\rm ph}$=150meV \cite{chen2023strong}, we examine the possibility of a pairing mechanism based on $A_1$, $B_1$ phonons. 
The mediated attractive interaction $J_{\rm A}$ is merely a few meV \cite{wu_theory_2018,liu_electron-k-phonon_2023,shi_moire_2024,wang_tbg_epc_2024}. 
However, we find $J_{\rm A}$ can overcome the much stronger $U$ if the system is close to a Mott insulator where the quenching of charge fluctuation significantly suppresses $U$. 
A prototype of this pairing mechanism is the A$_3$C$_{60}$ family of molecular superconductors \cite{capone_colloquium_2009,capone_strongly_2002,chakravarty_electronic_1991,auerbach_electron-vibron_1994}. 
For both systems, electron orbitals are local on the scale of super-lattice - giving rise to strong correlations - but are spread on the microscopic lattice and are coupled to atomic distortions. 
As the $A_1$, $B_1$ phonons lead to a dynamical valley-Jahn-Teller effect \cite{angeli_valley_2019,angeli_jahnteller_2020}, $J_{\rm A}$ plays a role similar to the anti-Hund's coupling \cite{blason_local_2022} induced by the Jahn-Teller-distortion in fullerene, which is also previously suggested in Ref.~\cite{dodaro_phases_2018}.

\begin{figure}[t]
\centering
\includegraphics[width=\linewidth]{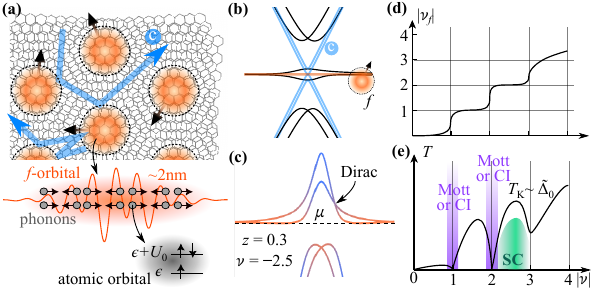}
\caption{Model. 
(a) Illustration for the effective $f$- and $c$-electrons. $f$-orbitals are located at AA stacking regions and dominate the flat bands. 
They are coupled to microscopic phonon modes via the dynamical valley Jahn-Teller effect. 
(b) Energy bands (black) as a result of hybridization between $f$-bands (orange) and itinerant Dirac $c$-bands (blue). 
(c) Heavy Fermi liquid bands at $\nu\!=\!-2.5$ with the quasi-particle weight $z\!=\!0.3$. 
Orange and blue colors represent contributions from $f$- and $c$-electrons, respectively. 
(d) and (e) sketch the $f$-occupation, $|\nu_f|$, and Kondo temperature, $T_{\rm K}$, as functions of the total flat band filling $|\nu|$ in DMFT calculations in the absence of $J_{\rm A,H}$, respectively. 
}
\label{fig:model}
\end{figure}

\paragraph{Approximations and methodology.} 
We use the topological heavy fermion model (THF) \cite{song_magic-angle_2022,shi2022heavy}, which has recently been applied to investigate the Kondo physics in MATBG \cite{chou_kondo_2023,zhou_kondo_2023,hu_symmetric_2023,hu_kondo_2023,datta_heavy_2023,rai_dynamical_2023,lau_topological_2023,chou_scaling_2023,merino_evidence_2024,calugaru_thermoelectric_2024,hu_link_2024}. 
It consists of effective local orbitals ($f_{\alpha\eta s}$) in AA stacking regions (\cref{fig:model}(a)), which dominate the flat bands, and itinerant Dirac $c$-electrons, which hybridize with $f$-orbitals to generate topology (\cref{fig:model}(b)) \cite{po_faithful_2019,song_all_2019,tarnopolsky_origin_2019,TBG2,ahn_failure_2019,wang_chiral_2021}. 
Here $\alpha$ (=1,2), $\eta$ (=$\pm$), $s$ (=$\up\down$) are the orbital, valley, and spin indices, respectively. 

\begin{figure}[t]
\centering
\includegraphics[width=\linewidth]{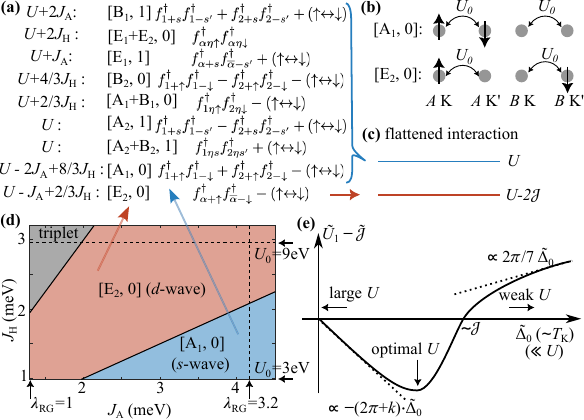}
\caption{Bare and renormalized interactions. (a) Two-particle eigenstates, labeled by $[\rho,j]$. $\rho$ denotes the $D_6$ representation, and $j$ denotes the total spin $j$. 
Pairing channels of the bare interaction coincide.
(b) Occupations of graphene sub-lattices ($A,B$) and valleys ($K,K'$) of the lowest $E_2$ and $A_1$ states. 
(c) The flattened interaction.
(d) Two-electron ground states in the parameter space of $J_{\rm A}$ and $J_{\rm H}$. 
As pairing channels of the bare interaction, the energies are repulsive due to $U$, hence they {\it cannot} form Cooper pairs. 
(e) The renormalized interaction $\td{U}_1 - \td{\mJ}$ for $d$-wave pairings as a function of the Kondo energy scale $\td\Delta_0 \!\sim\! T_{\rm K}$, which is assumed to be much smaller than the bare repulsion $U$. 
With other parameters fixed, $U$ expoentially suppresses $\td{\Delta}_0$, hence a smaller $\td{\Delta}_0$ also implies a larger $U$. 
}
\label{fig:interaction}
\end{figure}

Before detailed derivations, let us outline the chain of approximations and methodology employed in this work. After integrating out the fast $A_1,B_1$ phonons \cite{wang_tbg_epc_2024}, we obtain a multi-orbital Anderson lattice model where each impurity has eight flavors ($f_{\alpha\eta s}$), subject to a complex local interaction consisting of a Hubbard $U$ term (58meV), an anti-Hund's coupling $J_{\rm A}$ ($\sim$1meV), and a Hund's coupling $J_{\rm H}$ ($\sim$1meV) (\cref{fig:interaction}(a)).  
To analyze this unsolvable model, we assume the locality of correlation, treating each AA site as an Anderson impurity coupled to a bath that describes its environment. 
The locality of correlation is supported by the quantum-dot-like behavior \cite{wong_cascade_2020,choi_interaction-driven_2021} and evident local pairing gap (1-3meV) \cite{oh_evidence_2021,kim_evidence_2022} observed in experiments.  
It is also widely assumed in recent slave-particle \cite{chou_kondo_2023,lau_topological_2023}, dynamical mean-field theory (DMFT)  \cite{zhou_kondo_2023,hu_symmetric_2023,datta_heavy_2023,rai_dynamical_2023}, and Gutzwiller \cite{hu_link_2024} calculations that have reproduced key features of the experimental spectrum and compressibility \cite{hu_link_2024}. 
Kondo temperature $T_{\rm K}$ and $f$-occupation $\nu_f$ have been determined as functions  of the total filling $\nu$ \cite{zhou_kondo_2023,hu_symmetric_2023,datta_heavy_2023,rai_dynamical_2023,lau_topological_2023}, as sketched in \cref{fig:model}(e) and (d). 
Both $\nu$ and $\nu_f$ range from $-$4 to 4, with $\nu_f\!=\!\nu\!=\!0$ corresponding to the charge neutrality point. 
At $\nu\!=\!- 2- \delta\nu$, where the highest SC $T_c$ is observed, the ground state without $J_{\rm A,H}$ can be a heavy Fermi liquid characterized by $T_{\rm K}\!\sim$1-10K, $\nu_f\!\approx\!-2$, and a quasi-particle weight $z\!\sim$0.1-0.3.

We devote this work to investigating the pairing instability of the Fermi liquid at $\nu\!=\!-2-\delta\nu$ in the presence of $J_{\rm A,H}$. An immediate difficulty arises: Since $U \!\gg\! J_{\rm A,H}$, the bare interaction is positive definite and does not support any pairings \cite{TBG5,chowdhury2024_hongkong} in naive mean-field theories. In fact, this difficulty will appear in any attempt to explain the SC in MATBG through a weak attractive interaction, regardless of its origin. (The Luttinger mechanism may give rise to an attractive channel but will predict a much lower SC energy scale compared to the observed local pairing gap.) A crucial step in our analysis is that, under the so-called flattened interaction limit (explained later), which is justified for the Anderson impurity in the Fermi liquid phase with $T_{\rm K}\ll J_{\rm A}$,  we can obtain {\it exact} asymptotic behaviors of the fully renormalized local interaction. 
We further prove the existence of a pairing channel. This is particularly notable given that the bare interaction is positive definite. 

\begin{figure}
\includegraphics[width=\linewidth]{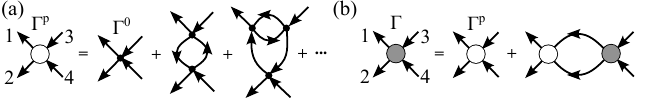}
\caption{Skeleton diagrams for 2PI (a) and 1PI (b) vertices. $\Gamma^0$ represents the anti-symmetrized bare vertex and lines represent fully addressed Green's functions. 
For local 2PI and 1PI vertices, all bare vertices in the skeleton diagrams are at the same site. 
}
\label{fig:diagram-main}
\end{figure}

A powerful theoretical tool that enables our analysis is the Ward identity \cite{nozieres_kondo_1980,coleman_heavy_2007,hewson_fermi_1993} that relates the local one-particle irreducible (1PI) vertex, representing the renormalized local interaction, to susceptibilities ($\chi$) of local conserved charges. 
The local 1PI vertex is given by {\it skeleton} diagrams (\cref{fig:diagram-main}) of bare vertices at the same site and fully dressed local Green's functions. 
The behavior of $\chi$ can be known once the local ground state manifold is determined. We then derive the asymptotic behaviors of the local 1PI vertex through the Ward identity and identify a pairing channel. 
This approach reproduces the Bethe ansatz result for the one-orbital Anderson impurity \cite{SM}, confirming its validity.

With the pairing channel identified, we next study the SC on the moir\'e lattice.
Consider the RPA pairing susceptibility $\chi_{\rm p} = \chi_{\rm p0}/ (1+ \chi_{\rm p0} \Gamma^{\rm p})$, where $\chi_{\rm p0}$ is the non-interacting susceptibility (bubble diagram) from the heavy quasi-particle excitations on the lattice, and $\Gamma^{\rm p}$ is the effective local interaction. 
Technically, $\Gamma^{\rm p}$ is given by the local 2PI vertex, connected to the local 1PI vertex via \cref{fig:diagram-main}(b), to avoid double counting in the ladder diagrams of $\chi_{\rm p}$, as standard in many approaches \cite{georges_infinite_d_1996,rohringer_diagrammatic_2018}. 
$\Gamma^{\rm p}$ replaces the bare interaction in weak-coupling RPA. 
Instead of examining the divergence of $\chi_{\rm p}$, we perform a straightforward mean-field calculation using the effective interaction $\Gamma^{\rm p}$ and a renormalized quasi-particle spectrum (\cref{fig:model}(c)). 
Quantitative results, including pairing symmetry, will be discussed later.


Due to the particle-hole symmetry of the model \cite{TBG2,TBG3,song_magic-angle_2022,wang_chiral_2021}, physics at $\nu\!=\!2 + \delta\nu$ is similar. 
In experiments, particle-hole asymmetry arises from various effects \cite{kang_pseudomagnetic_2023,herzog-arbeitman_heavy_2024} including non-local interlayer tunneling \cite{kwan_kekule_2021}.

\paragraph{Effective model.} 
We write the free action for an Anderson impurity as
\begin{equation} \label{eq:S0-main}
S_0 = -\sum_{\omega} \sum_{\alpha\eta s} 
    f_{\alpha\eta s}^\dagger (\omega) 
    (-i\omega + \ee_f - i \Delta(\omega))
    f_{\alpha\eta s}(\omega)\ .
\end{equation}
Here $\omega$ is the fermion Matsubara frequency, $\ee_f$ is the on-site energy, and $\Delta(\omega)$ is the hybridization function. 
In a Fermi liquid phase, $\Delta(\omega)$ can be well approximated by $\Delta_0 \sgn(\omega)$ for low energy physics. $\Delta_0$ should be understood as a phenomenological bare parameter that reproduces the correct $T_{\rm K}$ \footnote{$\Delta_0$ could be estimated from self-consistent hybridization function in DMFT calculation.}.  
Our analysis in this work does not directly depend on $\Delta_0$, but only on $T_{\rm K}$. 
The eight flavors have identical on-site terms because they are related by time-reversal ($\eta\to \! \ovl\eta$), spin ($s\to \ovl{s}$), and a $D_6$ point group ($\alpha\to\ovl{\alpha}$) symmetries \cite{song_all_2019,song_magic-angle_2022}. 
Here indices with a bar represent the opposite indices of the same degree of freedom.

We consider three interaction terms: an on-site Hubbard $U$ term (58meV) contributed by Coulomb repulsion of 2D electron gas \cite{song_magic-angle_2022}, an anti-Hund's coupling $J_{\rm A}\!\approx\!\lambda_{\rm RG} \!\times\! 1.3$meV contributed by electron-phonon coupling to $A_1,B_1$ phonons, and a Hund's coupling $J_{\rm H}\!\approx\!  0.33 \!\times\! 10^{-3} U_0$ contributed by Hubbard repulsion $U_0$ (3-9eV) at each carbon atom \cite{wehling_strength_2011,schuler_optimal_2013,gonzalez-arraga_electrically_2017,zhang_spin-polarized_2022}. 
The $J_{\rm A}$ term on THF basis was recently obtained in Refs.~[\onlinecite{wang_tbg_epc_2024}, \onlinecite{shi_moire_2024}].  
$\lambda_{\rm RG}\!\approx\!3.2$ is an enhancement factor due to renormalization effect \cite{basko_interplay_2008}. 
We also derive an analytical form of the $J_{\rm H}$ term \cite{SM}. 
We tabulate all the two-electron eigenstates and eigenenergies in \cref{fig:interaction}(a), which completely define the four-fermion interaction Hamiltonian. 
As the name suggests, $J_{\rm A}$ lowers the energies of inter-valley intra-orbital $s$-wave singlet ($A_1$ representation) and inter-valley inter-orbital $d$-wave singlets ($E_2$ representation) by $2J_{\rm A}$ and $J_{\rm A}$, respectively. 
Since $U_0$ disfavors double occupation on a carbon atom, and the $\alpha\!=\! 1,2$ $f$-orbitals are mainly distributed on the $A$ and $B$ graphene sub-lattices \cite{song_magic-angle_2022}, respectively, $J_{\rm H}$ disfavors double occupation on each $\alpha$ orbital alike. 
Consequently, the inter-orbital $d$-wave singlets are energetically less penalized ($\frac23 J_{\rm H}$) than the intra-orbital $s$-wave singlets ($\frac83 J_{\rm H}$).  

Varying $J_{\rm A}$ and $J_{\rm H}$ over a realistic range, we thus find a large region where the lowest two-electron states are $d$-wave (\cref{fig:interaction}(d)). 
In the absence of $U$, our single-site result fully aligns with the mean-field SC phase diagram of $s$-wave and $d$-wave pairings \cite{wu_theory_2018, liu_electron-k-phonon_2023}, where the full $\kk$-dependent interaction is employed. 
However, presence of the dominating $U$ blocks all pairing channels in the bare interaction. 
We thus aim to examine pairings in the renormalized interaction.
As will be shown, the $d$-wave pairing matches several unconventional features of the SC. 

\paragraph{Flattened interaction.} 
To study the pairing instability of the Fermi liquid phase at $\nu\!=\!-2-\!\delta\nu$, we argue that the complex interaction Hamiltonian defined by \cref{fig:interaction}(a) can be replaced by a simpler one if $T_{\rm K}$ is finite but sufficiently low. 
When the level splittings in \cref{fig:interaction}(a) far exceed the Kondo energy scale, {\it i.e.}, $J_{\rm A,H}\gg T_{\rm K}$, only the two-electron  ground states participate in Kondo screening \cite{nozieres_kondo_1980}.  
Correspondingly, high-energy states turn to virtual processes, whose splittings do not qualitatively change low energy physics.
Motivated by this observation, we introduce a flattened interaction (\cref{fig:interaction}(c)), where all pairing channels are set to energy $U$, except for the $d$-wave ground state that has the energy $U-2\mJ$, with $\mJ\!\sim\! J_{\rm A} \!\ll\! U$.
The flattened interaction enjoys a $\rm U(1)^{\times 4} \!\times\! SU(2)^{\times 2}$ symmetry generated by charge $\sigma^0 \tau^0 \spin^0$, valley $\tau^z$, orbital $\sigma^z$, angular momentum $\sigma^z \tau^z$, and two independent spin $\frac{\sigma^0\tau^0 \pm \sigma^z\tau^z}{2} \spin^{x,y,z}$ rotations \cite{SM}. 
Here $\sigma^{x,y,z}$, $\tau^{x,y,z}$, and $\spin^{x,y,z}$ are Pauli matrices for the orbital, valley, and spin degrees of freedom, respectively. 
The higher symmetry gives rise to Ward identities that help determine the renormalized interaction. 
Notably, the flattened interaction is still positive definite and does not support pairings in naive mean-field theories.

Ref.~\cite{SM} provides a more quantitative justification for the flattened interaction by a phenomenological susceptibility analysis. It shows that, if the original interaction is adopted, the breaking of $\rm U(1)^{\times4} \!\times\! SU(2)^{\times 2} $ symmetry in the renormalized theory is finite but weak. The flattened interaction also applies to the $T_{\rm K} \gg J_{\rm A,H}$ limit where any multiplet splitting becomes irrelevant at the Kondo energy scale.

Constrained by the $\rm U(1)^{\times 4} \!\times\! SU(2)^{\times 2}$ symmetry, a general parametrization of $S_I$ reads 
{\small
\begin{align} \label{eq:SI-main}
    S_I =&   \frac12 \int d\tau \sum_{\alpha\eta}  \bigg( 
    \Big( U_1+\frac{\mJ}2 \Big) {N}_{\alpha\eta} {N}_{\ovl\alpha\ovl\eta} 
    + U_2  {N}_{\alpha\eta} {N}_{\alpha\ovl\eta}  \nonumber\\
& 
    + U_3  {N}_{\alpha\eta} {N}_{\ovl\alpha\eta}
    + U_4  {N}_{\alpha\eta}^2 
    + 2\mJ \cdot {\bf S}_{\alpha\eta} \cdot {\bf S}_{\ovl\alpha\ovl\eta} \bigg) \ ,
\end{align}}%
where $\tau$ is the imaginary time, ${N}_{\alpha\eta}$ and ${\bf S}_{\alpha\eta}$ are respectively the charge and spin operators in the valley $\eta$ and orbital $\alpha$.
The bare flattened interaction is given by $U_1\!=\!U\!-\!\mJ$, $U_{2,3,4}\!=\!U$, but under renormalization, the values of $U_{1,2,3,4}$ and $\mJ$ flow. 
The inter-valley $d$-wave singlet (triplet) has the energy $U_1 \!- \! (+)  \mJ$.  

\paragraph{Quasi-particles in heavy Fermi liquid.}
In the Fermi liquid phase, the local Green's function has a quasi-particle part $z / [i\omega \!-\! \td{\ee}_f \!+\! i \td{\Delta}_0 \sgn(\omega) ]$ and a featureless incoherent part. 
Here $z \!=\! (1 - \partial_{i\omega}\Sigma(\omega))^{-1}|_{\omega=0}$ is the quasi-particle weight with $\Sigma(\omega)$ denoting the self-energy, 
$\td{\ee}_f = z(\ee_f + \Sigma(0))$ is the renormalized on-site energy, and $\td{\Delta}_0 \!=\!z\Delta_0 \!\sim\! T_{\rm K}$ is the renormalized hybridization.
A typical $T_{\rm K}$ is given by $D(\frac{32\Delta_0}{U})^{\frac18}\exp(-\frac{U}{32\Delta_0})$ if $J_{\rm A,H}\!=\!0$ \cite{coleman_introduction_2015}, where  $D$ is the bandwidth, and $T_{\rm K}$ will be further suppressed by finite $J_{\rm A,H}$. 
In this work we regard $T_{\rm K}\!\sim$1-10K, $\nu_f\!\approx\!-2$, and  $z\!\sim$0.1-0.3 \cite{zhou_kondo_2023,rai_dynamical_2023} as given quantities.
The ratio $\td{\ee}_f/\td{\Delta}_0 \!=\! \cot\delta_f$ is fixed by the occupation of $f$-electrons via the Friedel sum rule \cite{shiba_korringa_1975}, with $\delta_f \!=\! \pi (\nu_f+4)/8\approx\frac{\pi}4$. 

It is convenient to define the quasi-particle operator $\td{f} \!=\! z^{-\frac12} f$ \cite{hewson_renormalized_1993,hewson_renormalized_2001}, the interactions among which are given by $\td \Gamma \!=\! z^2 \Gamma$, with $\Gamma$ being the local 1PI vertex (\cref{fig:diagram-main}(b)). Due to the $\rm U(1)^{\times 4}\!\times\! SU(2)^{\times 2}$ symmetry, $\td\Gamma$ is parametrized in the same form as \cref{eq:SI-main}, and we denote the corresponding parameters (renormalized interactions) as $\td U_{1,2,3,4}$ and $\td\mJ$.

\paragraph{Renormalized interaction in the $\td\Delta_0 \!\ll\! \mJ \!\ll\! U$ limit.}
In this limit the Kondo temperature $T_{\rm K}\!\sim\! \td\Delta_0$  defines the {\it single} energy scale of the local Fermi liquid \cite{nozieres_kondo_1980,coleman_heavy_2007,hewson_fermi_1993}. 
Thus, the renormalized interactions $\td{U}_{1,2,3,4}$, $\td\mJ$ can be expressed in terms of $\td\Delta_0$. 

To derive $\td{U}_{1,2,3,4}$, $\td\mJ$, we make use of the Ward identities \cite{yamada_perturbation_1975,yamada_perturbation_1975-1,yoshimori_perturbation_1976} given by the $\rm U(1)^{\times 4}\!\times\! SU(2)^{\times 2}$ symmetry. 
They bridge the static susceptibilities $\chi^O$ of conserved charges $O$ to the renormalized interaction $\td\Gamma$ \cite{SM}
{\small
\begin{align} \label{eq:Ward-main}
\chi^O 
= \frac{\sin^2\delta_f}{\pi\td{\Delta}_0} \left[ \sum_{I} O_{I}^2 - \frac{\sin^2\delta_f}{\pi\td{\Delta}_0} \sum_{I, I'} \td\Gamma_{I,I';I',I} O_{I} O_{I'} \right] \ ,
\end{align}
}
where $O = \sum_{I} O_I f^\dagger_{I} f_{I}$ is chosen diagonal. Setting $O_I$ to be the electric charge $\sigma^0\tau^0\spin^0$ ($l=c$), spin  $\spin^z$ ($s$), valley  $\tau^z$ ($v$), and orbital $\sigma^z$ ($o$) operators and exploiting $\delta_f = \pi/4$, we obtain 
$\chi^{l} \!=\! \frac{4}{\pi \td{\Delta}_0} ( 1 -  \frac{1}{2\pi \td{\Delta}_0} \td{A}^{l} )$, with
$\td{A}^c \!=\! 2\td{U}_1 + 2\td{U}_2 + 2\td{U}_3 + \td{U}_4 + \td{\mJ}$,
$\td{A}^s \!=\! -\td{U_4} + \td{\mJ}$, 
$\td{A}^v \!=\! -2\td{U}_1 - 2\td{U}_2 + 2\td{U}_3 + \td{U}_4 - \td{\mJ}$,
$\td{A}^o \!=\! -2\td{U}_1 + 2\td{U}_2 - 2\td{U}_3 + \td{U}_4 - \td{\mJ}$, respectively. 
For the $d$-wave ground states (\cref{fig:interaction}(a)), since the electric charge, spin, valley, and orbital degrees of freedom are frozen, {\it i.e.}, they are constants in the two-fold ground state manifold, the corresponding susceptibilities are not contributed by the low-energy quasi-particles \cite{hewson_renormalized_1993,hewson_fermi_1993,nishikawa_renormalized_2010}.
Therefore, $\chi^{c,s,v,o}$ will not diverge as the quasi-particle density of states ($\sim\!\td\Delta_0^{-1}$) in the $\td\Delta_0\!\to\!0$ limit, which implies constraints $\td{A}^{c,s,v,o} = 2\pi\td{\Delta}_0$. 
Consequently, only one unknown parameter is left, which we choose as $\td{\mJ}$, and others are solved as
{\small
\begin{equation} \label{eq:renormalized-int-main}
\td{U}_1 = -2\pi \td\Delta_0 ,\quad 
\td{U}_{2,3} = 2\pi \td\Delta_0 - \frac{\td\mJ}2,\quad 
\td{U}_4 = -2\pi \td\Delta_0 + \td{\mJ}\ . 
\end{equation}}
$\td{U}_1$ has been determined to be negative, hence at least one of the renormalized pairing channels with energies, $\td U_1 \mp \td \mJ$ (given after \cref{eq:SI-main}), must be negative. 
Therefore, we have proven that the renormalized interaction must possess an attractive channel at $\nu_f\!\approx\!- 2$ in the $\td{\Delta}_0 \!\ll\! \mJ$ limit. 
Susceptibilities of other quantities suggest $\td\mJ = k \td\Delta_0$ with $k$ being a constant that ranges from 4.6 to 10.3 \cite{SM}. 
In this region, the inter-valley $d$-wave singlet pairing is attractive and more favored than other channels.

\paragraph{Renormalized interaction in the $\mJ \!\ll\! \td\Delta_0  \!\ll\! U$ limit.}
In this limit, $\mJ$ plays a minor role in the local Fermi liquid, and all two-electron states equally participate in the Kondo screening \cite{nozieres_kondo_1980}. 
With an approximate U(8) symmetry, $\td{U}_{1,2,3,4}$ remain equal under the renormalization, whereas $\td\mJ$ remains negligible. 
The U(8) Ward identity leads to $\td U_{1,2,3,4} \!=\! \frac{2\pi}7 \td\Delta_0$ \cite{nishikawa_renormalized_2010}. 

The universality is lost in the intermediate regime ($\td\Delta_0 \!\sim\! \mJ$) where various two-electron states participate in the Kondo screening with unequal weights. 
However, the behavior of $d$-wave pairing strength $\td{U}_1 \!-\! \td\mJ$ can be inferred by an interpolating sketch between the two limits (\cref{fig:interaction}(e)).
With a decreasing $\td{\Delta}_0$, $\td U_1 \!-\! \td\mJ$ should turn negative when $\td{\Delta}_0$ reaches the order of $\mJ$; when $\td{\Delta}_0$ is further lowered, $\td U_1 \!-\! \td\mJ$ must evolve non-monotonously to achieve the $\td{\Delta}_0\ll \mJ$ limit where $\td U_1 \!-\! \td\mJ$ vanishes linearly in $\td\Delta_0$. 
This suggests the existence of an optimal $\td\Delta_0$ for pairing. With the filling factor $\nu_f$ and the other bare parameters fixed, an increasing $U$ typically suppresses the $f$-charge fluctuation and hence reduces $\td{\Delta}_0$ \cite{coleman_introduction_2015}. 
Therefore, \cref{fig:interaction}(e) also suggests the existence of an optimal $U$ for pairing, as observed in $\rm A_{3} C_{60}$ \cite{capone_colloquium_2009}.

\begin{figure}[t]
\centering
\includegraphics[width=\linewidth]{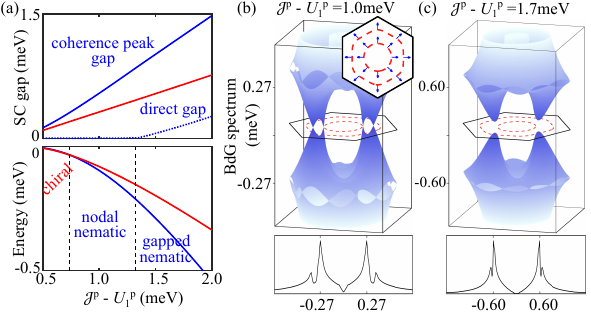}
\caption{Mean-field calculations of SC at $\nu\!=\!-\!2.5$ and $z=0.3$.
(a) Phase diagram of inter-valley inter-orbital $d$-wave singlet SC. 
The range of $\mJ^{\rm p} \!-\! U_1^{\rm p}$ corresponds to $T_{\rm K}$ ranging from 1.7K to 7.1K if $k\!=\!8$. 
(b) BdG bands of the nematic $d$-wave SC with a $p$-wave-like nodal structure.  The inset shows the $C_{2z}T$ sewing matrix phase $\phi_{\kk}$ on the Fermi surface. 
(c) BdG bands of the gapped nematic $d$-wave SC, where the gap function is still highly anisotropic.
The lower panels in (b) and (c) are the corresponding densities of states calculated using a Lorentz spread 0.004meV. 
} 
\label{fig:mf}
\end{figure}

\paragraph{Quasi-particle mean-field theory.}
We now investigate SC on the moir\'e lattice (THF model) using a mean-field theory with the effective interaction $z^2\Gamma^{\rm p}$ (local 2PI vertex) and the renormalized quasi-particle spectrum (\cref{fig:model}(c)). 
Through the ladder summation (\cref{fig:diagram-main}(b)), we find that $z^2\Gamma^{\rm p}$ has the same pairing channel as the local 1PI $z^2\Gamma$ but with a weaker potential $U_1^{\rm p} - \mJ^{\rm p} = - \frac{(2\pi+k)}{ 1 + \frac{2\pi+k}4 }\td\Delta_0$ in the $\td\Delta_0\ll \mJ$ limit \cite{SM}.

We carry out the calculation at $\nu\!=\!-2.5$ using $z\!=\!0.3$ and $\mJ^{\rm p} \!-\! U_1^{\rm p}$ in the range from 0.5 to 2meV (\cref{fig:mf}). 
Including the (much weaker) non-local interactions do not affect the results \cite{SM}. 
Since the $d$-wave pairings form the two-dimensional representation $E_2$, we find two possible phases.
One is a gapped chiral $d$-wave pairing $\td f_{\kk\alpha+\up}^\dagger \td f_{-\kk\ovl{\alpha}-\down}^\dagger \!-\! (\up\leftrightarrow\down)$ (for either $\alpha$=1 or 2). 
The other is a nematic $d$-wave pairing \cite{wu_theory_2018,liu_electron-k-phonon_2023,blason_local_2022,lothman_nematic_2022}
$ e^{-i\varphi} \td f_{\kk 1+\up}^\dagger \td f_{-\kk 2-\down}^\dagger 
\!+\! e^{i\varphi} \td f_{\kk 2+\up}^\dagger \td f_{-\kk 1-\down}^\dagger \!-\! (\up\leftrightarrow\down)$ that breaks the $C_{3z}$ symmetry. 
Here $\varphi$ sets the orientation of the nematic order. 
When $\mJ^{\rm p} \!-\! U_1^{\rm p}<$0.7meV, the chiral state has a slightly lower energy than the nematic state. 
When $\mJ^{\rm p} \!-\! U_1^{\rm p}>$0.7meV, the nematic state has a significantly lower energy than the chiral state. 

\paragraph{$p$-wave-like nodal SC.} 
An intermediate pairing strength leads to a $p$-wave-like nodal structure, as shown in \cref{fig:mf}(b). 
We now prove that the 2 (mod 4) nodes on each Fermi surface (FS) are guaranteed by the $\pi$ Berry's phase protected by $C_{2z}T$ symmetry. 
Suppose $\psi_{\kk + s}$ is the annihilation operator for Bloch states on a given FS in the $\eta\!=\!+$ valley, and $(C_{2z}T) \psi_{\kk + s}^\dagger (C_{2z}T)^{-1} \!=\! \psi_{\kk + s}^\dagger e^{i \phi_\kk}$.
Due to $(C_{2z}T) \td f_{\kk \alpha\eta s}^\dagger (C_{2z}T)^{-1} \!=\! \td f_{\kk \ovl\alpha\eta s}^\dagger$ \cite{song_magic-angle_2022,SM}, there must be $\psi^\dagger_{\kk + s}\sim \td f^\dagger_{\kk 1+s}  + e^{-i\phi_\kk} \td f^\dagger_{\kk 2+s} $.
Bloch states in the $\eta\!=\!-$ valley can be obtained by applying the time-reversal: $\psi^\dagger_{-\kk - s} \!\sim\! \td f^\dagger_{-\kk 1 - s}  \!+ e^{i\phi_\kk} \td f^\dagger_{-\kk 2-s} $.
Projecting the nematic $d$-wave pairing onto the FS, we obtain 
$ \cos(\phi_\kk + \varphi) \cdot \psi_{\kk+\up}^\dagger \psi_{-\kk-\down}^\dagger - (\up\leftrightarrow\down)$.
As the FS encloses an odd number of Dirac points (\cref{fig:model}(c)), $\phi_\kk$ must wind an odd ($2n\!+\!1$) multiple of $2\pi$ along the FS (\cref{fig:mf}(b)) \cite{ahn_failure_2019}, leaving $4n+2$ nodes at $\phi_\kk+\varphi=\pm\frac{\pi}2$. 
As detailed in Ref.~\cite{SM}, an alternative understanding of the pairing nodes is the Euler obstruction \cite{yu_euler_2022}.

As the pairing becomes stronger, nodes on the two FSs will merge, leading to a gapped phase (\cref{fig:mf}(c)). 
Spectrum of the gapped nematic SC remains highly anisotropic if the direct gap is significantly smaller than the pairing. 
Therefore, both the nodal and the gapped nematic $d$-wave SC can have a V-shaped density of states at an energy scale larger than the direct gap (0 in the nodal case). 
This is consistent with the V-shaped spectrum \cite{oh_evidence_2021} and nematicity \cite{cao_nematicity_2021} seen in experiments.

\paragraph{Discussion.} 
Our theory provides insights into the strong coupling features of SC in MATBG. The pairing potential $\mJ^{\rm p} \!-\! U_1^{\rm p}$ is a few times larger than $T_{\rm K}$, and the Fermi energy $E_{\rm F} \!\sim\! T_{\rm K}$ (\cref{fig:model}(c)). 
Therefore, $\mJ^{\rm p} \!-\! U_1^{\rm p} \gtrsim E_{\rm F}$, suggesting the SC is closer to a BEC state than a BCS state \cite{cao_unconventional_2018,lu_superconductors_2019}. Pairings are localized around ``moir\'e molecules" in AA-stacking regions, leading to a smaller phase stiffness - only contributed by hybridization with Dirac electrons in AB regions - compared to BCS pairings of delocalized states \cite{xie_topology-bounded_2020}. 
This may explain the large ratio between the pairing gap and $T_c$ \cite{oh_evidence_2021}.

In the intermediate regime where $T_{\rm K} \!\sim\! J_{\rm A}$, multiplet splittings breaking the $\rm U(1)^{\times 4}\!\times\! SU(2)^{\times 2}$ symmetry should be considered. The remaining Ward identities cannot fully constrain the renormalized interaction. However, the continuity (\cref{fig:interaction}(e)) suggests pairing is still possible. We leave this for future studies.

\begin{acknowledgements}
We are grateful to Xi Dai for helpful discussions about the $\rm A_3C_{60}$ family of superconductors. 
We thank B. Andrei Bernevig and Jiabin Yu for helpful discussions about nodal pairings. 
We also thank  Chang-Ming Yue, Xiao-Bo Lu, and Seung-Sup Lee for useful discussions. 
Z.-D. S., Y.-J. W. and G.-D. Z. were supported by National Natural Science
Foundation of China (General Program No. 12274005), National Key Research and Development Program of China (No. 2021YFA1401900), and Innovation Program for Quantum Science and Technology (No. 2021ZD0302403). B. L. is supported by the National Science Foundation under award DMR-2141966, and the National Science Foundation through Princeton University's Materials Research Science and Engineering Center DMR-2011750.
\end{acknowledgements}

\clearpage

\onecolumngrid

\tableofcontents
\appendix 

\footnotetext{Sections marked with a * symbol provide background reviews. Calculations supporting the results in the main text are given in sections without the * symbol.}

\clearpage

\section{Local interaction}
\label{sec:interaction}

\subsection{Local (moir\'e) orbitals *}

Here we summarize the relevant local interactions in a single AA-stacking region of magic-angle twisted bilayer graphene (MATBG). 
There are eight effective local $f$-orbitals in an AA-stacking region \cite{song_magic-angle_2022}: $f_{\alpha,\eta,s}$, where $\alpha=1,2$ is the orbital index, $\eta=\pm$ is the valley index, and $s=\up,\down$ is the spin index. 
A single AA-site has the time-reversal symmetry ($T$) and symmetries of the $D_6$ point group. 
The single-particle representations of these discrete symmetries are given in section S2A of the supplementary material of Ref.~\cite{song_magic-angle_2022}. 
They are 
\begin{equation} \label{eq:discrete-sym}
T = \sigma^0\tau^x\spin^0 K,\qquad 
C_{2z} = \sigma^x\tau^x\spin^0,\qquad 
C_{3z} = e^{i\frac{2\pi}3\sigma^z\tau^z\spin^0},\qquad 
C_{2x} = \sigma^x\tau^0\spin^0\ , 
\end{equation}
where  $K$ is the complex conjugation. 
Their actions on the second quantized operators can be obtained directly
\begin{equation} \label{eq:sym_f}
T f_{\alpha\eta s}^\dagger T^{-1} = f^\dagger_{\alpha \ovl{\eta} s},\qquad  
C_{2z} f_{\alpha\eta s}^\dagger C_{2z}^{-1} = f_{\ovl{\alpha} \ovl{\eta} s}^\dagger,\qquad 
C_{3z} f_{\alpha\eta s}^\dagger C_{3z}^{-1} =  e^{i\frac{2\pi}3\eta (-1)^{\alpha-1}} f_{\alpha\eta s}^\dagger ,\qquad 
C_{2x} f_{\alpha\eta s}^\dagger C_{2x}^{-1} =  f^\dagger_{\ovl{\alpha} \eta s}\ .
\end{equation}
In this work we {\it do not} distinguish the single-particle and the second-quantized representations of symmetry operators. 
If we write a unitary (anti-unitary) symmetry operator $g$ as a matrix $D^g$ ($D^g \cdot K$), as exampled in \cref{eq:discrete-sym}, then its action on second-quantized operators is defined by 
\begin{equation}
    g f^{\dagger}_{\alpha\eta s} g^{-1} =  \sum_{\alpha'\eta's'} D_{\alpha'\eta's',\alpha\eta s}^g f_{\alpha'\eta's'}^\dagger,\qquad 
    g f_{\alpha\eta s} g^{-1} = \sum_{\alpha'\eta's'} D_{\alpha'\eta's',\alpha\eta s}^{g*} f_{\alpha'\eta's'}\ . 
\end{equation}

We denote the fermion creation operator of the $p_z$ orbital at the carbon atom  belonging to the layer $l$ (=$+$ for the top layer and $-$ for the bottom layer) by $\psi_{l \rr s}^\dagger$.
Here $s$ is the spin index, $\rr \in \mathcal{L}_{l,1} + \mathcal{L}_{l,2}$ is the position of the atom, and $\mathcal{L}_{l,\beta}$ is the $\beta$-sub-lattice of the layer $l$, where we associate the A-, B-sub-lattices of graphene to $\beta=1$ and 2, respectively. 
Projected $\psi_{l \rr s}^\dagger$ can be written in terms of the $f$ operators as 
\begin{equation} \label{eq:graphene-basis}
\psi_{l\rr s}^\dagger = \sqrt{\Omega_{\rm G}} \sum_{\eta \alpha}  \sum_{\beta,  \mathcal{L}_{l,\beta} \ni \rr}  e^{-i\eta \mathbf{K}_l \cdot \rr} w_{l\beta,\alpha}^{(\eta)*} (\rr) f_{\alpha\eta s}^\dagger\ .
\end{equation}
Here $\pm \mathbf{K}_l$ are the momenta of Dirac points in the layer $l$, 
$\mathbf{K}_{+} = \frac{4\pi}{3a_{\rm G}} ( \cos\frac{\theta}2 , \sin\frac{\theta}2 )$, 
$\mathbf{K}_{-} = \frac{4\pi}{3a_{\rm G}} ( \cos\frac{\theta}2 , -\sin\frac{\theta}2 )$, 
with $a_{\rm G} = 2.46\mathring{\rm A}$ being the graphene lattice constant and $\theta\approx 1.08^\circ$  the magic twist angle. 
The length of $\mathbf{K}_l$ is given by $|\mathbf{K}_l| = \frac{4\pi}{3 a_{\rm G}} \approx 1.703\mathring{\rm A}^{-1}$. 
$\Omega_{\rm G} = \frac{\sqrt3}2 a_{\rm G}^2$ is the area of a graphene unit cell. 
$w_{l\beta,\alpha}^{(\eta)}(\rr)$ is the localized Wannier functions constructed in Ref.~\cite{song_magic-angle_2022}.  
The summation over the sub-lattice index $\beta$ on the right hand side is limited to the one containing $\rr$ on the left hand side, {\it i.e.}, $ \mathcal{L}_{l,\beta} \ni \rr $. 
Section S2A of the supplementary material of Ref.~\cite{song_magic-angle_2022}  provides a Gaussian approximation for the Wannier functions:
\begin{equation} \label{eq:WF-form1}
w_{l1,1}^{(\eta)}(\rr) = \frac{\alpha_1}{\sqrt2} \frac{1}{\sqrt{\pi\lambda_1^2}} e^{i \frac{\pi}4 l\eta  -\rr^2/(2\lambda_1^2)},\quad 
w_{l2,1}^{(\eta)}(\rr) = -l \frac{\alpha_2}{\sqrt2} \frac{x+i\eta y}{\lambda_2^2 \sqrt{\pi}} e^{i \frac{\pi}4 l\eta -\rr^2/(2\lambda_2^2)}, 
\end{equation}
\begin{equation} \label{eq:WF-form2}
w_{l1,2}^{(\eta)}(\rr) = l\frac{\alpha_2}{\sqrt2} \frac{x-i\eta y}{\lambda_2^2 \sqrt{\pi}} e^{-i \frac{\pi}4 l\eta - \rr^2/(2\lambda_2^2)},\qquad 
w_{l2,2}^{(\eta)}(\rr) = \frac{\alpha_1}{\sqrt2} \frac{1}{\sqrt{\pi\lambda_1^2}} e^{-i \frac{\pi}4 l\eta - \rr^2/(2\lambda_1^2)}\ .
\end{equation}
At the magic-angle, the parameters are estimated as 
\begin{equation} \label{eq:WF-parameter}
\alpha_1 = 0.8193,\qquad \alpha_2 = 0.5734,\qquad \lambda_1=0.1791 a_{\rm M}, \qquad \lambda_2 = 0.1910 a_{\rm M}\ . 
\end{equation}
where $a_{\rm M} =  \frac{4\pi}{3 k_{\theta} } = \frac{a_{\rm G}}{2\sin\frac{\theta}2} \approx 54.57 a_{\rm G} \approx 134.2\mathring{\rm A}$ is the moir\'e lattice constant.

\subsection{Coulomb interaction and intra-orbital Hund's coupling}
\label{sec:Hund}

The widely studied Coulomb interaction in MATBG is 
\begin{equation}
H_{I1} = \frac12 \sum_{\eta\eta'} \sum_{\rr} \sum_{\rr'}  V(\rr-\rr')  \brak{ \rho^{(\eta \eta)}(\rr)   \rho^{(\eta' \eta')}(\rr')   + \rho^{(\eta \ovl\eta)}(\rr)  \rho^{(\ovl\eta \eta')}(\rr')  }\ ,
\end{equation}
where 
\begin{align}
    \rho^{(\eta\eta')}(\rr) =& \Omega_{\rm G} \sum_{ \alpha \alpha' s} \sum_{l} \sum_{\beta, \mathcal{L}_{l \beta}\ni \rr} 
e^{-i(\eta  \mathbf{K}_l -\eta'  \mathbf{K}_l)\cdot \rr }
w_{l\beta, \alpha}^{(\eta)*}(\rr) w_{l\beta,\alpha'}^{(\eta')}(\rr) 
f_{\alpha\eta s}^\dagger f_{\alpha'\eta' s} 
\end{align}
and
\begin{equation}
    V(\rr) = U_{\xi} \sum_{n=-\infty}^\infty \frac{(-1)^n}{ \sqrt{ (\rr/\xi)^2 + n^2 } }
\end{equation}
is the double-gate-screened Coulomb interaction. 
Here $\xi$ is the distance between the two gates, $U_{\xi}=e^2/(4\pi \epsilon \xi)$, and $\epsilon\approx6$ is the dielectric constant. 
For $\xi=10$nm, there is $U_{\xi}=24$meV. 
This interaction respects a valley-U(1) symmetry. 
Bilinear terms due to the normal order form of operators are omitted because they can be absorbed into the chemical potential for the single-site problem. 
The Fourier transformation of $V(\rr)$ is 
\begin{equation}
    V(\rr) = \int \frac{d^2\mathbf{q}}{(2\pi)^2} V(\mathbf{q}) e^{-i \mathbf{q}\cdot\rr},\qquad 
    V(\mathbf{q}) = \pi \xi^2 U_\xi \frac{\tanh(\xi |\mathbf{q}|/2)}{\xi |\mathbf{q}|/2}\ . 
\end{equation}
The projected intra-valley scattering interaction between $\rho^{(\eta\eta)}$ and $\rho^{(\eta'\eta')}$ has been calculated in section S3B of the supplementary material of Ref.~\cite{song_magic-angle_2022}. 
It has the form
\begin{equation}
    H_{I1} = \frac{U}2 \sum_{\alpha\eta s} \sum_{\alpha'\eta' s'}
    f_{\alpha\eta s}^\dagger   f_{\alpha'\eta's'}^\dagger f_{\alpha'\eta's'} f_{\alpha\eta s} = U\frac{\hat N ( \hat N-1)}2, \qquad U\approx58\mathrm{meV}\ , 
\end{equation}
where $\hat N$ is the particle number operator. 
There is {\it no} other term, {\it e.g.}, Hund's coupling, contributed by the intra-valley scattering due to the symmetry of MATBG. 
One can see discussions around Eq. (S135) of the supplementary material of Ref.~\cite{song_magic-angle_2022} for the proof. 

Now we consider the inter-valley scattering interaction between $\rho^{(\eta \ovl\eta)}$ and $\rho^{(\ovl\eta\eta)}$.
Due to the large momentum transfer $\sim 2\mathbf{K}_l$, this interaction is strongly suppressed and usually neglected. 
Here we discuss it in more details. 
The relevant Fourier component for this interaction is $V(\pm 2\mathbf{K}_l + \mathbf{q}) \approx V(\pm 2\mathbf{K}_l) = \frac{\pi \xi U_{\xi}}{ |\mathbf{K}_l | } \approx 4.4 \times 10^{3} \mathrm{meV \!\cdot\! \mathring{A}^2}$. 
As $V(\pm 2\mathbf{K}_l+\mathbf{q})$ is almost $\mathbf{q}$-independent, it gives a $\delta$-like interaction on the microscopic graphene lattice. 
The projected interaction on the moir\'e orbitals can be estimated as $ \frac{4427 \mathrm{meV  \cdot \mathring{A}^2}} { \Omega_{\rm M}} \approx 0.3$meV, with $\Omega_{\rm M} = \frac{\sqrt3}2 a_{\rm M}^2 \approx 1.560\times 10^{4} \mathring{\rm A}^2$ being the moir\'e unit cell area. 
Therefore, this inter-valley scattering interaction contributed by $V(\mathbf{r})$ is indeed much weaker compared to the intra-valley one. 

Another usually omitted Coulomb interaction is the on-site Hubbard repulsion $U_0$ of the $p_z$ orbital of carbon atom. 
$U_{0}$ is estimated as large as 9.3eV \cite{wehling_strength_2011}, which, however, is still smaller than the critical values for the spin-liquid phase ($3.5t$) and anti-ferromagnetic phase ($(4.5\pm0.5)t$) of graphene as the hopping $t\approx2.8$eV is also large. 
We find that $U_0$ leads to a non-negligible Hund's coupling in MATBG. 
The microscopic interaction is
\begin{equation} \label{eq:HI2}
    H_{I2} = \frac{U_0}2 \sum_{ l ss'} \sum_{\rr} \psi_{l \rr s}^\dagger \psi_{l \rr s'}^\dagger \psi_{l\rr s'} \psi_{l\rr s}\ . 
\end{equation}
Projecting it onto the $f$-orbitals, we obtain 
{\small
\begin{align}
    H_{I2} =& \frac{U_0 \Omega_{\rm G}^2}2 \sum_{l\beta } \sum_{\rr \in \mathcal{L}_{l\beta}} \sum_{\alpha_{1,2,3,4}} \sum_{\eta_{1,2,3,4}}
    \delta_{\eta_1+\eta_2,\eta_3+\eta_4}
    w_{l\beta,\alpha_1}^{(\eta_1)*}(\rr) w_{l\beta,\alpha_2}^{(\eta_2)*}(\rr)
    w_{l\beta,\alpha_3}^{(\eta_3)}(\rr) w_{l\beta,\alpha_4}^{(\eta_4)}(\rr)
    f_{\alpha_1\eta_1 s}^\dagger  f_{\alpha_2\eta_2 s'}^\dagger  
    f_{\alpha_3\eta_3 s'}  f_{\alpha_4\eta_4 s} \nonumber\\
=& \frac{U_0 \Omega_{\rm G}}2 \sum_{l\beta } \sum_{\alpha_{1,2,3,4}} \sum_{\eta_{1,2,3,4}}
    \delta_{\eta_1+\eta_2,\eta_3+\eta_4} 
    f_{\alpha_1\eta_1 s}^\dagger  f_{\alpha_2\eta_2 s'}^\dagger  
    f_{\alpha_3\eta_3 s'}  f_{\alpha_4\eta_4 s}
    \int d^2\rr\ 
    w_{l\beta,\alpha_1}^{(\eta_1)*}(\rr) w_{l\beta,\alpha_2}^{(\eta_2)*}(\rr)
    w_{l\beta,\alpha_3}^{(\eta_3)}(\rr) w_{l\beta,\alpha_4}^{(\eta_4)}(\rr)\ . 
\end{align}}
Notice that $U_0 \Omega_{\rm G} \approx 4.7\times 10^{4} \mathrm{meV\!\cdot\!\mathring{A}^2}$ ($U_0$ taken as 9eV) is one order larger than $V(\pm \mathbf{K}_l)$. 
As the Wannier functions have the $C_{3z}$ eigenvalue $e^{i\frac{2\pi}3 \eta (-1)^{\alpha-1}}$, quasi-angular momentum conservation gives the constraint 
\begin{equation}
\eta_1 (-1)^{\alpha_1-1} + \eta_2 (-1)^{\alpha_2-1} =
\eta_3 (-1)^{\alpha_3-1} + \eta_4 (-1)^{\alpha_4-1} \mod 3\ . 
\end{equation}
Using the Gaussian wave-functions in \cref{eq:WF-form1,eq:WF-form2} and the valley-U(1), $C_{3z}$ symmetries, we find there are only two types of matrix elements in $H_{I2}$. 
We summarize them in the following table 
\begin{equation}
\begin{tabular}{c|c|c|c|c|c}
    & $\alpha_1$ & $\alpha_2$ & $\alpha_3$ & $\alpha_4$ & $\eta$'s \\
    \hline
    $J_{\rm H}$ & $\alpha_1$ & $\alpha_1$ & $\alpha_1$ & $\alpha_1$ & $\eta_1+\eta_2=\eta_3+\eta_4$   \\
    \hline
    $J_{\rm H}'$ & $\alpha_1 $& $\alpha_1$ & $\ovl \alpha_1$ & $\ovl \alpha_1$ & $\eta_1=\ovl\eta_2$, $\eta_3=\ovl \eta_4$ \\
    \hline 
    $J_{\rm H}'$ & $\alpha_1$ & $\ovl\alpha_1$ & $\ovl\alpha_1$ & $\alpha_1$ & $\eta_1=\eta_4$, $\eta_2=\eta_3$ \\
    \hline 
    $J_{\rm H}'$ & $\alpha_1$ & $\ovl\alpha_1$ & $\alpha_1$  & $\ovl\alpha_1$ & $\eta_1=\eta_3$, $\eta_2=\eta_4$ 
\end{tabular}  \ .
\end{equation}
The two parameters are 
\begin{align}
J_{\rm H} =& U_0 \Omega_{\rm G} \sum_{l\beta} \int d^2\rr |w_{l\beta,\alpha_1}^{(\eta_1)}(\rr)|^2 |w_{l\beta,\alpha_1}^{(\eta_2)}(\rr)|^2
= 2U_0 \Omega_{\rm G} \int d^2\rr  \brak{\pare{\frac{\alpha_1^2}{2\pi\lambda_1^2}}^2 e^{-2\frac{\rr^2}{\lambda_1^2}} + \pare{\frac{\alpha_2^2}{2\pi\lambda_2^4}}^2 \rr^4 e^{-2\frac{\rr^2}{\lambda_2^2}}  } \nonumber\\
=& U_0 \Omega_{\rm G} \brak{\frac{\alpha_1^4}{4\pi \lambda_1^2} + \frac{\alpha_2^4}{8\pi \lambda_2^2} }\ ,
\end{align}
\begin{align}
J_{\rm H}' =& U_0 \Omega_{\rm G} \sum_{l\beta} \int d^2\rr |w_{l\beta,\alpha_1}^{(\eta_1)}(\rr)|^2 |w_{l\beta,\ovl\alpha_1}^{(\eta_3)}(\rr)|^2
= 4U_0 \Omega_{\rm G} \int d^2\rr  \brak{\pare{\frac{\alpha_1^2}{2\pi\lambda_1^2}} e^{-\frac{\rr^2}{\lambda_1^2}} \times \pare{\frac{\alpha_2^2}{2\pi\lambda_2^4}} \rr^2 e^{-\frac{\rr^2}{\lambda_2^2}}  } \nonumber\\
=& U_0 \Omega_{\rm G} \frac{\alpha_1^2\alpha_2^2\lambda_1^2}{\pi(\lambda_1^2+\lambda_2^2)^2}\ . 
\end{align}
The projected $H_{I2}$ can be written as 
\begin{align} \label{eq:HI2-temp1}
H_{I2} =&  \sum_{\alpha ss'} \sum_{\eta_{1,2,3,4}}  \delta_{\eta_1+\eta_2,\eta_3+\eta_4} 
    \bigg[  \frac{J_{\rm H}}2 f_{\alpha\eta_1 s}^\dagger  f_{\alpha\eta_2 s'}^\dagger  
    f_{\alpha\eta_3 s'}  f_{\alpha\eta_4 s}  
    + \frac{J_{\rm H}'}2 \delta_{\eta_1\ovl\eta_2} \delta_{\eta_3\ovl\eta_4} \cdot 
    f_{\alpha \eta_1 s}^\dagger  f_{ \alpha \ovl\eta_1 s'}^\dagger      f_{\ovl \alpha \ovl \eta_4 s'}  f_{\ovl \alpha\eta_4 s} 
    \nonumber\\
    & \qquad\qquad + \frac{J_{\rm H}'}2 \delta_{\eta_1\eta_4} \delta_{\eta_2\eta_3} \cdot f_{\alpha\eta_1 s}^\dagger  f_{\ovl\alpha\eta_2 s'}^\dagger  
    f_{\ovl \alpha\eta_2 s'}  f_{\alpha\eta_1 s} 
+  \frac{J_{\rm H}'}2 \delta_{\eta_1\eta_3} \delta_{\eta_2\eta_4} \cdot f_{\alpha\eta_1 s}^\dagger  f_{\ovl\alpha\eta_2 s'}^\dagger  
    f_{ \alpha\eta_1 s'}  f_{\ovl \alpha\eta_2 s} \bigg] \ . 
\end{align}
The parameters in \cref{eq:WF-parameter} give $J_{\rm H} = 0.3590\times 10^{-3} U_0$, and $\mJ'_{\rm H}=0.1392\times 10^{-3} U_0$.
We can also calculate the two parameters using the numerical Wannier functions constructed in Ref.~\cite{song_magic-angle_2022}, which give 
\begin{equation} \label{eq:JHunds-U0}
    J_{\rm H} \approx 0.3284\times 10^{-3} U_0,\qquad J_{\rm H}' \approx 0.1029\times 10^{-3} U_0 \approx \frac13 J_{\rm H} \ . 
\end{equation}
One can see that the intra-orbital Hund's coupling is stronger than the inter-orbital Hund's coupling.

We can rewrite the $J_{\rm H}$ term in \cref{eq:HI2-temp1} as 
\begin{align}
& \sum_{\alpha ss'} \sum_{\eta_{1,2}} \brak{ \frac{J_{\rm H}}2 f_{\alpha\eta_1 s}^\dagger  f_{\alpha\eta_2 s'}^\dagger      f_{\alpha\eta_2 s'}  f_{\alpha\eta_1 s}  
+ \frac{J_{\rm H}}2 f_{\alpha\eta_1 s}^\dagger  f_{\alpha\eta_2 s'}^\dagger      f_{\alpha\eta_1 s'}  f_{\alpha\eta_2 s}  } \nonumber\\
=&  - \frac{J_{\rm H}}2 \hat N + \frac{J_{\rm H}}2 \sum_{\alpha} \hat N_{\alpha}^2  + J_{\rm H} \hat N - 
\frac{J_{\rm H}}2  \sum_{\alpha ss'} \sum_{\eta_{1,2}} f_{\alpha\eta_1 s}^\dagger   f_{\alpha\eta_1 s'} f_{\alpha\eta_2 s'}^\dagger      f_{\alpha\eta_2 s}  \ . 
\end{align}
Using the relation $\sum_{\mu=0,x,y,z} \spin^{\mu}_{s_1s_2} \spin^{\mu}_{s_3s_4} = 2\delta_{s_1 s_4} \delta_{s_2s_3}$, the above expression equals
\begin{align}
\frac{J_{\rm H}}2 \hat{N} + \frac{J_{\rm H}}4 \sum_{\alpha} \hat{N}_{\alpha}^2 - J_{\rm H} \sum_{\alpha} \hat{\mathbf{S}}_\alpha\cdot \hat{\mathbf{S}}_\alpha \ ,
\end{align}
where
\begin{equation}
    \hat N_{\alpha} = \sum_{\eta s} f_{\alpha \eta s}^\dagger f_{\alpha\eta s}, \qquad 
    \hat{\mathbf{S}}_{\alpha} = \frac12 \sum_{\eta ss'} f_{\alpha \eta s}^\dagger \boldsymbol{\spin}_{ss'} f_{\alpha\eta s'}\ ,
\end{equation}
and $N=\sum_{\alpha} N_{\alpha}$. 
Similarly, we can rewrite the third and forth terms in \cref{eq:HI2-temp1} as 
\begin{equation}
    \frac{J_{\rm H}'}4 \sum_{\alpha} \hat N_{\alpha} \hat N_{\ovl\alpha} - J_{\rm H}' \sum_{\alpha} \hat{\bf S}_{\alpha} \cdot \hat{\bf S}_{\ovl \alpha}\ . 
\end{equation}
However, the second term in \cref{eq:HI2-temp1} cannot be written in terms of charge and spin operators. 
In summary, $H_{I2}$ can be written as 
\begin{align}
H_{I2} = \frac{J_{\rm H}}2 \hat N + \sum_{\alpha} \brak{ \frac{J_{\rm H}}4 \hat{N}_{\alpha}^2 + \frac{J_{\rm H}'}4 \hat{N}_{\alpha} \hat{N}_{\ovl\alpha} - J_{\rm H} \hat{\bf S}_{\alpha}^2 - J_{\rm H}' \hat{\bf S}_{\alpha} \cdot \mathbf{S}_{\ovl \alpha}   }  
  + \frac{J_{\rm H}'}2 \sum_{\alpha ss'\eta\eta'}
      f_{\alpha \eta s}^\dagger  f_{ \alpha \ovl\eta s'}^\dagger      f_{\ovl \alpha \ovl \eta' s'}  f_{\ovl \alpha\eta' s} \ . 
\end{align}

\subsection{\texorpdfstring{$A_1$, $B_1$}{A1, B1}-phonon mediated inter-valley anti-Hund's coupling *}

Ref.~\cite{wu_theory_2018} studied the $A_1$, $B_1$ phonon mediated inter-valley attractive interaction
\begin{equation}
    H_{I3} = - g \sum_l \sum_{\eta \eta' ss'} \sum_{\beta\beta'} \int d^2\rr 
    \pare{ 1 - \eta \eta' }
    \psi_{l \beta\eta s}^\dagger (\rr) \psi_{l \beta' \eta' s'}^\dagger(\rr) 
    \psi_{l \ovl\beta' \ovl\eta' s'}(\rr) \psi_{l \ovl\beta \ovl\eta  s} (\rr)  \ . 
\end{equation}
$\psi_{l\beta\eta s}^\dagger(\rr)$ is a continuous version of \cref{eq:graphene-basis} that is limited to the valley $\eta$ and sub-lattice $\beta$
\begin{equation}
    \psi_{l\alpha\eta s}^\dagger (\rr) = \sum_{\alpha}   e^{-i\eta \mathbf{K}_l \cdot \rr} w_{l\beta,\alpha}^{(\eta)*} (\rr) f_{\alpha\eta s}^\dagger\ .
\end{equation}
$g= \lambda_{\rm RG} \times 6.9\times 10^3 \mathrm{meV\!\cdot\!\mathring{A}^2}$ is the coupling constant and $\lambda_{\rm RG}\approx 3.2$ is an enhancement factor due to the renormalization effect from higher energy ($\gtrsim200$meV) electron and phonon states \cite{basko_interplay_2008}.
This renormalization effect is also considered in Ref.~\cite{kwan_electron-phonon_2023} to obtain the T-IVC (time-reversal inter-valley coherent) state at the fillings $\nu=\pm2$. 
The projected $H_{I3}$ is calculated in Ref.~\cite{wang_tbg_epc_2024}.
Here we directly give the results
\begin{equation}
    H_{I3} =
     -\frac{J_{\rm d}}{2} \sum_{\alpha\beta \eta s s'} 
    f_{\ovl{\alpha} \ovl{\eta} s}^\dagger f_{\ovl{\beta}\eta s'}^\dagger f_{\beta \ovl{\eta} s'} f_{\alpha\eta s} 
    - \frac{J_{\rm e}}{2} \sum_{\alpha\eta ss'}  
    f_{\alpha\ovl{\eta}s}^\dagger f_{\alpha\eta s'}^\dagger f_{\alpha\ovl{\eta}s'} f_{\alpha\eta s} \ ,
\end{equation}
with $J_{\rm d}=\lambda_{\rm RG} \times 1.3$meV, $J_{\rm e}=\lambda_{\rm RG} \times 1.2$meV. 
One can see that $H_{I3}$ favors inter-valley spin-singlets. 
Thus we call $J_{\rm d,e}$ the inter-valley anti-Hund's couplings. 
As $|J_{\rm d}-J_{\rm e}|\ll J_{\rm d}$, in this work we assume $J_{\rm d}=J_{\rm e}=\lambda_{\rm RG} \times 1.3$meV and  rewrite the interaction  as 
\begin{align} 
H_{I3} =& 
-\frac{J_{\rm A}}{2} \sum_{\alpha\eta s s'} 
   f_{\ovl{\alpha} \ovl{\eta} s}^\dagger f_{\alpha \eta s'}^\dagger f_{ \ovl{\alpha} \ovl{\eta} s'} f_{\alpha\eta s} 
- \frac{J_{\rm A}}2 \sum_{\alpha\eta s s'} 
   f_{\ovl{\alpha} \ovl{\eta} s}^\dagger f_{\ovl\alpha \eta s'}^\dagger f_{ {\alpha} \ovl{\eta} s'} f_{\alpha\eta s} 
- \frac{J_{\rm A}}{2} \sum_{\alpha\eta ss'}  
   f_{\alpha\ovl{\eta}s}^\dagger f_{\alpha\eta s'}^\dagger f_{\alpha\ovl{\eta}s'} f_{\alpha\eta s} \nonumber\\
=& 
-\frac{J_{\rm A}}{2} \sum_{\alpha\beta\eta s s'} 
   f_{\beta \ovl{\eta} s}^\dagger f_{\alpha \eta s'}^\dagger f_{ \beta \ovl{\eta} s'} f_{\alpha\eta s} 
- \frac{J_{\rm A}}2 \sum_{\alpha\eta s s'} 
   f_{\ovl{\alpha} \ovl{\eta} s}^\dagger f_{\ovl\alpha \eta s'}^\dagger f_{ {\alpha} \ovl{\eta} s'} f_{\alpha\eta s} \ ,
\end{align}
$J_{\rm A}=\lambda_{\rm RG} \times 1.3$meV, where the subscript A stands for anti-Hund's coupling. 
Using the relation $\sum_{\mu=0,x,y,z} \spin^{\mu}_{s_1s_2} \spin^{\mu}_{s_3s_4} = 2\delta_{s_1 s_4} \delta_{s_2s_3}$, we can rewrite $H_{I3}$ as 
\begin{equation}
    H_{I3} = \frac{J_{\rm A}}4 \sum_{\eta} \hat N_{\eta} \hat N_{\ovl\eta} + J_{\rm A} \sum_{\eta} \hat {\mathbf{S}}_{\eta} \cdot \hat{\mathbf{S}}_{\ovl \eta} 
    - \frac{J_{\rm A}}{2} \sum_{\alpha\eta ss'}  
   f_{\ovl\alpha\ovl{\eta}s}^\dagger f_{\ovl\alpha\eta s'}^\dagger f_{\alpha\ovl{\eta}s'} f_{\alpha\eta s} \ ,
\end{equation}
with 
\begin{eqnarray}
    \hat{N}_\eta = \sum_{\alpha s} f^\dagger_{\alpha\eta s} f_{\alpha\eta s},\qquad 
    \hat{\mathbf{S}}_\eta =\frac{1}{2} \sum_{\alpha ss'} f^\dagger_{\alpha\eta s} \boldsymbol{\spin}_{ss'} f_{\alpha\eta s'} \ ,
\end{eqnarray}
being total charge and spin operators in the valley $\eta$, respectively.

\subsection{The \texorpdfstring{$\rm U(1)^{\times3} \times SU(2)$}{U(1)3xSU(2)} symmetry and the two-electron states} \label{sec:original}

Adding up $H_{I1,2,3}$, we have 
\begin{align} \label{eq:HI123}
H_{I1} + H_{I2} + H_{I3} =& U \frac{\hat N (\hat N-1)}2 + \frac{J_{\rm H}}2 \hat{N} + \frac{J_{\rm H}}4 (\hat{N}_1^2 + \hat{N}_2^2) + \frac{J_{\rm H}'}2 \hat{N}_1 \hat{N}_2 + \frac{J_{\rm A}}2 \hat{N}_+ \hat{N}_- \nonumber\\ 
& - J_{\rm H} (\hat{\bf S}_1^2 + \hat{\bf S}_2^2) - J_{\rm H}' ( \hat{\bf S}^2 - \hat{\bf S}_1^2 - \hat{\bf S}_2^2 ) + J_{\rm A} ( \hat{\bf S}^2 -  \hat{\bf S}_+^2 - \hat{\bf S}_-^2 ) \nonumber\\
& - \frac{J_{\rm A}-J_{\rm H}'}2 \sum_{\alpha\eta ss'} f_{\ovl{\alpha} \ovl{\eta} s}^\dagger f_{\ovl\alpha \eta s'}^\dagger f_{ {\alpha} \ovl{\eta} s'} f_{\alpha\eta s}
+ \frac{J_{\rm H}'}2 \sum_{\alpha\eta ss'} f_{\ovl{\alpha} {\eta} s}^\dagger f_{\ovl\alpha \ovl\eta s'}^\dagger f_{ {\alpha} \ovl{\eta} s'} f_{\alpha\eta s}\ . 
\end{align}
We have made use of the relations 
$2\hat{\bf S}_1 \cdot \hat{\bf S}_2 = \hat{\bf S}^2 - \hat{\bf S}_1^2 - \hat{\bf S}_2^2$ and 
$2\hat{\bf S}_+ \cdot \hat{\bf S}_- = \hat{\bf S}^2 - \hat{\bf S}_+^2 - \hat{\bf S}_-^2$, where $\hat{\bf S}$ is the total spin operator, in the derivation. $\hat{\bf S}_\pm$ here are $\hat{\bf S}_\eta$ for $\eta=\pm$ and one should not confuse them with the raising and lowering operators. 

We note that this interaction enjoys a $\rm U(1)^{\times 3} \times SU(2)$ continuous symmetry. The three U(1) factors are generated by 
\begin{align}
    \sigma^0 \tau^0 \zeta^0 \ \textrm{(charge)}, 
    \qquad \sigma^0 \tau^z \zeta^0  \ \textrm{(valley)}, 
    \qquad \sigma^z \tau^z \zeta^0  \ \textrm{(orbital angular momentum)}, 
\end{align}
and the SU(2) factor is generated by 
\begin{align}
    \sigma^0 \tau^0 \zeta^{x,y,z}  \ \textrm{(global spin rotation)} \ . 
\end{align}
In particular, the orbital angular momentum U(1) emerges due to the following reason: although in the lattice model, the orbital angular momentum is only conserved modulo 3 (generated by $C_{3z} = e^{i\frac{2\pi}{3} \sigma^z\tau^z}$), on the effective Anderson impurity, any bilinear terms and quadratic terms that respect $C_{3z}$ must also respect a continuous rotation symmetry generated by $\sigma^z \tau^z$. 

We now study all the two-electron eigenstates, $F^\dagger_{i} |0\rangle$ ($i=1,\cdots,28$) of \cref{eq:HI123}, where $F_i^\dagger$ is a bilinear of two $f^\dagger$ operators, and also stands for an eigen-channel for two-particle scatterings. 
We normalize $F_i^\dagger$ such that $\langle 0 | F_i F_i^\dagger | 0 \rangle = 1$, and label them with $[\rho, j]$, where $\rho$ denotes a representation of the $D_6$ space group, and $j=0,1$ denotes the total SU(2) spin. 
Note that the valley U(1) charge is also conserved, but as it does not commute with $C_{2z}\in D_6$, we have not analyzed how it is represented yet. 
As it will turn out, all the levels that are to be obtained belong to one of the two following cases: Either 1) $\rho$ is by itself an irreducible representation (irrep) of $D_6$, and transforms trivially under valley U(1) (zero valley charge); or 2) $\rho$ is a summation of two irreps of $D_6$, but they are related to each other by a valley U(1) action. 
For both cases, $[\rho, j]$ forms an irrep of the entire symmetry group, with both $D_6$ and the valley U(1) taken into consideration. 

To solve the two-particle levels of \cref{eq:HI123}, notice that the third row of \cref{eq:HI123} only acts on two electrons that are in the same orbital and the opposite valleys. 
The states
\begin{equation} \label{eq:d-wave-tmp}
[E_2, 0], \qquad E=U - J_{\rm A} + 2J_{\rm H}',\qquad \frac{f_{\alpha+\up}^\dagger f_{\ovl\alpha-\down}^\dagger -  f_{\alpha+\down}^\dagger f_{\ovl\alpha-\up}^\dagger}{\sqrt2} \ket{0},\qquad (\alpha=1,2)\ , 
\end{equation}
have $N_{\alpha}=N_{\eta}=1$, $\hat{\bf S}^2 = 0$, $\hat{\bf S}_{\alpha}^2 = \hat{\bf S}_{\eta}^2 = \frac12(1+\frac12)$, and do not feel the third row of \cref{eq:HI123}, which would render $N_\alpha$ no longer good quantum numbers. 
Its energy can be directly calculated as 
$E=U + J_{\rm H} + \frac12 J_{\rm H} + \frac12 J_{\rm H}' + \frac12 J_{\rm A}- \frac32 J_{\rm H} + \frac32 J_{\rm H}' - \frac32 J_{\rm A}$, and can be verified to form the $[E_2, 0]$ irrep. 
The states
\begin{equation}
[E_1, 1], \qquad E=U + J_{\rm A} , \qquad 
    f_{\alpha+\up}^\dagger f_{\ovl\alpha-\up}^\dagger\ket{0},\quad 
    f_{\alpha+\down}^\dagger f_{\ovl\alpha-\down}^\dagger\ket{0},\quad 
    \frac{f_{\alpha+\up}^\dagger f_{\ovl\alpha-\down}^\dagger +  f_{\alpha+\down}^\dagger f_{\ovl\alpha-\up}^\dagger}{\sqrt2} \ket{0},\qquad (\alpha=1,2)\ , 
\end{equation}
have $N_{\alpha}=N_{\eta}=1$, $\hat{\bf S}^2 = 1(1+1)$, $\hat{\bf S}_{\alpha}^2 = \hat{\bf S}_{\eta}^2 = \frac12(1+\frac12)$, and its energy can be calculated as 
$E=U + J_{\rm H} + \frac12 J_{\rm H} + \frac12 J_{\rm H}' + \frac12 J_{\rm A} - \frac32 J_{\rm H} - \frac12 J_{\rm H}' + \frac12 J_{\rm A}$. 
Similarly, we can obtain other eigenstates that are annihilated by the third row of \cref{eq:HI123} 
\begin{equation} 
[E_1+E_2, 0], \qquad E = U + 2J_{\rm H}, \qquad 
f_{\alpha\eta \up}^\dagger f_{\alpha\eta \down}^\dagger \ket{0},\qquad (\alpha=1,2\; \eta=\pm) \ , 
\end{equation}
\begin{equation}
[A_1+B_1, 0], \qquad E = U + 2 J_{\rm H}' \qquad 
\frac{f_{1\eta \up}^\dagger f_{2\eta \down}^\dagger - f_{1\eta \down}^\dagger f_{2\eta \up}^\dagger}{\sqrt2} \ket{0},\qquad (\eta=\pm) \ , 
\end{equation}
\begin{equation} \label{eq:triplets-1}
[A_2+B_2, 1], \qquad E =  U, \qquad
f_{1\eta \up}^\dagger f_{2\eta \up}^\dagger\ket{0},\quad
f_{1\eta \down}^\dagger f_{2\eta \down}^\dagger\ket{0},\quad
\frac{f_{1\eta \up}^\dagger f_{2\eta \down}^\dagger + f_{1\eta \down}^\dagger f_{2\eta \up}^\dagger}{\sqrt2} \ket{0},\qquad (\eta=\pm)  \ . 
\end{equation}
The singlet states $f_{\alpha+\up}^\dagger f_{\alpha-\down}^\dagger - f_{\alpha+\down}^\dagger f_{\alpha-\up}^\dagger$ would have the energy $E=U+2J_{\rm H} - J_{\rm A}$ if the third row of \cref{eq:HI123} vanished.  
The third row of \cref{eq:HI123} scatters the singlet state with  $\alpha=1$ to the singlet state with $\alpha=2$
{\small
\begin{align}
- \frac{J_{\rm A}-J_{\rm H}'}2 \sum_{\alpha\eta ss'} f_{\ovl{\alpha} \ovl{\eta} s}^\dagger f_{\ovl\alpha \eta s'}^\dagger f_{ {\alpha} \ovl{\eta} s'} f_{\alpha\eta s}
  \frac{f_{1+\up}^\dagger f_{1-\down}^\dagger - f_{1+\down}^\dagger f_{1-\up}^\dagger}{\sqrt2} \ket{0} =
  - \pare{J_{\rm A}-J_{\rm H}'}  \frac{f_{2+\up}^\dagger f_{2-\down}^\dagger - f_{2+\down}^\dagger f_{2-\up}^\dagger}{\sqrt2} \ket{0}
\end{align}
\begin{align}
\frac{J_{\rm H}'}2 \sum_{\alpha\eta ss'} f_{\ovl{\alpha} {\eta} s}^\dagger f_{\ovl\alpha \ovl \eta s'}^\dagger f_{ {\alpha} \ovl{\eta} s'} f_{\alpha\eta s}
  \frac{f_{1+\up}^\dagger f_{1-\down}^\dagger - f_{1+\down}^\dagger f_{1-\up}^\dagger}{\sqrt2} \ket{0} 
  = J_{\rm H}' \frac{f_{2+\up}^\dagger f_{2-\down}^\dagger - f_{2+\down}^\dagger f_{2-\up}^\dagger}{\sqrt2} \ket{0}
\end{align}}
and vice versa. 
The two singlet states then form the bonding and anti-bonding states
\begin{equation}  \label{eq:s-wave-tmp}
[A_1, 0], \qquad E = U + 2J_{\rm H} - 2J_{\rm A} + 2J_{\rm H}' ,\qquad 
\frac{f_{1+\up}^\dagger f_{1-\down}^\dagger + f_{2+\up}^\dagger f_{2-\down}^\dagger - f_{1+\down}^\dagger f_{1-\up}^\dagger - f_{2+\down}^\dagger f_{2-\up}^\dagger}{2} \ket{0}, 
\end{equation}
\begin{equation} \label{eq:B2-wave}
[B_2, 0], \qquad E = U + 2J_{\rm H} - 2J_{\rm H}' ,\qquad 
\frac{f_{1+\up}^\dagger f_{1-\down}^\dagger - f_{2+\up}^\dagger f_{2-\down}^\dagger - f_{1+\down}^\dagger f_{1-\up}^\dagger + f_{2+\down}^\dagger f_{2-\up}^\dagger}{2} \ket{0} \ . 
\end{equation}
The triplet states $f_{\alpha+\up}^\dagger f_{\alpha-\up}^\dagger$, $f_{\alpha+\down}^\dagger f_{\alpha-\down}^\dagger$, 
$f_{\alpha+\up}^\dagger f_{\alpha-\down}^\dagger + f_{\alpha+\down}^\dagger f_{\alpha-\up}^\dagger $
would have the energy $E=U+J_{\rm A}$ if the third row of \cref{eq:HI123} vanished.  
The third row of \cref{eq:HI123} scatters the triplet states with  $\alpha=1$ to the triplet states with $\alpha=2$, {\it e.g.},
{\small
\begin{align}
- \frac{J_{\rm A}-J_{\rm H}'}2 \sum_{\alpha\eta ss'} f_{\ovl{\alpha} \ovl{\eta} s}^\dagger f_{\ovl\alpha \eta s'}^\dagger f_{ {\alpha} \ovl{\eta} s'} f_{\alpha\eta s}
  \cdot f_{1+\up}^\dagger f_{1-\up}^\dagger \ket{0} 
= (J_{\rm A}-J_{\rm H}')\cdot f_{2+\up}^\dagger f_{2-\up}^\dagger \ket{0} 
\end{align}
\begin{align}
\frac{J_{\rm H}'}2 \sum_{\alpha\eta ss'} f_{\ovl{\alpha} {\eta} s}^\dagger f_{\ovl\alpha \ovl \eta s'}^\dagger f_{ {\alpha} \ovl{\eta} s'} f_{\alpha\eta s}
  \cdot f_{1+\up}^\dagger f_{1-\up}^\dagger \ket{0} 
= J_{\rm H}' \cdot  f_{2+\up}^\dagger f_{2-\up}^\dagger \ket{0} 
\end{align}
}
and vice versa. 
The two triplet states then form the bonding and anti-bonding states
\begin{equation} \label{eq:triplets-2}
[A_2, 1], \quad E = U ,\quad 
\frac{f_{1+\up}^\dagger f_{1-\up}^\dagger - f_{2+\up}^\dagger f_{2-\up}^\dagger }{\sqrt2} \ket{0},\quad 
\frac{f_{1+\down}^\dagger f_{1-\down}^\dagger - f_{2+\down}^\dagger f_{2-\down}^\dagger }{\sqrt2} \ket{0},\quad 
\frac{f_{1+\up}^\dagger f_{1-\down}^\dagger - f_{2+\up}^\dagger f_{2-\down}^\dagger + (\up\leftrightarrow\down)}{2} \ket{0}, 
\end{equation}
\begin{equation} \label{eq:B1_0-wave}
[B_1, 1], \quad  E = U  + 2J_{\rm A},\quad 
\frac{f_{1+\up}^\dagger f_{1-\up}^\dagger + f_{2+\up}^\dagger f_{2-\up}^\dagger }{\sqrt2} \ket{0},\quad 
\frac{f_{1+\down}^\dagger f_{1-\down}^\dagger + f_{2+\down}^\dagger f_{2-\down}^\dagger }{\sqrt2} \ket{0},\quad 
\frac{f_{1+\up}^\dagger f_{1-\down}^\dagger + f_{2+\up}^\dagger f_{2-\down}^\dagger + (\up\leftrightarrow\down)}{2} \ket{0} \ .
\end{equation}

Assuming $J_{\rm H}' = \frac13 J_{\rm H}$ (\cref{eq:JHunds-U0}), we find that the two-electron ground states must be one of \cref{eq:triplets-1,eq:triplets-2,eq:d-wave-tmp,eq:s-wave-tmp}. 
To be concrete, the ground states are
\begin{equation} \label{eq:d-wave}
\ket{\Phi_{d,\alpha}} = \frac{f_{\alpha+\up}^\dagger f_{\ovl\alpha-\down}^\dagger -  f_{\alpha+\down}^\dagger f_{\ovl\alpha-\up}^\dagger}{\sqrt2} \ket{0},\quad (\alpha=1,2),\quad 
E = U - J_{\rm A} + \frac23 \mJ_{H},\quad 
\text{if}\;  \frac12 J_{\rm A} < J_{\rm H} < \frac32 J_{\rm A}\ ,
\end{equation}
\begin{equation} \label{eq:s-wave}
\ket{\Phi_{s}} = \frac{ f_{1+\up}^\dagger f_{1-\down}^\dagger +  f_{2+\up}^\dagger f_{2-\down}^\dagger - (\up\leftrightarrow\down) }{2} \ket{0},\quad 
E = U - 2J_{\rm A} + \frac83 \mJ_{H},\quad 
\text{if}\;  J_{\rm H}  < \frac12 J_{\rm A} \ ,
\end{equation}
and the triplets (\cref{eq:triplets-1,eq:triplets-2}) if $J_{\rm H}> \frac32 J_{\rm A}$. 
According to the crystalline symmetries in \cref{eq:discrete-sym}, the singlets $\Phi_{d,\alpha}$ in \cref{eq:d-wave} transform as $d_{x^2-y^2}$ and $d_{xy}$ orbitals under operations in the $D_6$ point group, belonging to the $E_2$ representation; the singlet $\Phi_s$ in \cref{eq:s-wave} transforms as the $s$ orbital and belongs to the $A_1$ representation. 
Using the parameters $U_0=9$eV \cite{wehling_strength_2011}, $\lambda_{\rm RG}=3.2$ \cite{basko_interplay_2008}, there are $J_{\rm A} = 4.16$meV, $J_{\rm H}=2.96$meV, and the ground states are the $d$-wave states. 

\subsection{Flattened interaction and the
\texorpdfstring{$\rm U(1)^{\times 4} \!\times\! SU(2)^{\times2}$}{U(1)4xSU(2)2} symmetry}
\label{sec:flatten-HI}

\cref{eq:HI123} is difficult to address analytically when coupled to a bath of itinerant electrons. 
To simplify the problem, we study an alternative interaction Hamiltonian $H_I$ that captures the main features of \cref{eq:HI123}. 
We require $H_I$ to satisfy the following conditions:
\begin{enumerate}
\item $H_I$ should reproduce the correct two-electron ground states $\ket{\Phi_{d,\alpha}}$ (\cref{eq:d-wave}), which are {\it inter-valley inter-orbital spin singlets} that belong to the $E_2$ representation of the point group $D_6$.
\item The two-electron spectrum of $H_I$ should be positive definite due to the large Coulomb repulsion $U$. In other words, $H_I$ itself does {\it not} have a pairing channel. 
\item The gap between the two-electron ground states and excited states should be at the order of $J_{\rm A}$. 
\item $H_I$ should have as high as possible symmetry for the sake of analytical convenience. 
\end{enumerate}
We find the following $H_I$ can match these requirements 
\begin{align} \label{eq:HI-def0}
H_I =&  \frac12 \sum_{\alpha \eta s s'} \big[ 
    U_1\cdot f_{\alpha\eta s}^\dagger f_{\ovl\alpha\ovl\eta s'}^\dagger f_{\ovl\alpha\ovl\eta s'} f_{\alpha\eta s}
    +U_2 \cdot f_{\alpha\eta s}^\dagger f_{\alpha\ovl\eta s'}^\dagger f_{\alpha\ovl\eta s'} f_{\alpha\eta s} 
    +U_3 \cdot f_{\alpha\eta s}^\dagger f_{\ovl\alpha\eta s'}^\dagger f_{\ovl\alpha\eta s'} f_{\alpha\eta s} 
    \nonumber\\
& \qquad  
    + U_4 \cdot f_{\alpha\eta s}^\dagger f_{\alpha\eta s'}^\dagger f_{\alpha\eta s'} f_{\alpha\eta s} 
    -\mJ \cdot  f_{\alpha\eta s}^\dagger f_{\ovl\alpha\ovl\eta s'}^\dagger f_{\alpha\eta s'} f_{\ovl\alpha\ovl\eta s} \big]\ ,
\end{align}
where $\mJ\sim J_{\rm A}$ is an inter-valley inter-orbital anti-Hund's coupling that favors the $\ket{\Phi_{d,\alpha}}$ states, $U_1=U-\mJ$, $U_2=U_3=U_4=U$. 
Making use of the relation $\sum_{\mu=0,x,y,z} \spin^{\mu}_{s_1s_2} \spin^{\mu}_{s_3s_4} = 2\delta_{s_1 s_4} \delta_{s_2s_3}$, we can rewrite $H_I$ as 
\begin{equation} \label{eq:HI-def}
    H_I = - \frac{U_4}{2} \hat{N} +  \frac12\sum_{\alpha\eta}  \pare{ 
    \pare{U_1+\frac12\mJ} \hat{N}_{\alpha\eta} \hat{N}_{\ovl\alpha\ovl\eta} 
    + U_2 \cdot \hat{N}_{\alpha\eta} \hat{N}_{\alpha\ovl\eta} 
    + U_3 \cdot \hat{N}_{\alpha\eta} \hat{N}_{\ovl\alpha\eta} 
    + U_4 \cdot \hat{N}_{\alpha\eta}^2 
    + 2\mJ \cdot \hat{\bf S}_{\alpha\eta} \cdot \hat{\bf S}_{\ovl\alpha\ovl\eta}
    }\ ,
\end{equation}
where 
\begin{equation}
    \hat{N}_{\alpha\eta} = \sum_{ s} f^\dagger_{\alpha\eta s} f_{\alpha\eta s},\qquad 
    \hat{\mathbf{S}}_{\alpha\eta} =\frac{1}{2} \sum_{ ss'} f^\dagger_{\alpha\eta s} \boldsymbol{\spin}_{ss'} f_{\alpha\eta s'} \ ,
\end{equation}
are the charge and spin in the orbital $\alpha$ and valley $\eta$, respectively, and $\hat{N}$ is the total charge operator. 
In order to calculate the eigenstates of $H_I$, we expand $H_I$ as 
\begin{align}\label{eq:HI-expand}
H_I =& - \frac{U_4}2 \hat{N} + \pare{ U_1 + \frac12 \mJ } \pare{ \hat N_{1+} \hat N_{2-} + \hat N_{2+} \hat N_{1-}  }
+ U_2 \pare{ \hat N_{1+} \hat N_{1-} + \hat N_{2+} \hat N_{2-}   } 
+ U_3 \pare{ \hat N_{1+} \hat N_{2+} + \hat N_{1-} \hat N_{2-}   }  \nonumber\\
& + \frac{U_4}2 (\hat N_{1+}^2 + \hat N_{1-}^2 + \hat N_{2+}^2 + \hat N_{2-}^2 ) 
+ \mJ (\hat{\bf S}_{1+} + \hat{\bf S}_{2-} )^2 + \mJ   (\hat{\bf S}_{1-} + \hat{\bf S}_{2+})^2
  - \mJ (   \hat{\bf S}_{1+}^2 + \hat{\bf S}_{1-}^2 + \hat{\bf S}_{2+}^2 + \hat{\bf S}_{2-}^2 )\ . 
\end{align}
The energies can be directly read from this expression once the good quantum numbers are known.

$H_I$ has an $\rm U(1)^{\otimes 4} \!\times\! SU(2)^{\otimes2}$ symmetry. 
The four U(1) factors are generated by 
\begin{equation} \label{eq:U1-generators}
    \sigma^0\tau^0\spin^0\; \text{(charge)},   \quad 
    \sigma^0\tau^z\spin^0\; \text{(valley)},   \quad 
    \sigma^z\tau^0\spin^0\; \text{(orbital)},  \quad
    \sigma^z\tau^z\spin^0\; \text{(angular momentum)}\ .  
\end{equation}
We call $\sigma^z\tau^z\spin^0$  angular momentum because $\sigma^z\tau^z$ is the quasi-angular momentum of the $C_{3z}$ operator (\cref{eq:discrete-sym}), which is now promoted to a continuous rotation symmetry. 
The two SU(2) factors are generated by 
\begin{equation}\label{eq:SU2-generators}
    \frac{\sigma^0\tau^0+\sigma^z\tau^z}2 \spin^{x,y,z},\qquad 
    \frac{\sigma^0\tau^0-\sigma^z\tau^z}2 \spin^{x,y,z},
\end{equation}
respectively. 
They are independent spin rotations in the $\sigma^z\tau^z=1$ and $-1$ flavors, respectively. 

We now calculate all the two-electron eigenstates of \cref{eq:HI-def}. 
First, we classify the two-electron states into angular momentum sectors.
If the total angular momentum is 0, there must be one state in the $\sigma^z\tau^z=1$ flavor and one in the $\sigma^z\tau^z=-1$ flavor. 
Due to the two independent SU(2) rotations, all these states should be spin-degenerate. 
We find the energies and eigenstates are 
\begin{equation} \label{eq:2particle-1}
    E = U_2,\qquad f_{\alpha + s}^\dagger f_{\alpha - s'}^\dagger \ket{0},\qquad (\alpha=1,2,\quad s,s'=\up\down)\ ,
\end{equation}
\begin{equation}\label{eq:2particle-2}
    E = U_3,\qquad f_{1 \eta s}^\dagger f_{2 \eta s'}^\dagger \ket{0},\qquad (\eta=+,-,\quad s,s'=\up\down)\ . 
\end{equation}
The degeneracy between different orbitals and valleys are protected by the discrete symmetries (\cref{eq:discrete-sym}). 
If the total angular momentum is 2, the two electrons must occupy $(\alpha,\eta)=(1,+)$ or $(2,-)$ flavors. 
If both particles occupy the same $\alpha,\eta$, they must form a singlet; otherwise they form a singlet and a triplet. 
The same analyses also apply to the sector with total angular momentum -2. 
We find the energies and eigenstates as 
\begin{equation}\label{eq:2particle-3}
    E = U_4,\qquad  f_{\alpha \eta \up}^\dagger f_{\alpha \eta \down}^\dagger\ket{0},\qquad (\alpha=1,2,\quad \eta=+,-)\ ,
\end{equation}
\begin{equation} \label{eq:2particle-4}
    E = U_1-\mJ ,\qquad  \frac{f_{\alpha \eta \up}^\dagger f_{\ovl\alpha \ovl\eta \down}^\dagger - f_{\alpha \eta \down}^\dagger f_{\ovl\alpha \ovl\eta \up}^\dagger }{\sqrt2} \ket{0},
    \qquad ((\alpha,\eta)=(1,+),(1,-) ) \ ,
\end{equation}
\begin{equation}\label{eq:2particle-5}
    E = U_1+\mJ ,\qquad  f_{\alpha \eta \up}^\dagger f_{\ovl\alpha \ovl\eta \up}^\dagger \ket{0},\quad  
    f_{\alpha \eta \down}^\dagger f_{\ovl\alpha \ovl\eta \down}^\dagger \ket{0},\quad 
    \frac{f_{\alpha \eta \up}^\dagger f_{\ovl\alpha \ovl\eta \down}^\dagger + f_{\alpha \eta \down}^\dagger f_{\ovl\alpha \ovl\eta \up}^\dagger }{\sqrt2} \ket{0},
    \qquad ((\alpha,\eta)=(1,+),(1,-) ) \ . 
\end{equation}
The degeneracies between different valleys and orbitals are protected by the discrete symmetries in \cref{eq:discrete-sym}. 
As there is no accidental degeneracy in the two-electron spectrum, \cref{eq:HI-def} is already in the most generic form allowed by the $\rm U(1)^{\otimes 4} \!\times\! SU(2)^{\otimes 2}$ symmetry. 

We choose $U_1=U-\mJ$, $U_2=U_3=U_4=U$, $\mJ\sim J_{\rm A}$ such that the inter-valley inter-orbital singlets have the energy $U-2\mJ$, while all the other states have the energy $U$.
Such parameterized $H_I$ can be thought as a ``flattened'' interaction where all the excited two-electron states are made degenerate. 

One can also design a flattened interaction $H_I'$ with the intra-orbital $s$-wave singlet (\cref{eq:s-wave-tmp}) and $B_2$ singlet (\cref{eq:B2-wave}) as the ground states. 
For a single site problem, $H_I'$ is equivalent to $H_I$ because they are related by the gauge transformation $f_{\alpha - s} \to f_{\ovl \alpha - s}$, $f_{\alpha + s} \to f_{\alpha + s}$.
Thus, all the formal discussions on the single impurity problem with $H_I$ also apply to the one with $H_I'$.
However, $H_I$ and $H_I'$ lead to physically different pairings because the lattice model is not unchanged under this gauge transformation.
$H_I'$ can be used to study the $s$-wave pairing. 
(The $B_2$ pairing should have a higher energy than $s$-wave pairing on the lattice model because it cannot open a full gap.)

\subsection{Negligible inter-site interactions}   \label{sec:inter-site}

In this subsection, we show that the inter-site interactions are all negligible. 
We have considered three different microscopic origins of interactions: 1) long-range gated Coulomb repulsion of 2D electron gas, 2) the atomic Hubbard $U_0$, and 3) the $A_1$, $B_1$-phonons. 

For 1), \cite{song_magic-angle_2022} have shown that inter-site interaction must take the Hubbard form as the on-site $U$, namely, only in a density-density manner, which only counts the $f$-electron number of each site. The strength is estimated as $U^{\rm inter-site} \approx 2.3\mrm{meV}$.  
As the giant on-site $U$ has already frozen the $f$-charge of each site to a fixed number, $U^{\rm inter-site}$ will act as an identity in all low-energy configurations, hence irrelevant. 
This is to be contrasted with the on-site $J_{\rm A,H}$, which does split the low-energy configuration with a fixed $f$-electron number, hence strongly influencing the Kondo physics. 
Correspondingly, the quasi-particle interactions should be around the same order as $z^2 U^{\rm inter-site}$, without a significant renormalization to the corresponding vertex, which is also to be contrasted with the (anti-)Hund's splittings, which receive receives strong renormalization away from $z^2 J_{\rm A,H}$, to the universal values determined by $T_{\rm K}$. 
With $z=0.2\sim0.3$, $z^2 U^{\rm inter-site} \approx 0.1\sim0.2\mrm{meV}$, which is weak. Moreover, it acts in inter-site pairing channels only, hence cannot affect the energy of on-site pairings. 
In Refs.~\cite{chou_kondo_2023,zhou_kondo_2023,hu_symmetric_2023,hu_kondo_2023,datta_heavy_2023,rai_dynamical_2023,lau_topological_2023,chou_scaling_2023}, $U^{\rm inter-site}$ is either omitted or decomposed in the Hartree channel such that it only shifts the energy levels of $f$ electrons. 

For 2), we remark that the atomic Hubbard is extremely short-ranged, hence the inter-site strength should be vanishingly weak, due to the highly localized nature of $f$-orbitals. 
We carry out a specific estimation below, referring to \cref{eq:HI2}. 
In general, each $\psi^\dagger_{l\rr s}$ operator can overlap with $f$-orbitals $f^\dagger_{\RR\alpha\eta s}$ on different AA-sites $\RR$, with the overlap given by 
\begin{align} \label{eq:graphene-basis-2}
    \psi^\dagger_{l\rr s} = \sqrt{\Omega_{\rm G}} \sum_{\RR} \sum_{\eta \alpha} \sum_{\beta, \mcl{L}_{l,\beta} \ni \rr} e^{-i\eta \mbf{K}_l \cdot (\rr - \RR)} w^{(\eta)*}_{l\beta, \alpha}(\rr - \RR) f^\dagger_{\RR\alpha\eta s}
\end{align}
where the overlap $w^{(\eta)}_{l\beta, \alpha}(\rr-\RR)$ dacays in a Gaussian form with respect to $|\rr - \RR|$. \cref{eq:graphene-basis} has only kept the closest AA-site to $\rr$. 
Inserting \cref{eq:graphene-basis-2} into \cref{eq:HI2}, we obtain
\begin{align}
    H_{I2} &= \frac{U_0 \Omega_G^2}{2} \sum_{ss'} \sum_{\alpha_{1,2,3,4}} \sum_{\eta_{1,2,3,4}} \delta_{\eta_1+\eta_2, \eta_3+\eta_4} f^\dagger_{\RR_4 \alpha_4 \eta_4 s} f^\dagger_{\RR_3 \alpha_3 \eta_3 s'} f_{\RR_2 \alpha_2 \eta_2 s'} f_{\RR_1 \alpha_1 \eta_1 s} \\ \nonumber 
    &~ \left[ \sum_\rr \sum_{l\beta} e^{i \mbf{K}_l \cdot (\eta_1 \RR_1 + \eta_2 \RR_2 - \eta_3\RR_3 - \eta_4\RR_4 ) } w^{(\eta_4)*}_{l\beta, \alpha_4}(\rr-\RR_4) w^{(\eta_3)*}_{l\beta,\alpha_3}(\rr - \RR_3) w^{(\eta_2)}_{l\beta,\alpha_2}(\rr-\RR_2) w^{(\eta_1)}_{l\beta,\alpha_1}(\rr-\RR_1) \right]
\end{align}
where we used the following fact that, if and only if $\eta_1 + \eta_2 = \eta_3 + \eta_4$, all the fast oscillating factors $e^{i\mbf{K}_l \cdot \rr}$ can neatly cancel, so that the summation over $\rr$ can produce a finite value.  

It suffices to check those inter-site terms that conserve the charge of each site separately, as otherwise two charge excitations that each costs energy $U$ would be involved. 
Moreover, as the Wannier functions $w^{(\eta)}_{l\beta, \alpha}(\rr-\RR)$ decay in a Gaussian form with increasing $|\rr-\RR|$, it suffices to check the strengths of the nearest-neighbor bonds. 
Two different cases can arise - A) $\RR_1=\RR_4=0$, $\RR_2=\RR_3=\RR$, with $|\RR|=a_M$, or B) $\RR_1=\RR_3=0$, $\RR_2=\RR_4=\RR$, with $|\RR|=a_M$. 
The type A) preserves the spin SU(2) of each site, and takes a density-density-like form, while type B) swaps the spin. 
The strengths are of the same order of magnitude, hence it suffices to evaluate one, \textit{e.g.} the case A). 
The strength is given by
\begin{align}
    & U_0 \Omega_G^2 \left|\sum_\rr \sum_{l\beta} e^{i \mbf{K}_l \cdot (\eta_1 \RR_1 + \eta_2 \RR_2 - \eta_3\RR_3 - \eta_4\RR_4 ) } w^{(\eta_4)*}_{l\beta, \alpha_4}(\rr) w^{(\eta_3)*}_{l\beta,\alpha_3}(\rr - \RR) w^{(\eta_2)}_{l\beta,\alpha_2}(\rr-\RR) w^{(\eta_1)}_{l\beta,\alpha_1}(\rr) \right| \\\nonumber
    &\leq  U_0 \Omega_G^2 \sum_\rr \sum_{l\beta} \left| w^{(\eta_4)*}_{l\beta, \alpha_4}(\rr) w^{(\eta_3)*}_{l\beta,\alpha_3}(\rr - \RR) w^{(\eta_2)}_{l\beta,\alpha_2}(\rr-\RR) w^{(\eta_1)}_{l\beta,\alpha_1}(\rr) \right| 
\end{align}
Referring to \cref{eq:WF-form1} to \cref{eq:WF-form2}, each Wannier factor can be (approximately) upperbounded by $|w^{(\eta)}_{l\beta, \alpha}(\rr-\RR)| \lesssim \frac{1}{\sqrt{2\pi} \lambda_2} e^{-\frac{(\rr-\RR)^2}{2\lambda_2^2}} $. 
The integration over $\rr$ can be decomposed to along two directions - perpendicular or parallel to $\RR$.  The perpendicular integration produces a factor of order $\frac{1}{\lambda_2}$, but the parallel integration is strongly suppressed by the localization of the Wannier functions ($\frac{\lambda_2}{a_M} = 0.1910$), as 
\begin{align}
    \frac{1}{2\pi \lambda_2^2} \int \mrm{d}y ~ e^{-\frac{y^2}{\lambda_2^2} - \frac{(y-a_M)^2}{\lambda_2^2}} \approx 2 \times 10^{-7} \times \frac{1}{\lambda_2}
\end{align}
Altogether, the nearest-neighbor strength is approximately $ U_0 \frac{\Omega_G^2}{\lambda_2^2} \times 2\times 10^{-7} < 10^{-5}\mrm{meV}$, which is indeed vanishingly weak. 

For 3), the $A_1, B_1$-phonons can propagate for a finite range on the microscopic lattice, but the inter-site strengths will still be much weaker than the on-site strengths $\lambda_{\rm RG} \times 1.3\mrm{meV}$. 
The detailed estimation of the phonon-mediated nearest-neighbor interaction can be found in \cite{wang_tbg_epc_2024}, and we summarize the results here - the density-density-like and spin-exchanging interaction strengths are $\lambda_{\rm RG} \times 0.005 \mrm{meV}$ and $\lambda_{\rm RG} \times 0.01 \mrm{meV}$, respectively. 
They are even smaller than the typical thermal fluctuation at 1K ($k_B \times 1\mrm{K} \approx 0.09\mrm{meV}$), hence their effects can be completely negligible as well.

\clearpage
\section{Anderson impurity problem with the flattened interaction} \label{sec:flatten}

\subsection{Effective action *}
\label{sec:action}

The effective free Hamiltonian of the single impurity problem is \cite{zhou_kondo_2023}  
\begin{equation}
    H_0 = \ee_f \sum_{\eta \alpha s} f_{\alpha\eta s}^\dagger f_{\alpha\eta s} 
    + \int_{-D}^D d\ee \cdot {\ee} \cdot d_{\alpha\eta s}^\dagger (\ee) d_{\alpha\eta s}(\ee) 
    + \int_{-D}^D d\ee \cdot h(\ee) \cdot d_{\alpha\eta s}^\dagger (\ee) f_{\alpha\eta s} + \mrm{H.c.} \ , 
\end{equation}
where $d_{\alpha\eta s}(\ee)$, satisfying $\{d_{\alpha\eta s}(\ee), d^\dagger_{\alpha'\eta's'}(\ee')\}=\delta_{\alpha\alpha'}\delta_{\eta\eta'}\delta_{ss'}\delta(\ee-\ee')$, is an auxiliary bath reproducing the (retarded) hybridization function 
\begin{equation}
    \Delta^{(\rm ret)}(\omega) = \pi \int_{-D}^D d\ee \cdot \delta(\omega-\ee) h^2(\ee) = \pi h^2(\omega)  . 
\end{equation}
Hence, one can choose $h(\ee) = \sqrt{\frac{\Delta^{\rm (ret)}(\ee)}{\pi}}$.

It is convenient to calculate susceptibilities and vertex functions in the path integral formalism.
By introducing the Grassmann variables $f_{\alpha\eta s}(\tau)$, where $\tau$ is the imaginary time, and their Fourier transformations 
\begin{equation}
    f_{\alpha\eta s}(\tau) = \sqrt{T} \sum_{\omega} f_{\alpha\eta s}(\omega)e^{-i\omega\tau},\qquad 
    f_{\alpha\eta s}(\omega) = \sqrt{T} \int_{0}^{1/T}d\tau\cdot f_{\alpha\eta s}(\tau)e^{i\omega\tau}\ ,
\end{equation}
with $T$ being the temperature and $\omega=(2n+1)\pi T$ the fermion Matsubara frequency, the partition function can be written as 
\begin{equation}
    \mathcal{Z} = \int D[f^\dagger f] e^{-S_0 - S_I}\ .
\end{equation}
Here $S_0$ is the free part 
\begin{equation} \label{eq:S0-8B-def}
    S_0 = \sum_{\omega} \sum_{\alpha\eta s} f_{\alpha\eta s}^\dagger(\omega) (-i\omega + \ee_f -i \Delta(\omega)) f_{\alpha \eta s}(\omega)\ ,
\end{equation}
The $i\Delta(\omega)$ term in $S_0$ is obtained by integrating out the auxiliary bath
\begin{equation}
    \Delta(\omega) = -i \pi\cdot \int_{-D}^D d\ee \cdot \frac{h^2(\ee) }{-i\omega + \ee} \ . 
\end{equation}
One can see that in the low-frequency regime ($|\omega|\ll D$), there is $\Delta(\omega) \approx \Delta_0 \cdot \sgn(\omega)$. 
As the frequency dependence is not relevant, in the rest of this manuscript we consider a flat hybridization function, {\it i.e.},
\begin{equation}
    \Delta(\omega) \equiv \Delta_0 \cdot \sgn(\omega)\ . 
\end{equation}
$S_I$ is the interaction part
\begin{align}\label{eq:SI-8B-def}
S_I =& \frac{T}2 \sum_{\nu} \sum_{\alpha\eta} \bigg[   
 \pare{U_1+\frac{\mJ}2} {N}_{\alpha\eta}(\nu) {N}_{\ovl\alpha\ovl\eta}(-\nu)
    + U_2 \cdot {N}_{\alpha\eta}(\nu) {N}_{\alpha\ovl\eta}(-\nu)
    + U_3 \cdot {N}_{\alpha\eta}(\nu) {N}_{\ovl\alpha\eta}(-\nu) \nonumber\\
&\qquad\qquad 
    + U_4 \cdot {N}_{\alpha\eta}(\nu){N}_{\alpha\eta}(-\nu)
    + 2\mJ \cdot {\bf S}_{\alpha\eta}(\nu) \cdot {\bf S}_{\ovl\alpha\ovl\eta}(-\nu) \bigg]\ . 
\end{align}
where $\nu=2\pi T n$ is bosonic Matsubara frequency, and 
\begin{equation}
    {N}_{\alpha\eta}(\nu) = T \sum_{\omega} \sum_{s} f^\dagger_{\alpha \eta s}(\omega+\nu) f_{\alpha\eta s}(\omega), \qquad 
    {\bf S}_\eta(\nu) = \frac{T}2 \sum_{\omega} \sum_{ s s' } f^\dagger_{\alpha \eta s}(\omega+\nu) \boldsymbol{\spin}_{ss'} f_{\alpha\eta s'}(\omega)\ ,
\end{equation}
are the charge and spin in the orbital valley flavor $(\alpha,\eta)$, respectively. 
It may be worth mentioning that the bilinear term (the first term) in \cref{eq:HI-def} does not appear in $S_I$.
When one writes the partition function as a path integral of Grassmann variables, one should write the Hamiltonian in the normal ordered form with respect to the vacuum. 
Thus $S_I$ should be first written in a normal ordered form where all $f$ are on right hand side of $f^\dagger$. 
Then one rewrites $S_I$ in terms of $\mathbf{S}_\eta$ and $N_\eta$. 
However, interchanging Grassmann variables does not yield the bilinear terms as interchanging second-quantized fermion operators. 

For later convenience, we formally rewrite $S_0$ (\cref{eq:S0-8B-def}) and $S_I$ (\cref{eq:SI-8B-def,eq:SI-8B-def}) as 
\begin{align}
    S_0 = \sum_{1} f^\dagger(1) (-i\omega_1 + \ee_f - i\Delta(\omega_1)) f(1)
\end{align}
and 
\begin{equation} \label{eq:SI-formal}
    S_I = \frac{T}{4}  \sum_{1,2,3,4} \delta_{\omega_1+\omega_2,\omega_3+\omega_4} \Gamma^0(1,2;3,4)
    f^\dagger(1) f^\dagger(2) f(3) f(4)\ , 
\end{equation}
respectively, where $f(1)$, $f(2)$ are shorthand for $ f_{I_1}(\omega_1)$, $f_{I_2}(\omega_2)$, {\it etc.}, and $I_{i=1,2,3,4}=(\alpha_i,\eta_i,s_i)$ are composite indices.  
The Arabic numbers can be regarded as composite indices including frequencies. 
$\Gamma^0(1,2;3,4)\equiv \Gamma^0_{I_1,I_2;I_3,I_4}(\omega_1,\omega_2;\omega_3,\omega_4)$ is the anti-symmetrized bare vertex function, which satisfies
\begin{equation}
    \Gamma^0(1,2;3,4) =
    - \Gamma^0(2,1;3,4) = 
    - \Gamma^0(1,2;4,3)
    =  \Gamma^0(2,1;4,3)\ .
\end{equation}
It should give the same interaction as \cref{eq:SI-8B-def} (or \cref{eq:HI-def0}). 
We can read the (not anti-symmetrized) vertex function from \cref{eq:HI-def0}
\begin{align}
& 4 \brak{ \frac{\Gamma^0_{U1}}2 \delta_{\alpha_1\ovl\alpha_2} \delta_{\eta_1\ovl\eta_2} 
    + \frac{\Gamma^0_{U2}}2 \delta_{\alpha_1\alpha_2} \delta_{\eta_1\ovl\eta_2}  
    + \frac{\Gamma^0_{U3}}2 \delta_{\alpha_1\ovl\alpha_2} \delta_{\eta_1\eta_2}  
    + \frac{\Gamma^0_{U4}}2 \delta_{\alpha_1\alpha_2} \delta_{\eta_1\eta_2}  
    } \delta_{\alpha_1 \alpha_4} \delta_{\alpha_2\alpha_3} \delta_{\eta_1\eta_4} \delta_{\eta_2\eta_3}
    \delta_{s_1s_4} \delta_{s_2 s_3} 
\nonumber\\ 
- & 4 \frac{\Gamma^0_{\mJ}}2 \delta_{\alpha_1\ovl\alpha_2} \delta_{\eta_1 \ovl \eta_2} \cdot \delta_{\alpha_2 \alpha_4} \delta_{\alpha_1\alpha_3} \delta_{\eta_2\eta_4} \delta_{\eta_1\eta_3}
    \delta_{s_1s_4} \delta_{s_2 s_3}  \ ,
\end{align}
where the factor 4 is due to the factor $\frac14$ in \cref{eq:SI-formal} and $\Gamma^0_{U1,2,3,4} = U_{1,2,3,4}$ and $\Gamma^0_{\mJ} = \mJ$.
After anti-symmetrization, we have 
\begin{align}
\Gamma_{I_1,I_2;I_3,I_4}^0(0,0;0,0) =& \brak{  
    {\Gamma^0_{U1}} \cdot\delta_{\alpha_1\ovl\alpha_2} \delta_{\eta_1\ovl\eta_2} 
    + {\Gamma^0_{U2}} \cdot\delta_{\alpha_1\alpha_2} \delta_{\eta_1\ovl\eta_2}  
    + {\Gamma^0_{U3}} \cdot\delta_{\alpha_1\ovl\alpha_2} \delta_{\eta_1\eta_2}  
    + {\Gamma^0_{U4}} \cdot \delta_{\alpha_1\alpha_2} \delta_{\eta_1\eta_2}  
    } \nonumber\\
    &\times [\delta_{\alpha_1 \alpha_4} \delta_{\alpha_2\alpha_3} \delta_{\eta_1\eta_4} \delta_{\eta_2\eta_3} \delta_{s_1s_4} \delta_{s_2 s_3} 
    - \delta_{\alpha_2 \alpha_4} \delta_{\alpha_1\alpha_3} \delta_{\eta_2\eta_4} \delta_{\eta_1\eta_3} \delta_{s_2s_4} \delta_{s_1 s_3} 
    ] \nonumber\\
& - \Gamma^0_{\mJ} \cdot \delta_{\alpha_1\ovl\alpha_2} \delta_{\eta_1 \ovl \eta_2} 
    [ \delta_{\alpha_2 \alpha_4} \delta_{\alpha_1\alpha_3} \delta_{\eta_2\eta_4} \delta_{\eta_1\eta_3}
    \delta_{s_1s_4} \delta_{s_2 s_3}
    - \delta_{\alpha_1 \alpha_4} \delta_{\alpha_2\alpha_3} \delta_{\eta_1\eta_4} \delta_{\eta_2\eta_3}
    \delta_{s_2s_4} \delta_{s_1 s_3}]\ . 
\end{align}

\subsection{Quasi-particle and vertex function *}
\label{sec:vertex}

The Matsubara Green's function is defined as
\begin{equation}
\mG(\omega) = -\inn{ f_{\alpha\eta s}(\omega) f^\dagger_{\alpha\eta s}(\omega) } = -\inn{ e^{-S_I} f_{\alpha\eta s}(\omega) f^\dagger_{\alpha\eta s}(\omega) }_{0C}\ ,
\end{equation}
where $\inn{\cdots }$ is the average weighted by $e^{-S_0-S_I}$, and $\inn{\cdots }_{0C}$ is the average weighted by $e^{-S_0}$, and with only the connected diagrams included. 
Due to the $\rm U(1)^{\otimes 4} \!\times\! SU(2)^{\otimes 2}$ symmetry and the discrete symmetries (\cref{eq:discrete-sym}), $\mG$ must be proportional to an identity in the $\alpha$, $\eta$, $s$ indices. 
One can formally write $\mG$ as 
\begin{equation}
    \mG(\omega) = \frac{1}{i\omega  + i\Delta_0\cdot\sgn(\omega) - \ee_f - \Sigma(\omega)}\ . 
\end{equation}
The self-energy satisfies the Hermitian condition $\Sigma(\omega) = \Sigma^*(-\omega)$. 
If the ground state is a Fermi liquid, then $\Sigma(\omega)$ behaves as 
\begin{equation}
    \Sigma(\omega) = \Sigma(0) + (1-z^{-1}) i\omega + \mathcal{O}(\omega^2)
\end{equation}
around zero-frequency, where $\Sigma(0)$ is a real number and 
\begin{equation}
z = ( 1 - \partial_{\omega} \Sigma(\omega))^{-1} |_{\omega=0}
\end{equation}
is the quasi-particle weight. 
According to the Friedel sum rule \cite{shiba_korringa_1975,yoshimori_perturbation_1976}, the total occupation is related to the zero-frequency Green's function 
\begin{equation}
    \frac{\nu_f+4}8 = \frac1{\pi} \Im  \ln \mG(0^+) = \frac1{\pi} \arccos \frac{\ee_f + \Sigma(0)}{ \sqrt{(\ee_f + \Sigma(0))^2 + \Delta_0^2} }\ .
\end{equation}
For later convenience, we define the scattering phase shift 
\begin{equation}
    \delta_f = \frac{\nu_f+4}8 \pi = \arccos \frac{\ee_f + \Sigma(0)}{ \sqrt{(\ee_f + \Sigma(0))^2 + \Delta_0^2} }. 
\end{equation}
It is 0 when the $f$ orbital is empty ($\ee_f+\Sigma(0)\to \infty$), $\frac{\pi}2$ at half-filling ($\ee_f+\Sigma(0)=0$), and $\pi$ when $f$ orbital is fully occupied ($\ee_f+\Sigma(0)\to -\infty$). 
The density of states (per flavor) at the Fermi level, or the spectral height at zero-energy, is 
\begin{equation}\label{eq:A(0)-Friedel}
A(0) = - \frac1{\pi} \Im[\mG(0^+)] = \frac1{\pi} \frac{\Delta_0}{ (\ee_f + \Sigma(0))^2 + \Delta_0^2 } = \frac{\sin^2\delta_f}{\pi\Delta_0} \ . 
\end{equation}
One can see that $A(0)$ is completely determined by $\delta_f$ and the hybridization $\Delta_0$, and is independent of the interaction.

The full anti-symmetrized vertex function $\Gamma(1,2;3,4)$ is defined by the following skeleton diagrams up to infinite orders of the bare interaction 
\begin{equation}
\includegraphics[width=0.7\linewidth]{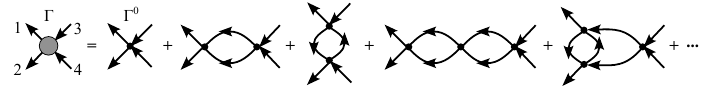}\ .
\end{equation}
Here the black dots represent the bare interaction (denoted as $\Gamma^0$) and the solid lines are the full Green's functions with self-energy corrections. 
For now we are only interested in the zero frequency part of $\Gamma(00;00)$. 
Notice that $\Gamma^0$, read from $H_I$ in \cref{eq:HI-def0}, is already in the most generic form allowed by the symmetries because $H_I$ has no accidental degeneracy.
Thus, $\Gamma(00;00)$ must have the same form as $\Gamma^0$, {\it i.e.},
\begin{align} \label{eq:Gamma-symm}
\Gamma_{I_1,I_2;I_3,I_4}(0,0;0,0) =& \brak{  
    {\Gamma_{U1}} \cdot\delta_{\alpha_1\ovl\alpha_2} \delta_{\eta_1\ovl\eta_2} 
    + {\Gamma_{U2}} \cdot\delta_{\alpha_1\alpha_2} \delta_{\eta_1\ovl\eta_2}  
    + {\Gamma_{U3}} \cdot\delta_{\alpha_1\ovl\alpha_2} \delta_{\eta_1\eta_2}  
    + {\Gamma_{U4}} \cdot \delta_{\alpha_1\alpha_2} \delta_{\eta_1\eta_2}  
    } \nonumber\\
    &\times [\delta_{\alpha_1 \alpha_4} \delta_{\alpha_2\alpha_3} \delta_{\eta_1\eta_4} \delta_{\eta_2\eta_3} \delta_{s_1s_4} \delta_{s_2 s_3} 
    - \delta_{\alpha_2 \alpha_4} \delta_{\alpha_1\alpha_3} \delta_{\eta_2\eta_4} \delta_{\eta_1\eta_3} \delta_{s_2s_4} \delta_{s_1 s_3} 
    ] \nonumber\\
- & \Gamma_{\mJ} \cdot \delta_{\alpha_1\ovl\alpha_2} \delta_{\eta_1 \ovl \eta_2} 
    [ \delta_{\alpha_2 \alpha_4} \delta_{\alpha_1\alpha_3} \delta_{\eta_2\eta_4} \delta_{\eta_1\eta_3}
    \delta_{s_1s_4} \delta_{s_2 s_3}
    - \delta_{\alpha_1 \alpha_4} \delta_{\alpha_2\alpha_3} \delta_{\eta_1\eta_4} \delta_{\eta_2\eta_3}
    \delta_{s_2s_4} \delta_{s_1 s_3}]\ ,
\end{align}
except that the bare values $\Gamma^0_{U1,2,3,4}$, $\Gamma^0_{\mJ}$ are replaced by renormalized values  $\Gamma_{U1,2,3,4}$, $\Gamma_{\mJ}$.

It is convenient to define an effective field theory for the low-energy quasi-particle excitations \cite{hewson_renormalized_1993,hewson_renormalized_2001}.  
We define the quasi-particle operator as $\td{f} = z^{-\frac12} f$ and rewrite the action as $S_{\rm qp} + S_{\rm c}$, with
\begin{equation} \label{eq:Sqp}
    S_{\rm qp} = \sum_{\omega} \td{f}^\dagger_I(\omega) (-i\omega - i \td{\Delta}_0 \cdot \sgn(\omega) + \td{\ee}_f ) \td{f}_I(\omega)
    + \frac14\sum_{1,2,3,4} \delta_{\omega_1+\omega_2,\omega_3+\omega_4} \td{\Gamma}^0(1,2;3,4)  
    \td{f}^\dagger(1)\td{f}^\dagger(2) \td{f}(3) \td{f}(4)\ ,
\end{equation}
\begin{equation}
    \td{\Delta}_0 = z\Delta_0,\qquad \td{\ee}_f = z \ee_f + z\Sigma(0),\qquad 
    \td{\Gamma}_{I_1,I_2;I_3,I_4}^0(\omega_1\omega_2;\omega_3\omega_4) = z^2 \Gamma_{I_1,I_2;I_3,I_4}(00;00)\ , 
\end{equation}
\begin{equation} \label{eq:Sc}
S_{\rm c} = \sum_{\omega} \td{f}^\dagger_I(\omega) ( -i \lambda_0 \omega + \lambda_1 ) \td{f}_I(\omega)
 + \frac14 \sum_{1234} \Lambda(1,2;3,4) \td{f}^\dagger(1)\td{f}^\dagger(2) \td{f}(3) \td{f}(4)\ ,
\end{equation}
\begin{equation}
    \lambda_0 = z-1,\qquad \lambda_1 = -z\Sigma(0),\qquad 
    \Lambda(1,2;3,4) = z^2 \Gamma^0_{I_1,I_2;I_3,I_4}(00;00) - \td{\Gamma}^0_{I_1,I_2;I_3,I_4}(00;00)\ . 
\end{equation}
The decomposition of $S$ into $S_{\rm qp}$ and $S_{\rm c}$ is exact. 
$S_{\rm qp}$ describes the low-energy Fermi liquid fixed point, and $S_{\rm c}$ - the counter term - guarantees that there is no further renormalization to $S_{\rm qp}$ \cite{hewson_renormalized_1993,hewson_renormalized_2001}. 
One can define the renormalized interaction parameters as 
\begin{equation}
\td{U}_{1,2,3,4} = z^2 \Gamma_{U1,2,3,4},\qquad 
\td{\mJ} = z^2 \Gamma_{\mJ}\ . 
\end{equation}
$\lambda_{0,1}$ and $\Lambda$ can be expressed in powers of $\td{\Gamma}^0$, {\it i.e.}, $\lambda_i = \sum_n \lambda_i^{(n)}$ and  $\Lambda_i = \sum_n \Lambda_i^{(n)}$, where $n$ represents the order of $\td{\Gamma}^0$. 
These expressions should be determined order by order in such a way that they cancel all the further renormalizations, {\it i.e.}, the quasi-particle self-energy $\td{\Sigma}(\omega)$ and the quasi-particle full vertex function $\td{\Gamma}$ satisfies
\begin{equation} \label{eq:no-further-renormalization}
    \td{\Sigma}(0) = 0, \qquad  
    \frac{\partial \td{\Sigma}}{\partial i\omega} \bigg|_{\omega=0}=0,\qquad 
    \td{\Gamma}(00;00) = \td{\Gamma}^0 = z^2\Gamma(00;00)\ . 
\end{equation}
We will carry out this perturbation calculation in \cref{sec:perturbation}. 
One should be aware that $\td{\Sigma}$ can have higher-order frequency-dependence such as $\omega^2$ terms and $\td{\Gamma}$ can also have frequency dependence.

\subsection{Susceptibilities}
\label{sec:chi}

We now consider the susceptibility
\begin{equation} \label{eq:chi-def}
    \chi^{AB}(\nu) = \int_{0}^{\beta} d\tau \cdot \inn{\hat{A}(\tau) \hat{B}(0)  } e^{i\nu\tau }
\end{equation}
where $\nu$ is the bosonic Matsubara frequency, 
$\hat{A} = \sum_{I_1 I_2} f_{I_1}^\dagger A_{I_1,I_2} f_{I_2}$,
$\hat{B} = \sum_{I_1 I_2} f_{I_1}^\dagger B_{I_1,I_2} f_{I_2}$. 
We will only focus on the static response, {\it i.e.}, $\chi^{AB}(\nu\to 0^+)$. 
It describes the response $\delta \inn{\hat{A}}$ with respect to a static external field $-h^B \hat{B}$, {\it i.e.}, $\chi^{AB}(0^+) = \frac{\partial \inn{\hat{A}}}{ \partial h^B}$. 

\begin{table}[t]
\centering
\begin{tabular}{c|c|c|c|c|c}
\hline
    & $\sigma^{0,z}\tau^{0,z}$ & $\sigma^{\pm}\tau^{0,z}$ & $\sigma^{0,z}\tau^{\pm}$ & $\sigma^{\pm}\tau^{\pm}$ &  $\sigma^{\pm}\tau^{\mp}$ \\
\hline\hline     
    $\tau^z$ charge & 0 & 0 & $\pm2$ & $\pm2$ & $\mp2$ \\ 
    $\sigma^z$ charge & 0 & $\pm2$ & 0 & $\pm2$ & $\pm2$ \\
    $\sigma^z\tau^z$ charge & 0 & $\pm2$ & $\pm2$ & 0 & 0 \\
\hline
    Independent & {$\sigma^{0,z}\tau^{0,z}$} & $\sigma^x \tau^{0,z}$ & $\sigma^{0,z}\tau^x$ & \multicolumn{2}{c}{$\sigma^x\tau^x$} \\
\hline
    \end{tabular}
    \caption{Nine independent charge neutral spin-0 channels in the susceptibility of the four-orbital Anderson impurity model with the flattened interaction. }
    \label{tab:chi-neutral-singlet}
\end{table}

The symmetries can help us identify independent channels of the susceptibilities, since $\chi^{AB}$ should be block-diagonal in the irreducible representations. 
We first classify the spin-0 operators $\sigma^{0,x,y,z} \tau^{0,x,y,z} \spin^0 $. 
According to the U(1) symmetry generators in \cref{eq:U1-generators}, these operators can be labeled by (i) valley ($\tau^z$) charge, (ii) orbital ($\sigma^z$) charge, and (iii) angular momentum ($\sigma^z\tau^z$). 
We summarize the U(1) charges of the sixteen operators in \cref{tab:chi-neutral-singlet}. 
Operators in different columns do not couple to each other due to the  U(1) charges conservation. 
For each column, the susceptibility is in principle a matrix. 
We can make use of the discrete symmetries in \cref{eq:discrete-sym} to further diagonalize the susceptibility for each column:
\begin{enumerate}
\item The four operators $\sigma^0\tau^0$, $\sigma^0\tau^z$, $\sigma^z\tau^0$, $\sigma^z\tau^z$ in the first column form four different one-dimensional representations of the $D_2$ point group generated by $C_{2x}=\sigma^x$, $C_{2z}=\sigma^x\tau^x$. 
    Therefore, each of them is an eigenmode of the susceptibility. 
    (We do not need to consider the $C_{3z}$ symmetry here because it is already promoted to the continuous rotation generated by the angular momentum $\sigma^z\tau^z$.)
\item The four operators in the second column can be recombined to the hermitian matrix basis $\sigma^{x,y} \tau^{0,z}$. The four hermitian operators do not couple to each other because they belong to four different one-dimensional representations of the $D_2$ point group. 
    The susceptibilities for $\sigma^{y}\tau^{0,z}$ must be same as as that of $\sigma^{x}\tau^{0,z}$ due to the orbital U(1) symmetry generated by $\sigma^z$.
    Therefore, we only need to consider the two representative channels $\sigma^{x}\tau^{0,z}$.
\item Analyses for the second column also apply to the third column, except that the roles of valley and orbital are exchanged.  
\item The four operators in the fourth and fifth columns can be recombined to the hermitian matrix basis $\sigma^{x,y} \tau^{x,y}$. Again, they belong to four different one-dimensional representations of the $D_2$ point group and do not couple to each other. Due to the valley-U(1) ($\tau^z$) and orbital-U(1) ($\sigma^z$), we only need to consider the susceptibility of $\sigma^x\tau^x$. 
\end{enumerate}

We second classify the spin-1 operators. 
Due to the global SU(2) symmetry, we only need to consider the $\sigma^{\mu} \tau^{\nu} \spin^z$ ($\mu,\nu=0,x,y,z$) operators. 
As the valley-U(1), orbital-U(1), angular momentum, and $D_2$ point group operators are independent of spin, all the discussions in the last paragraph apply to $\sigma^{\mu} \tau^{\nu} \spin^z$. 
Therefore, there are at most nine independent spin-1 channels, which are given by the operators in the last row of \cref{tab:chi-neutral-singlet} multiplied by $\spin^z$. 
We now show that the spin-1 operators only contribute three new independent channels:
\begin{enumerate}
\item Spin-1 operators descended from the first column of \cref{tab:chi-neutral-singlet} are $\sigma^{0,z}\tau^{0,z}\spin^z$.
    Nevertheless, the $\sigma^0\tau^0\spin^z$ and $\sigma^z\tau^z\spin^z$ operators are related to each other by the successive global and relative SU(2) rotations $ e^{i\frac{\pi}4 \sigma^0 \tau^0 \spin^y} \cdot e^{-i\frac{\pi}4 \sigma^z\tau^z \spin^y} $ (\cref{eq:SU2-generators}), thus they must have the same susceptibility. 
    Similarly, the other two operators $\sigma^z\tau^0\spin^z$ and $\sigma^0\tau^z\spin^z$ are also related by $ e^{i\frac{\pi}4 \sigma^0 \tau^0 \spin^y} \cdot e^{-i\frac{\pi}4 \sigma^z\tau^z \spin^y} $. 
    Therefore, we can choose $\sigma^0\tau^0\spin^z$ and $\sigma^0\tau^z\spin^z$ as the representative operators. 
\item Spin-1 operators descended from the second column of \cref{tab:chi-neutral-singlet} are $\sigma^x \tau^{0,z} \spin^{z}$. 
    The successive orbital U(1) and relative SU(2) rotations $e^{i\frac{\pi}4 \sigma^z\tau^0\spin^0} \cdot e^{-i\frac{\pi}4\sigma^z \tau^z \spin^z}$ transform them to spin-0 operators $\sigma^x\tau^{z,0}\spin^0$. Thus these operators do not contribute new channels. 
\item Spin-1 operators descended from the third column of \cref{tab:chi-neutral-singlet} do not contribute new channels for the same reason as above. 
\item The only spin-1 operator descended from the last two columns of \cref{tab:chi-neutral-singlet} is $\sigma^x\tau^x\spin^z$. 
\end{enumerate}

In summary, there are twelve independent channels in the susceptibility.
We use the notation $O^{\mu\nu\rho} = \sigma^\mu \tau^\nu \spin^\rho$ to represent these operators. 
The twelve eigenmodes can be chosen as 
\begin{equation}\label{eq:chi-channels}
\begin{aligned}
    & O^{000},\quad 
    O^{0z0},\quad 
    O^{z00},\quad 
    O^{zz0},\quad 
    O^{00z},\quad 
    \\
    & 
    O^{0zz},\quad 
    O^{x00},\quad 
    O^{xz0},\quad 
    O^{0x0},\quad 
    O^{zx0},\quad 
    O^{xx0},\quad 
    O^{xxz}\ . 
\end{aligned}
\end{equation}
Operators in the first row are generators of the continuous symmetry group, whereas operators in the second row are not. 

We now apply perturbation calculation of $\chi$ to the first order of $S_I$. 
Applying Fourier transformation to \cref{eq:chi-def}, we obtain 
\begin{equation}
    \chi^{O}(\nu) = T \sum_{1234} 
    \delta_{\omega_1-\omega_2,-\nu} \delta_{\omega_3-\omega_4,\nu}  
    \Inn{e^{-S_I} f^\dagger(1) O(1,2) f(2) f^\dagger(3) O(3,4) f(4) }_{0C} \ ,
\end{equation}
where $i=1,2,3,4$ are composite indices and $O(i,j')=O_{I_i, I_{j}}$. 
To the zeroth order of $S_I$, the susceptibility is 
{
\begin{align}
\chi^{O(0)}(\nu) =& T \sum_{1234} 
    \delta_{\omega_1-\omega_2,-\nu} \delta_{\omega_3-\omega_4,\nu}  
    \wick{   \c1 f^\dagger(1) O(1,2) \c2 f(2) \c2 f^\dagger(3) O(3,4) \c1 f(4) } 
=  -T \sum_{12} \delta_{\omega_1-\omega_2,-\nu} O(1,2) O(2,1) 
    \mG^0 (\omega_1) \mG^0(\omega_2)  \nonumber \\
=& T\cdot \tr[O^2] \sum_{\omega}     
    \frac1{\omega+\Delta(\omega)+i\ee_f} \cdot 
    \frac1{\omega+\nu+\Delta(\omega+\nu)+i\ee_f}\ ,
\end{align}}
where $\tr[\cdots]$ represents the trace over $\alpha,\eta,s$. 
To the zeroth order of $\nu$ we have 
\begin{align}
\chi^{O(0)}(0^+) =& \tr[O^2] \int_{-\infty}^\infty \frac{d\omega}{2\pi} \frac1{(\omega + \Delta(\omega) + i\ee_f)^2 }
=\tr[O^2]\int_{0}^{\infty} \frac{d\omega}{2\pi} \pare{\frac1{(\omega + \Delta_0 + i\ee_f)^2 } + \frac1{(\omega + \Delta_0 - i\ee_f)^2 } } \nonumber\\
=& \tr[O^2] \frac{\Delta_0} { \pi \cdot( \Delta_0^2 + \ee_f^2) } 
=\tr[O^2] \cdot \frac{\sin^2\delta_f}{\pi \Delta_0}\ ,
\end{align}
where $\delta_f = \pi (\nu_f+4)/8$ is the scattering phase shift. 
To the first order of $S_I$, we have 
\begin{align}
\chi^{O(1)}(0^+) =& - T^2 \sum_{1234} \sum_{5678} 
    \delta_{\omega_1-\omega_2,-\nu} \delta_{\omega_3-\omega_4,\nu}
    O(1,2) O(3,4) \times\frac14 
    \delta_{\omega_5 + \omega_6, \omega_7+\omega_8} \Gamma^0(5,6;7,8) \nonumber\\
& \times  \Inn{ f^\dagger(1)  f(2) f^\dagger(3) f(4) f^\dagger(5) f^\dagger(6)  f(7) f(8) }_{0C} \ .
\end{align}
Applying Wick's theorem, there is 
\begin{align}
\chi^{O(1)}(\nu) =& - T^2  \sum_{1234} \sum_{5678} 
    \delta_{\omega_1-\omega_2,-\nu} \delta_{\omega_3-\omega_4,\nu}
    O(1,2) O(3,4) \times 
    \frac14 \delta_{\omega_5 + \omega_6, \omega_7+\omega_8} \Gamma^0(5,6;7,8) \nonumber\\
& \times  4 \times \wick{ \c1 f^\dagger(1)  \c2 f(2) \c3 f^\dagger(3) \c4 f(4) 
                 \c2 f^\dagger(5) \c4 f^\dagger(6)  \c3 f(7) \c1 f(8) } \ .
\end{align}
The factor 4 comes from equivalent contractions, which are equal to one another because $\Gamma^0$ is fully anti-symmetrized. 
The contraction equals 
\begin{equation}
    (-1)^2 \delta_{52} \delta_{64} \delta_{73} \delta_{81} \cdot 
    \mG^0(\omega_1) \mG^0(\omega_2) \mG^0(\omega_3) \mG^0(\omega_4) \ . 
\end{equation}
Hence, 
\begin{align}
\chi^{O(1)}(\nu) =& - T^2  \sum_{1234} 
    \delta_{\omega_1-\omega_2,-\nu} \delta_{\omega_3-\omega_4,\nu}
    O(1,2) O(3,4)  \cdot 
    \Gamma^0(2,4;3,1) \cdot \mG^0(\omega_1) \mG^0(\omega_2) \mG^0(\omega_3) \mG^0(\omega_4) 
\end{align}
Since $O$ and $\Gamma^0$ does not depend on the frequency, we can separate the summations over frequency and matrix indices. 
Introducing the factor
\begin{equation} \label{eq:kappa}
    \kappa[O,\Gamma^0] = \frac1{\tr[O^2]} \sum_{I_{1,2,3,4}} O_{I_1, I_2} 
    O_{I_3, I_4} \Gamma^0_{I_2, I_4; I_3, I_1}(00;00)
    =  \frac1{\tr[O^2]} \sum_{I_{1,2,3,4}} O_{I_4, I_1} 
    O_{I_3, I_2} \Gamma^0_{I_1, I_2; I_3, I_4}(00;00)\ ,
\end{equation}
we have 
\begin{equation}
\chi^{O(1)}(0^+) = - \tr[O^2] \cdot \kappa[O,\Gamma^0] \int \frac{d\omega_1}{2\pi} \frac{d\omega_3}{2\pi} 
    \frac{ 1 } {  (  \omega_1+\Delta(\omega_1)+i\ee_f )^2  } 
    \cdot \frac{ 1 } {  (  \omega_3+\Delta(\omega_3)+i\ee_f )^2  } 
= - \tr[O^2] \kappa[O,\Gamma^0] \pare{ \frac{\sin^2\delta_f}{\pi\Delta_0} }^2 \ . 
\end{equation}
Adding up the zeroth and first-order contributions, we obtain 
\begin{equation} \label{eq:chi-bare}
\chi^{O} \approx \chi^{O(0)} + \chi^{O(1)}
= \tr[O^2] \frac{\sin^2\delta_f}{\pi\Delta_0}\pare{  1 - \kappa[O,\Gamma^0] \cdot \frac{\sin^2\delta_f}{\pi\Delta_0}  }\ . 
\end{equation}

\cref{eq:chi-bare} is not very useful as $\Gamma^0$ is usually large and cannot be treated as a perturbation. 
However, we can obtain the quasi-particle susceptibility by replacing $\Delta_0$ and $\Gamma^0$ with the renormalized values $\td{\Delta}_0 = z\Delta_0$ and $\td{\Gamma}^0 = z^2\Gamma(00;00)$
\begin{equation} \label{eq:chi-qp}
\td{\chi}^{O} 
\approx \tr[O^2] \frac{\sin^2\delta_f}{\pi \td{\Delta}_0}\pare{  1 - \kappa[O,\td{\Gamma}^0] \cdot \frac{\sin^2\delta_f}{\pi \td{\Delta}_0}  }\ .  
\end{equation}
Considering $\td{\Gamma}^0$ in \cref{eq:Sqp} is already renormalized, which means it is the full quasi-particle vertex function at zero frequency $\td{\Gamma}(00;00)$, there will be no higher order correction to the above equation.
But \cref{eq:chi-qp} is also not obviously exact because it omits the frequency dependence of the $\td{\Gamma}$.
It also omits the frequency dependence of the quasi-particle self-energy $\td{\Sigma}(\omega)$ (beyond quasi-particle weight correction), {\it e.g.}, $\omega^2$ terms in $\td{\Sigma}(\omega)$. 
Nevertheless,  we will prove in the next subsection through Ward identities that \cref{eq:chi-qp} equals the {\it exact susceptibilities} of the {\it bare} particles if $O$ is a generator of the continuous symmetry group. 

We now determine the factor $\kappa[O,\Gamma^0]$ (\cref{eq:kappa}) for the twelve channels in \cref{eq:chi-channels}. 
In the first six channels there are $I_1=I_4$ and $I_2=I_3$ in the expression \cref{eq:kappa}. 
The involved vertex function elements are 
{\small
\begin{equation}
\td{\Gamma}^0_{I_1 I_2; I_2 I_1}(00;00) = 
    \td{U}_1 \cdot \delta_{\alpha_1 \ovl \alpha_2} \delta_{\eta_1 \ovl\eta_2} 
    + \td{U}_2 \cdot \delta_{\alpha_1  \alpha_2} \delta_{\eta_1 \ovl\eta_2} 
    + \td{U}_3 \cdot \delta_{\alpha_1  \ovl \alpha_2} \delta_{\eta_1 \eta_2} 
    + \td{U}_4 \cdot \delta_{\alpha_1   \alpha_2} \delta_{\eta_1 \eta_2} (1 - \delta_{s_1 s_2})
    + \td{\mJ} \cdot \delta_{\alpha_1 \ovl \alpha_2} \delta_{\eta_1 \ovl\eta_2} \delta_{s_1s_2}
\end{equation}}
It is direct to obtain
\begin{equation}\label{eq:kappa000}
    \kappa[O^{000},\td{\Gamma}^0] = 2\td{U}_1 + 2\td{U}_2 + 2\td{U}_3 + \td{U}_4 + \td{\mJ}\ ,
\end{equation}
\begin{equation}
    \kappa[O^{0z0},\td{\Gamma}^0] = -2\td{U}_1 - 2\td{U}_2 + 2\td{U}_3 + \td{U}_4 - \td{\mJ}\ ,
\end{equation}
\begin{equation}
    \kappa[O^{z00},\td{\Gamma}^0] = -2\td{U}_1 + 2\td{U}_2 - 2\td{U}_3 + \td{U}_4 - \td{\mJ}\ , 
\end{equation}
\begin{equation}
    \kappa[O^{zz0},\td{\Gamma}^0] = 2\td{U}_1 - 2\td{U}_2 - 2\td{U}_3 + \td{U}_4 + \td{\mJ}\ ,
\end{equation}
\begin{equation}
    \kappa[O^{00z},\td{\Gamma}^0] =  - \td{U}_4 + \td{\mJ}\ , 
\end{equation}
\begin{equation}
    \kappa[O^{0zz},\td{\Gamma}^0] =  - \td{U}_4 - \td{\mJ}\ . 
\end{equation}
For the seventh and eighth operators $O^{x00}$, $O^{xz0}$, there are 
$\alpha_4=\ovl{\alpha}_1$, $\eta_4=\eta_1$, $s_4=s_1$,
$\alpha_3=\ovl{\alpha}_2$, $\eta_3=\eta_2$, $s_3=s_2$ 
in the expression \cref{eq:kappa}. The involved vertex function elements are 
\begin{equation}
\td{\Gamma}^0_{I_1,I_2;I_3,I_4}(00;00) = 
- \td{U}_3 \cdot \delta_{\alpha_1\ovl{\alpha}_2} \delta_{\eta_1 \eta_2} \delta_{s_2 s_1}\ .
\end{equation}
It is direct to obtain
\begin{equation}
    \kappa[O^{x00},\td{\Gamma}^0] = \kappa[O^{xz0},\td{\Gamma}^0] = - \td{U}_3\ . 
\end{equation}
For the ninth and tenth operators $O^{0x0}$, $O^{zx0}$, the roles of orbital and valley are exchanged, thus the $\kappa$ factors should be the same except the inter-orbital repulsion $\td{U_3}$ is replaced by the inter-valley repulsion $\td{U}_2$:
\begin{equation}
    \kappa[O^{0x0},\td{\Gamma}^0] = \kappa[O^{zx0},\td{\Gamma}^0] = - \td{U}_2\ . 
\end{equation}
For the eleventh and twelfth operators $O^{xx0}$, $O^{xxz}$, there are 
$\alpha_4=\ovl{\alpha}_1$, $\eta_4=\ovl\eta_1$, $s_4=s_1$,
$\alpha_3=\ovl{\alpha}_2$, $\eta_3=\ovl\eta_2$, $s_3=s_2$ 
in the expression \cref{eq:kappa}. The involved vertex function elements are 
\begin{equation}
\td{\Gamma}^0_{I_1,I_2;I_3,I_4}(00;00) = 
- \td{U}_1 \cdot \delta_{\alpha_1\ovl{\alpha}_2} \delta_{\eta_1 \ovl\eta_2} \delta_{s_1 s_2}
- \td{\mJ} \cdot \delta_{\alpha_1\ovl{\alpha}_2} \delta_{\eta_1 \ovl\eta_2}\ .
\end{equation}
It is direct to obtain 
\begin{equation}
    \kappa[O^{xx0},\td{\Gamma}^0] = - \td{U}_1 - 2\td{\mJ},\qquad 
    \kappa[O^{xxz},\td{\Gamma}^0] = -\td{U}_1\ . 
\end{equation}
For later reference, we explicitly write down the quasi-particle susceptibilities here. 
Taking $O=O^{000}$, $O^{0z0}$, $O^{z00}$, $O^{zz0}$, $O^{00z}$, we obtain the charge, valley, orbital, angular momentum, and spin susceptibilities
\begin{equation} \label{eq:chi-c}
    \td{\chi}^c \approx \chi^{c} = 8 \frac{\sin^2\delta_f}{\pi \td{\Delta}_0}\pare{  1 - \frac{\sin^2\delta_f}{\pi \td{\Delta}_0} ( 2\td{U}_1 + 2\td{U}_2 + 2\td{U}_3 + \td{U}_4 + \td{\mJ} ) },
    \qquad \text{(exact)}\ , 
\end{equation}
\begin{equation}
    \td{\chi}^v \approx \chi^{v} = 8 \frac{\sin^2\delta_f}{\pi \td{\Delta}_0}\pare{  1 - \frac{\sin^2\delta_f}{\pi \td{\Delta}_0} ( -2\td{U}_1 - 2\td{U}_2 + 2\td{U}_3 + \td{U}_4 - \td{\mJ} ) },
    \qquad \text{(exact)}\ , 
\end{equation}
\begin{equation} 
    \td{\chi}^o \approx \chi^{o} = 8 \frac{\sin^2\delta_f}{\pi \td{\Delta}_0}\pare{  1 - \frac{\sin^2\delta_f}{\pi \td{\Delta}_0} ( -2\td{U}_1 + 2\td{U}_2 - 2\td{U}_3 + \td{U}_4 - \td{\mJ} ) },
    \qquad \text{(exact)}\ ,
\end{equation}
\begin{equation} 
    \td{\chi}^a \approx \chi^{a} = 8 \frac{\sin^2\delta_f}{\pi \td{\Delta}_0}\pare{  1 - \frac{\sin^2\delta_f}{\pi \td{\Delta}_0} ( 2\td{U}_1 - 2\td{U}_2 - 2\td{U}_3 + \td{U}_4 + \td{\mJ} ) },
    \qquad \text{(exact)}\ ,
\end{equation}
\begin{equation} \label{eq:chi-s}
    \td{\chi}^s \approx \chi^{s} = 8 \frac{\sin^2\delta_f}{\pi \td{\Delta}_0}\pare{  1 - \frac{\sin^2\delta_f}{\pi \td{\Delta}_0} ( -\td{U}_4 + \td{\mJ} ) },
    \qquad \text{(exact)}\ . 
\end{equation}
As the above five channels are given by generators of the continuous symmetry, they are exact susceptibilities of the bare particles, as will be proven in the next subsection. 
The remaining channels $O^{0zz}$, $O^{x00}$, $O^{xz0}$, $O^{0x0}$, $O^{zx0}$, $O^{xx0}$, $O^{xxz}$ give 

\begin{equation} \label{eq:chi-0zz}
    \td{\chi}^{0zz} \approx 8 \frac{\sin^2\delta_f}{\pi \td{\Delta}_0}\pare{  1 - \frac{\sin^2\delta_f}{\pi \td{\Delta}_0} ( - \td{U}_4 - \td{\mJ} ) }\ ,
\end{equation}
\begin{equation}
    \td{\chi}^{x00} = \td{\chi}^{xz0} \approx 8 \frac{\sin^2\delta_f}{\pi \td{\Delta}_0}\pare{  1 - \frac{\sin^2\delta_f}{\pi \td{\Delta}_0} ( - \td{U}_3 ) }\ ,
\end{equation}
\begin{equation}
    \td{\chi}^{0x0} = \td{\chi}^{zx0} \approx 8 \frac{\sin^2\delta_f}{\pi \td{\Delta}_0}\pare{  1 - \frac{\sin^2\delta_f}{\pi \td{\Delta}_0} ( - \td{U}_2 ) }\ ,
\end{equation}
\begin{equation}
    \td{\chi}^{xx0} \approx 8 \frac{\sin^2\delta_f}{\pi \td{\Delta}_0}\pare{  1 - \frac{\sin^2\delta_f}{\pi \td{\Delta}_0} ( - \td{U}_1 - 2\td{\mJ} ) }\ ,
\end{equation}
\begin{equation} \label{eq:chi-xxz}
    \td{\chi}^{xxz} \approx 8 \frac{\sin^2\delta_f}{\pi \td{\Delta}_0}\pare{  1 - \frac{\sin^2\delta_f}{\pi \td{\Delta}_0} ( - \td{U}_1  ) }\ . 
\end{equation}

\subsection{Ward identities and exact susceptibilities *}
\label{sec:ward}

We consider a perturbation 
\begin{equation}
    \Delta S = \sum_{1} f^\dagger (1) V(1) f(1)
\end{equation}
where $V(1)=V_{\alpha_1 \eta_1 s_1}$ is a diagonal matrix and is independent of frequency. 
Then the first order correction to the Green's function can be written as 
\begin{align}
\delta \mG(1) =& -\inn{f(1)f^\dagger(1) (-\Delta S) e^{-S_I}   }_{0C} \nonumber\\
=& \sum_{2} \wick{ \c1 f(1) \c2 f^\dagger(1) \c1 f^\dagger(2) V(2) \c2 f(2) }  - 4\times \frac{T}4 \sum_{23456} 
    \wick{\c1 f(1)  \c2 f^\dagger(1)  \c1 f^\dagger(3) \c3 f^\dagger(4) \c4 f(5) \c2 f(6)
    \c4 f^\dagger(2)  \c3 f(2)
    }
     \cdot \Gamma^0(3,4;5,6) \cdot V(2) + \cdots \nonumber \\ 
=& [\mG^0(1)]^2 \cdot V(1) + T \sum_{2} [\mG^0(1)]^2 \cdot \Gamma^0(1,2;2,1) \cdot [\mG^0(2)]^2 \cdot V(2)  + \cdots 
\end{align}
The subscript $0C$ in the first row represents the connected diagrams, and the factor $4$ in the second term of the second row comes from four equivalent contractions. 
Summing over all the terms to infinite order, we obtain the {\it exact} result
\begin{equation} \label{eq:dG-general}
    \delta \mG(1) = [\mG(1)]^2 \cdot V(1) + T \sum_{2} [\mG(1)]^2 \cdot \Gamma(1,2;2,1) \cdot [\mG(2)]^2 \cdot V(2)\ ,
\end{equation}
which can be represented by the skeleton diagram
\begin{equation}
    \includegraphics[width=0.35\linewidth]{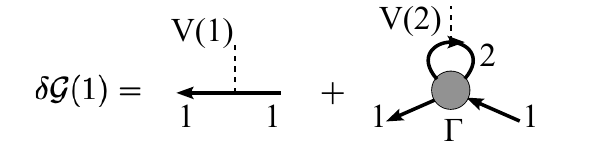}
\end{equation}
Here bare Green's functions and vertex function are replaced by the full Green's functions and vertex function respectively. 
Writing $\delta \mG(1) = - [\mG(1)]^2 \cdot (- V(1) - \delta \Sigma(1) )$, we obtain the perturbed self-energy 
\begin{equation}
    \delta \Sigma_{I_1}(\omega_1) = \sum_{I_{2}}\int \frac{d\omega_2}{2\pi} \cdot 
    \Gamma_{I_1, I_2; I_2, I_1}(\omega_1,\omega_2;\omega_2,\omega_1) \cdot [\mG(\omega_2)]^2 \cdot  V_{I_2} \ .  
\end{equation}
Here $I_i = (\alpha_i,\eta_i,s_i)$ is a composite index. 
We consider the perturbation term $V_I = - h^O \cdot O_I$ with $O$ being a diagonal matrix, then there is 
\begin{equation}\label{eq:dSigma-dh}
    \sum_{I_1} \frac{\partial \Sigma_{I_1} (\omega_1)}{\partial h^O} O_{I_1}
= - \sum_{I_1 I_2} \int \frac{d\omega_2}{2\pi} \cdot O_{I_1} O_{I_2} \cdot  \Gamma_{I_1,I_2;I_2,I_1}(\omega_1,\omega_2;\omega_2,\omega_1) \cdot [\mG(\omega_2)]^2 \ . 
\end{equation}

We then consider a gauge transformation $f_{I}'(\omega) = f_{I}(\omega+\nu\cdot O_I)$. 
On the one hand, there must be 
\begin{align} \label{eq:dG-tmp}
& \mG_{I_1}'(\omega_1) - \mG(\omega_1) = \mG(\omega_1+\nu\cdot O_{I_1})- \mG(\omega_1)
= -[\mG(\omega_1)]^2 \cdot \frac{\partial}{\partial \omega_1 }  \pare{   
    [\mG^{0} (\omega_1)]^{-1} - \Sigma(\omega_1) } \cdot \nu \cdot O_{I_1} + \mathcal{O}(\nu^2) \nonumber\\
=& -[\mG(\omega_1)]^2 \cdot \frac{\partial [\mG^{0} (\omega_1)]^{-1} }{\partial \omega_1} \cdot \nu \cdot O_{I_1}
+ [\mG(\omega_1)]^2 \frac{\partial \Sigma(\omega)}{\partial i\omega} \cdot i\nu\cdot O_{I_1} + \mathcal{O}(\nu^2) 
\end{align}
Here $\mG^{0}(\omega_1)=(i\omega_1+i\Delta_0\cdot\sgn(\omega_1)-\ee_f)^{-1}$ is the free propagator. 
On the other hand, one can rewrite the action $S[f^\dagger,f]$ in terms of $S[f^{\prime\dagger},f']+ \Delta S[f^{\prime\dagger},f']$, with the perturbation
\begin{equation}
    \Delta S = \sum_{1} f^{\prime\dagger} (1) (  -i \nu \cdot O(1)  - i \nu \cdot 2\Delta_0 \delta(\omega_1) O(1)   ) f'(1) = \sum_{1} f^{\prime\dagger} (1) \pare{ -\nu \cdot O(1)  \partial_{\omega_1} [\mG^{0}(1)]^{-1} } f'(1) \ . 
\end{equation}
There is no other perturbation term in $\Delta S$ {\it as long as $O$ is a generator of the continuous symmetry} such that it leaves all the instantaneous terms ($\omega$-independent terms) invariant. 
Substituting $V(1)= -\nu \cdot O(1) \partial_{\omega_1} [\mG^{0}(1)]^{-1} $ into \cref{eq:dG-general} and comparing it to \cref{eq:dG-tmp}, one obtains 
\begin{align}\label{eq:Sigma-temp1}
    \frac{\partial \Sigma(\omega_1)}{\partial i\omega_1} \cdot i\nu\cdot O_{I_1}
=& - \sum_{I_2} \int \frac{d\omega_2}{2\pi}  \Gamma_{I_1, I_2; I_2, I_1}(\omega_1,\omega_2;\omega_2,\omega_1) \cdot [\mG(\omega_2)]^2 \cdot \nu\cdot O_{I_2} \cdot \partial_{\omega_2} [\mG^{0} (\omega_2)]^{-1}
\end{align}
We do not explicitly write $\partial_{\omega_2} [\mG^{0}(\omega_2)]^{-1}$ as $i + 2i \Delta_0 \delta(\omega_2)$ because it will lead to the ill-defined term $[\mG(\omega_2)]^2 \delta(\omega_2)$, where $\mG(\omega_2)$ is discontinuous at $\omega_2=0$. 
This problem comes from abuse of the $\delta$-function. 
Here we avoid this problem by rewriting $[\mG(\omega_2)]^2\partial_{\omega_2} [\mG^{0}(\omega_2)]^{-1}$ as 
\begin{align}
[\mG(\omega_2)]^2 \partial_{\omega_2} [\mG^{0}(\omega_2)]^{-1}
= [\mG(\omega_2)]^2 \pare{ \partial_{\omega_2} [\mG(\omega_2)]^{-1} + \partial_{\omega_2}\Sigma(\omega_2)   }
= -\partial_{\omega_2}\mG(\omega_2) + [\mG(\omega_2)]^2 \partial_{\omega_2}\Sigma(\omega_2)
\end{align}
The derivative $\partial_{\omega_2}\mG(\omega_2)$ consists of a continuous part at $\omega_2\neq0$ and a $\delta$-function peak at $\omega_2=0$ that is responsible for the discontinuity of  $\mG(\omega_2)$, {\it i.e.,}
\begin{equation}\label{eq:G-derivative}
    \partial_{\omega_2}\mG(\omega_2) = -[\mG(\omega_2)]^2 (i - \partial_{\omega_2}\Sigma(\omega_2)) + (\mG(0^+) - \mG(0^-)) \delta(\omega_2)\ .
\end{equation}
This expression is correct in the sense that integrating it gives the correct antiderivative.  
Then there is
\begin{equation}
    [\mG(\omega_2)]^2 \partial_{\omega_2} [\mG^{0}(\omega_2)]^{-1}
= -(\mG(0^+) - \mG(0^-)) \delta(\omega_2) + i  [\mG(\omega_2)]^2\ .
\end{equation}
Substituting this expression into \cref{eq:Sigma-temp1} yields 
\begin{align}
\frac{\partial \Sigma(\omega_1)}{\partial i\omega_1} \cdot O_{I_1}
=& \frac{\mG(0^+)-\mG(0^-)}{2\pi i} \sum_{I_2} \Gamma_{I_1, I_2; I_2, I_1}(\omega_1,0;0,\omega_1) \cdot O_{I_2} - \sum_{I_2} \int \frac{d\omega_2}{2\pi}  \Gamma_{I_1, I_2; I_2, I_1}(\omega_1,\omega_2;\omega_2,\omega_1) \cdot [\mG(\omega_2)]^2 \cdot O_{I_2}
\end{align}
According to the Friedel sum rule (\cref{eq:A(0)-Friedel}), there is 
\begin{equation}
    \frac{\mG(0^+)-\mG(0^-)}{2\pi i} = - A(0) = - \frac{\sin^2(\delta_f)}{\pi\Delta_0}
\end{equation}
and hence
\begin{equation} \label{eq:dSigma-dw}
    \frac{\partial \Sigma(\omega_1)}{\partial i\omega_1} \cdot O_{I_1}
= - \frac{\sin^2(\delta_f)}{\pi \Delta_0} \sum_{I_2} \Gamma_{I_1, I_2; I_2, I_1}(\omega_1,0;0,\omega_1) \cdot O_{I_2} - \sum_{I_2} \int \frac{d\omega_2}{2\pi}  \Gamma_{I_1, I_2; I_2, I_1}(\omega_1,\omega_2;\omega_2,\omega_1) \cdot [\mG(\omega_2)]^2 \cdot O_{I_2}\ . 
\end{equation}
Comparing \cref{eq:dSigma-dw} to \cref{eq:dSigma-dh}, we have 
\begin{equation} \label{eq:Ward-1}
    \frac{\partial \Sigma(\omega_1)}{\partial i\omega_1} \sum_{I_1}  O_{I_1}^2 
= \sum_{I_1} \frac{\partial \Sigma_{I_1} (\omega_1)}{\partial h^O} O_{I_1}
  - \frac{\sin^2(\delta_f)}{\pi \Delta_0} \sum_{I_1 I_2} \Gamma_{I_1, I_2; I_2, I_1}(\omega_1,0;0,\omega_1) \cdot O_{I_1} O_{I_2}\ .
\end{equation}

Now we consider the response $ \ovl{O} = \sum_{I} O_I \inn{ f^\dagger_I f_I }$ 
\begin{align}
 \ovl{O} = - \sum_{I} O_I \inn{ f_I(-0^+) f_I^\dagger(0) } 
 = T \sum_{I} O_I \sum_{\omega} e^{i\omega 0^+} \mG_I(\omega) 
\end{align}
to the external field $-h^O O_I$. 
{\small
\begin{align}
\chi^O = \frac{\partial \ovl{O}}{\partial h^O} 
= - T \sum_{I} O_I \sum_{\omega} e^{i\omega 0^+} [\mG(\omega)]^2 
    \frac{\partial}{\partial h^O} [  
    h^O O_I - \Sigma_I(\omega) ]
= T \sum_{I} O_I \sum_{\omega} e^{i\omega 0^+} [\mG(\omega)]^2 
[ - O_I + \frac{\partial \Sigma_I(\omega)}{\partial h^O} ]\ .
\end{align}}
Making use of the Ward identity \cref{eq:Ward-1}, we can replace the $h$-derivative by the $\omega$-derivative 
\begin{align}
\chi^O
=& T \sum_I O_I^2 \sum_{\omega} e^{i\omega 0^+} [\mG(\omega)]^2 
    \brak{  - 1 + \frac{\partial \Sigma(\omega)}{\partial i\omega}
    } 
  +T \frac{\sin^2(\delta_f)}{\pi \Delta_0}  \sum_{I I_2}
  \sum_{\omega} e^{i\omega 0^+} [\mG(\omega)]^2 \cdot \Gamma_{I,I_2;I_2,I}(\omega,0;0,\omega) O_I O_{I_2}
\end{align}
$[\mG_I(\omega)]^2 \brak{  - 1 + \frac{\partial \Sigma(\omega)}{\partial i\omega}} $ is the continuous part of the Green's function derivative (\cref{eq:G-derivative}), and it can be rewritten as 
\begin{equation}
[\mG_I(\omega)]^2  \brak{  - 1 + \frac{\partial \Sigma(\omega)}{\partial i\omega}  }
= \frac{\partial \mG(\omega)}{\partial i\omega} - \frac{\mG(0^+)-\mG(0^-)}{i}\delta(\omega)
=  \frac{\partial \mG(\omega)}{\partial i\omega} + 2\pi \frac{\sin^2\delta_f}{\pi \Delta_0} \delta(\omega)\ . 
\end{equation}
Thus, 
\begin{equation}
\chi^O = T \sum_I O_I^2 \sum_{\omega} e^{i\omega 0^+} \frac{\partial \mG(\omega)}{\partial i\omega} + \frac{\sin^2\delta_f}{\pi \Delta_0} \sum_I O_I^2 +T \frac{\sin^2(\delta_f)}{\pi \Delta_0}  \sum_{I I_2}
\sum_{\omega} e^{i\omega 0^+} [\mG(\omega)]^2 \cdot \Gamma_{I,I_2;I_2,I}(\omega,0;0,\omega) O_I O_{I_2}\ ,
\end{equation}
where the first term vanishes. 
Making use of the identity \cref{eq:dSigma-dh}, we rewrite the above equation as
\begin{align}
\chi^O
=& \frac{\sin^2\delta_f}{\pi \Delta_0} \sum_I O_I^2
 - \frac{\sin^2(\delta_f)}{\pi \Delta_0} \cdot \sum_I \frac{\partial \Sigma_I(0)}{\partial h^O} O_I 
=\frac{\sin^2(\delta_f)}{\pi \Delta_0} \sum_I \pare{  O_I^2 -  \frac{\partial \Sigma_I(0)}{\partial h^O} O_I }\ . 
\end{align} 
Now we have expressed the exact susceptibility in terms of the derivatives of  zero-frequency self-energy. 
To relate the susceptibility to the vertex function, we replace the $\partial_{h^O} \Sigma_{I}(0)$ by the $\omega$-derivative via \cref{eq:Ward-1}
\begin{align}
\chi^O = \frac{\sin^2(\delta_f)}{\pi \Delta_0} \sum_I O_I^2 \pare{ 1 - \frac{\Sigma(\omega)}{\partial i\omega}  }_{\omega=0}
 - \pare{\frac{\sin^2(\delta_f)}{\pi \Delta_0}}^2 \sum_{I_1 I_2} 
    \Gamma_{I_1 I_2; I_2 I_1}(00;00) \cdot O_{I_1} O_{I_2} \ . 
\end{align}
Notice that $ \pare{ 1 - \frac{\Sigma(\omega)}{\partial i\omega}  }_{\omega=0}$ equals inverse of the quasi-particle weight $z^{-1}$, there is 
\begin{equation}
\chi^O = \frac{\sin^2(\delta_f)}{z \cdot \pi \Delta_0} 
\pare{  \sum_I O_I^2  - \frac{\sin^2(\delta_f)}{z \cdot \pi \Delta_0}  \sum_{I_1 I_2} z^2 \Gamma_{I_1 I_2; I_2 I_1}(00;00) \cdot O_{I_1} O_{I_2} }\ .
\end{equation}
We now have expressed the {\it exact} susceptibility in terms of quasi-particle weight $z$, hybridization $\Delta_0$, filling $\delta_f = \pi(\nu_f+4)/8$, and {\it full vertex function at zero-frequency}. 
It is worth emphasizing that this identity applies only when $O$ is a generator of the continuous symmetry. 
Applying it to the $O$ operators discussed in the last subsection, we find the five susceptibilities in Eqs.~(\ref{eq:chi-c}) to (\ref{eq:chi-s}) are exact.

\subsection{Pairing susceptibilities and irreducible vertex in Cooper channel}
\label{sec:irreducible}

Following the same procedure in \cref{sec:chi}, we can calculate the pairing susceptibilities $\chi_{\rm p}$.

One can diagonalize the pairing susceptibilities by enumerating distinct two-electron representations, which have been discussed in detail in \cref{sec:flatten-HI}.  
There are five non-degenerate two-electron levels (\cref{eq:2particle-1,eq:2particle-2,eq:2particle-3,eq:2particle-4,eq:2particle-5}).
To proceed, we choose the following five operators to represent the five channels
\begin{equation}
\sum_{I_1 I_2} f_{I_1}^\dagger O^{0xy}_{I_1 I_2} f_{I_2}^\dagger 
\qquad \text{(intra-orbital inter-valley singlet)}\ ,
\end{equation}
\begin{equation}
\sum_{I_1 I_2} f_{I_1}^\dagger O^{x0y}_{I_1 I_2} f_{I_2}^\dagger
\qquad \text{(inter-orbital intra-valley singlet)}\ ,
\end{equation}
\begin{equation}
\sum_{I_1 I_2} f_{I_1}^\dagger O^{00y}_{I_1 I_2} f_{I_2}^\dagger
\qquad \text{(intra-orbital intra-valley singlet)}\ ,
\end{equation}
\begin{equation}
\sum_{I_1 I_2} f_{I_1}^\dagger O^{xxy}_{I_1 I_2} f_{I_2}^\dagger
\qquad \text{(inter-orbital inter-valley singlet)}\ ,
\end{equation}
\begin{equation}
\sum_{I_1 I_2} f_{I_1}^\dagger O^{xy0}_{I_1 I_2} f_{I_2}^\dagger
\qquad \text{(inter-orbital inter-valley triplet)}\ ,
\end{equation}
where $O^{\mu\nu\rho} = \sigma^\mu \tau^\nu \spin^\rho$.
Each of them is chosen as a linear combination of a few degenerate two-electron states in \cref{eq:2particle-1,eq:2particle-2,eq:2particle-3,eq:2particle-4,eq:2particle-5}. 

To apply \cref{eq:chi-def} to pairing susceptibility, we should define $\hat{A} = \sum_{I_1 I_2} f_{I_2} O^*_{I_1,I_2} f_{I_1}$,
$\hat{B} = \sum_{I_1 I_2} f_{I_1}^\dagger O_{I_1,I_2} f^\dagger_{I_2}$, for $O$ matrices given in the above equations. 
Notice that $O$ is always anti-symmetric. 
Fouriering Eq. (\ref{eq:chi-def}), 
\begin{equation}
    \chi^O_{\rm p}(\nu) = T \sum_{1234} \delta_{\omega_1+\omega_2, \nu} \delta_{\omega_3+\omega_4, \nu} \Inn{ e^{-S_I} f(2) O^*(1,2) f(1)  f^\dagger(3) O(3,4) f^\dagger(4) }_{0C} \ . 
\end{equation}
At the zeroth order of $S_I$, 
\begin{align}
\chi^{O(0)}_{\rm p}(\nu) &= T \sum_{1234} \delta_{\omega_1+\omega_2, \nu} \delta_{\omega_3+\omega_4, \nu} 
    \pare{ \wick{  \c1 f(2) O^*(1,2) \c2 f(1)  \c2 f^\dagger(3) O(3,4) \c1 f^\dagger(4) } - \wick{ \c1 f(2) O^*(1,2) \c2 f(1)  \c1 f^\dagger(3) O(3,4) \c2 f^\dagger(4) } } \nonumber \\
&= \sum_{12} \left[ O^*(1,2) O(1,2) - O^*(1,2) O(2,1) \right] T \sum_{\omega_1} \mG^0(\omega_1) \mG^0(-\omega_1+\nu)  \nonumber \\
&\overset{\nu\to0^+}{=}  \sum_{12} 2 \cdot  O^*(1,2) O(1,2) \cdot \frac{\tan \delta_f}{\pi \Delta_0} \left( \frac{\pi}{2} - \delta_f \right) 
\end{align}
where we used
\begin{eqnarray}
    T \sum_{\omega} \frac{1}{i\omega - \epsilon_f + i\Delta(\omega)} \frac{1}{-i\omega - \epsilon_f + i\Delta(-\omega)} &=& 2 T \sum_{\omega>0} \frac{1}{(\omega + \Delta_0)^2 + (\epsilon_f)^2} = \frac{1}{\pi} \int_0^{\infty} \mrm{d}\omega \frac{1}{(\omega+\Delta_0)^2 + (\epsilon_f)^2} \\\nonumber
    &=& \frac{1}{\pi\epsilon_f} \left( \frac{\pi}{2} - \arctan\frac{\Delta_0}{\epsilon_f} \right) = \frac{\tan \delta_f}{\pi \Delta_0} \left( \frac{\pi}{2} - \delta_f \right)
\end{eqnarray}
and $O(1,2) = -O(2,1)$. 
Thus the zeroth order susceptibility can be written as 
\begin{equation}
    \chi^{O(0)}_{\rm p} = 2 \cdot \tr \left[ O^\dagger O \right] \cdot \frac{\tan \delta_f}{\pi \Delta_0} \left( \frac{\pi}{2} - \delta_f \right) \ . 
\end{equation}
At the first order of $S_I$, 
\begin{eqnarray}
    \chi^{O(1)}_{\rm p}(\nu) &=& - T^2 \sum_{1234} \delta_{\omega_1+\omega_2, \nu} \delta_{\omega_3+\omega_4, \nu} O^*(1,2) O(3,4) \sum_{5678}  \frac{1}{4} \delta_{\omega_5+\omega_6, \omega_7+\omega_8} \Gamma^0(5,6;7,8) \\\nonumber
    &~& \times \Inn{ f(2) f(1) f^\dagger(3) f^\dagger(4) f^\dagger(5) f^\dagger(6) f(7) f(8) }_{0C} \\\nonumber
    &=& - T^2 \sum_{1234} \delta_{\omega_1+\omega_2, \nu} \delta_{\omega_3+\omega_4, \nu} O^*(1,2) O(3,4) \sum_{5678}  \frac{1}{4} \delta_{\omega_5+\omega_6, \omega_7+\omega_8} \Gamma^0(5,6;7,8) \\\nonumber
    &~& \times 4 \times \wick{ \c1 f(2) \c2 f(1) \c3 f^\dagger(3) \c4 f^\dagger(4) \c1 f^\dagger(5) \c2 f^\dagger(6) \c3 f(7) \c4 f(8) } 
\end{eqnarray}
where the $\times 4$ factor is contributed by equivalent contractions. The contraction gives 
\begin{eqnarray}
    (-1)^2 \delta_{25} \delta_{16} \delta_{37} \delta_{48} \mG^0(\omega_1) \mG^0(\omega_2) \mG^0(\omega_3) \mG^0(\omega_4)\ ,
\end{eqnarray}
thereby, 
\begin{eqnarray}
    \chi^{O(1)}_{\rm p}(0^+) &=& - T^2 \sum_{1234} O^*(1,2) O(3,4) \Gamma^0(2,1;3,4) \sum_{\omega_1, \omega_3} \mG^0(\omega_1) \mG^0(-\omega_1) \mG^0(\omega_3) \mG^0(-\omega_3) \\\nonumber
    &=& - \left[\sum_{1234} O^*(1,2) \Gamma^0(2,1;3,4) O(3,4) \right] \left( \frac{\tan \delta_f}{\pi\Delta_0} (\frac{\pi}{2} - \delta_f) \right)^2\ .
\end{eqnarray}

Replacing ${\Delta}_0$, $\Gamma^0$ by the effective parameters of quasi-particles $\td{\Delta}_0$, $\td{\Gamma}^0$, we obtain the quasi-particle contributed pairing susceptibilities 
\begin{equation} \label{eq:chi-pairing}
\td{\chi}^O_{\rm p} \approx 2\cdot \tr [O^\dagger O] \cdot \frac{\tan \delta_f}{\pi \td\Delta_0} \pare{ \frac{\pi}{2} - \delta_f } 
    \brak{ 1 - \kappa_{\rm p} [O,\td{\Gamma}^0] \cdot  \frac{\tan \delta_f}{\pi \td\Delta_0} \pare{ \frac{\pi}{2} - \delta_f }  }
\end{equation}
where 
\begin{equation}
\kappa_{\rm p} [O,\td{\Gamma}^0]  = \frac1{2\cdot \tr[O^\dagger O]} \sum_{I_{1,2,3,4}} O^*_{I_2,I_1} \td{\Gamma}^0_{I_1,I_2;I_3,I_4}(0,0;0,0) O_{I_3,I_4} \ . 
\end{equation}
One should be aware that $\td{\Gamma}^0$ equals the exact full vertex function $\td{\Gamma}$ of quasi-particles at zero frequency, but the frequency dependence of $\td{\Gamma}$ has been omitted in the above equation.
We have also omitted the frequency dependence of the quasi-particle self-energy $\td{\Sigma}(\omega)$ (beyond quasi-particle weight correction), {\it e.g.}, $\omega^2$ terms in $\td{\Sigma}(\omega)$. 
Therefore, \cref{eq:chi-pairing} is an {\it approximate} result. 

We find $\kappa_{\rm p} [O,\td{\Gamma}^0]$ is nothing but the energy  (in terms of bare $\td{\Gamma}^0$) of the two-electron states defined by $O$. 
Readers may directly verify this. 
Here we prove this statement. 
$O$ defines a two-electron state $\ket{\Phi}= \sum_{IJ} O_{IJ} f_I^\dagger f_J^\dagger \ket{0}$.
Its norm is given by $\inn{\Phi|\Phi} =\sum_{IJ} O_{IJ}^*O_{IJ} - O_{IJ}^*O_{JI} = 2\cdot \tr[O^\dagger O]$. 
Then $\kappa_{\rm p} [O,\td{\Gamma}^0]$ is by definition $\inn{\Phi| H_{\rm eff} |\Phi} / \inn{\Phi|\Phi}$ with $H_{\rm eff}$ being the effective Hamiltonian defined by $\td{\Gamma}^0$. 
For the two-electron states given in \cref{eq:2particle-1,eq:2particle-2,eq:2particle-3,eq:2particle-4,eq:2particle-5}, where $O$ is chosen as 
$O^{0xy}$, $O^{x0y}$, $O^{00y}$, $O^{xxy}$,  $O^{xy0}$, respectively, the corresponding pairing susceptibilities are
\begin{equation} \label{eq:chip-1}
\td{\chi}_{\rm p}^{0xy} 
\approx 16 \frac{\tan \delta_f}{\pi \td\Delta_0} \pare{ \frac{\pi}{2} - \delta_f } 
    \brak{ 1 - \frac{\tan \delta_f}{\pi \td\Delta_0} \pare{ \frac{\pi}{2} - \delta_f } \td{U}_2  }\ ,
\end{equation}
\begin{equation}\label{eq:chip-2}
\td{\chi}_{\rm p}^{x0y} 
\approx 16 \frac{\tan \delta_f}{\pi \td\Delta_0} \pare{ \frac{\pi}{2} - \delta_f } 
    \brak{ 1 - \frac{\tan \delta_f}{\pi \td\Delta_0} \pare{ \frac{\pi}{2} - \delta_f }
    \td U_3    }\ ,
\end{equation}
\begin{equation}\label{eq:chip-3}
\td{\chi}_{\rm p}^{00y} 
\approx 16 \frac{\tan \delta_f}{\pi \td\Delta_0} \pare{ \frac{\pi}{2} - \delta_f } 
    \brak{ 1 - \frac{\tan \delta_f}{\pi \td\Delta_0} \pare{ \frac{\pi}{2} - \delta_f }
    \td U_4    }\ ,
\end{equation}
\begin{equation}\label{eq:chip-4}
\td{\chi}_{\rm p}^{xxy} 
\approx 16 \frac{\tan \delta_f}{\pi \td\Delta_0} \pare{ \frac{\pi}{2} - \delta_f } 
    \brak{ 1 - \frac{\tan \delta_f}{\pi \td\Delta_0} \pare{ \frac{\pi}{2} - \delta_f }
    \pare{ \td{U}_1 - \td{\mJ} }    }\ ,
\end{equation}
\begin{equation}\label{eq:chip-5}
\td{\chi}_{\rm p}^{xy0} 
\approx 16 \frac{\tan \delta_f}{\pi \td\Delta_0} \pare{ \frac{\pi}{2} - \delta_f } 
    \brak{ 1 - \frac{\tan \delta_f}{\pi \td\Delta_0} \pare{ \frac{\pi}{2} - \delta_f }
    \pare{ \td{U}_1 + \td{\mJ} }    }\ . 
\end{equation}

We consider to decompose $\td{\chi}_{\rm p}$ into a summation of ladder diagrams
\begin{equation}\label{eq:irreducible}
\includegraphics[width=0.9\linewidth]{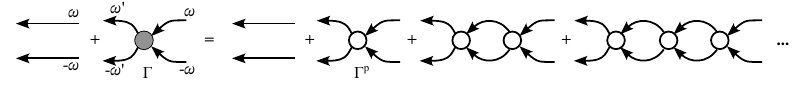}
\end{equation}
where $\Gamma^{\rm p}$ is the two-particle irreducible vertex in the pairing channel.
$\Gamma^{\rm p}$ cannot be divided into two by cutting two left-going Green's function lines. 
Due to the $\rm U(1) \!\times\! U(1) \!\times\! U(2) \!\times\! U(2)$ symmetry, $\Gamma^{\rm p}$ must have the same form as $\Gamma$, and can be parameterized by $\Gamma^{\rm p}_{U1,2,3,4}$, ${\Gamma}^{\rm p}_{\mJ}$. 
We define the effective interaction in Cooper channel as $U_{1,2,3,4}^{\rm p} = z^2 \Gamma^{\rm p}_{U1,2,3,4}$,  $\mJ^{\rm p} = z^2\Gamma^{\rm p}_\mJ$. 
Neglecting the frequency dependencies of $\Gamma^{\rm p}$ and the renormalized self-energy, the pairing susceptibilities can be written in terms of effective interactions as 
\begin{equation} 
\td{\chi}_{\rm p}^{0xy} 
\approx 16 \frac{\tan \delta_f}{\pi \td\Delta_0} \pare{ \frac{\pi}{2} - \delta_f } 
    \brak{ 1 + \frac{\tan \delta_f}{\pi \td\Delta_0} \pare{ \frac{\pi}{2} - \delta_f }
    U_2^{\rm p}    }^{-1} \ ,
\end{equation}
\begin{equation}
\td{\chi}_{\rm p}^{x0y} 
\approx 16 \frac{\tan \delta_f}{\pi \td\Delta_0} \pare{ \frac{\pi}{2} - \delta_f } 
    \brak{ 1 + \frac{\tan \delta_f}{\pi \td\Delta_0} \pare{ \frac{\pi}{2} - \delta_f }
    U_3^{\rm p}   }^{-1} \ ,
\end{equation}
\begin{equation}
\td{\chi}_{\rm p}^{00y} 
\approx 16 \frac{\tan \delta_f}{\pi \td\Delta_0} \pare{ \frac{\pi}{2} - \delta_f } 
    \brak{ 1+ \frac{\tan \delta_f}{\pi \td\Delta_0} \pare{ \frac{\pi}{2} - \delta_f }
    U_4^{\rm p}   }^{-1} \ ,
\end{equation}
\begin{equation}
\td{\chi}_{\rm p}^{xxy} 
\approx 16 \frac{\tan \delta_f}{\pi \td\Delta_0} \pare{ \frac{\pi}{2} - \delta_f } 
    \brak{ 1 + \frac{\tan \delta_f}{\pi \td\Delta_0} \pare{ \frac{\pi}{2} - \delta_f }
    \pare{ U_1^{\rm p} - \mJ^{\rm p} }    }^{-1}\ ,
\end{equation}
\begin{equation}
\td{\chi}_{\rm p}^{xy0} 
\approx 16 \frac{\tan \delta_f}{\pi \td\Delta_0} \pare{ \frac{\pi}{2} - \delta_f } 
    \brak{ 1 + \frac{\tan \delta_f}{\pi \td\Delta_0} \pare{ \frac{\pi}{2} - \delta_f }
    \pare{ U_1^{\rm p} + \mJ^{\rm p} }    }^{-1}\ .
\end{equation}
Equaling these equations to Eqs.~(\ref{eq:chip-1}) to (\ref{eq:chip-5}), we obtain the effective interactions (in the pairing channel) in terms of the renormalized interactions 
\begin{equation} \label{eq:Ueff1}
U^{\rm p}_2 = \frac{\td U_2}{1 - \frac{\tan \delta_f}{\pi \td\Delta_0} \pare{ \frac{\pi}{2} - \delta_f } \td{U}_2} \ ,\qquad 
U^{\rm p}_3 = \frac{\td U_3}{1 - \frac{\tan \delta_f}{\pi \td\Delta_0} \pare{ \frac{\pi}{2} - \delta_f } \td{U}_3}\ , \qquad
U^{\rm p}_4 = \frac{\td U_4}{1 - \frac{\tan \delta_f}{\pi \td\Delta_0} \pare{ \frac{\pi}{2} - \delta_f } \td{U}_4}\ ,
\end{equation}
\begin{equation} \label{eq:Ueff2}
U^{\rm p}_1  - {\mJ}^{\rm p} = \frac{\td U_1  - \td{\mJ} }{1 - \frac{\tan \delta_f}{\pi \td\Delta_0} \pare{ \frac{\pi}{2} - \delta_f } (\td U_1  - \td{\mJ}) } \ ,\qquad 
U^{\rm p}_1  + {\mJ}^{\rm p} = \frac{\td U_1  + \td{\mJ} }{1 - \frac{\tan \delta_f}{\pi \td\Delta_0} \pare{ \frac{\pi}{2} - \delta_f } (\td U_1  + \td{\mJ}) }\ . 
\end{equation}

\subsection{Asymptotic behavior of exact vertex function}
\label{sec:mott-limit}
\subsubsection{Bethe ansatz as a benchmark}
As reviewed in this subsubsection, Ward identity has been used to extract local 1PI previously \cite{hewson_renormalized_1993}, which reproduces the Bethe ansatz results on the one-orbital half-filled Anderson impurity as shown by Ref.~\cite{pandis_development_2015}.
The charge and spin susceptibilities of this model \cite{kawakami_ground_1982,wiegmann_exact_1983} are 
\begin{align}
\chi^c = \frac1{4\pi U} \int_{-\infty}^{\infty} {\rm d}y \cdot e^{- \frac{\pi Uy^2}{2\Delta_0} } \frac{ (U/2\Delta_0)^{\frac32} }{ (y+\frac12)^2 (U/2\Delta_0)^2 + \frac14 } \ . 
\end{align}
\begin{align}
    \chi^s = \frac{1}{4\pi U}\left[2\pi\sqrt{\frac{U}{2\Delta_0}} e^{\frac{\pi(U^2-4\Delta_0^2)}{8U\Delta_0}}+\int_{-\infty}^{\infty}{\rm d} y  e^{-\frac{\pi Uy^2}{2\Delta_0}}\frac{(U/2\Delta_0)^{\frac32}}{(iy+\frac{1}{2})^2(U/2\Delta_0)^2+1/4}\right] \,
\end{align}
Following \cref{eq:chi-qp} and discussions thereafter, they should be 
\begin{align}
    \chi^c =  2\frac{1}{\pi \tilde\Delta_0}\left(1-\frac{\tilde{U}}{\pi \tilde\Delta_0}\right),\chi^s =  2\frac{1}{\pi \tilde\Delta_0}\left(1+\frac{\tilde{U}}{\pi \tilde\Delta_0}\right)\label{eq:chics-half}
\end{align}
when expressed in terms of renormalized interactions. We then have $\tilde{\Delta}_0 = \frac{4}{\pi(\chi^c+\chi^s)},\tilde{U}=\frac{\pi^2\tilde{\Delta}_0^2(\chi^s-\chi^c)}{4}$. In large $U$ limit, $\chi^c\ll \chi^s$, for which we easily see that $\chi^c\ll \tilde{\Delta}_0^{-1}=\pi(\chi^c+\chi^s)/4$. Combined with \cref{eq:chics-half}, this implies $\tilde{U}=\pi\tilde{\Delta}_0$. We can also explicitly verify 

\begin{align}
    \lim_{U\to\infty}\frac{\tilde{U}}{\pi\tilde{\Delta}_0} = \lim_{U\to\infty}\frac{\pi\tilde{\Delta}_0(\chi^s-\chi^c)}{4}=\lim_{U\to\infty}\frac{\chi^s-\chi^c}{\chi^s+\chi^c}= 1\, .
\end{align}

\subsubsection{The \texorpdfstring{$U\gg \mJ \gg  T_{\rm K}$}{U>>J>>TK} limit}

\paragraph{The $\nu_f=\pm2$ states.}
We focus on the doped correlated insulators at the total fillings $\nu=\pm(2+\delta\nu)$, where $0\!<\!\delta\nu\!<\!1$.  
According to the (approximate) particle-hole symmetry of the problem, we only consider the $\nu=-2$ case. 
According to the calculations in Ref.~\cite{zhou_kondo_2023,hu_kondo_2023,datta_heavy_2023,rai_dynamical_2023}, the strong repulsion interaction will fix $\nu_f$ around $-2$ for $-3<\nu<-2$.  
The two-electron states are already discussed in \cref{sec:flatten-HI}. 
In the $T_{\rm K} \sim \td \Delta_0 \ll \mJ$ limit, only the ground states  (\cref{eq:2particle-4}) of $H_I$ 
\begin{equation}
   \frac{ f_{\alpha \eta \up}^\dagger f_{\ovl \alpha \ovl\eta \down}^\dagger -  f_{\alpha \eta \down}^\dagger f_{\ovl \alpha \ovl\eta \up}^\dagger }{\sqrt2} \ket{0},\qquad (\alpha\eta=1+,1-),\qquad E = U-2\mJ\ .
\end{equation}
will participate in the Kondo screening \cite{nozieres_kondo_1980}. 
(See section 6.4 of Ref.~\cite{nozieres_kondo_1980} for the discussion about splittings in atomic levels.)
They transform as $d_{x^2-y^2}$ and $d_{xy}$ orbitals under the symmetry operations in $D_6$ and form the $E_2$ representation. 
We have shown that Eqs.~(\ref{eq:chi-c}) to (\ref{eq:chi-s}) are the {\it exact} susceptibilities of {\it bare} particles. 
The charge ($\sigma^0\tau^0\spin^0$), the valley-charge ($\sigma^0\tau^z\spin^0$), the orbital-charge ($\sigma^z\tau^0\spin^0$), and the total spin ($\sigma^0\tau^0\spin^z$) take fixed values 2, 0, 0, 0 in the ground state manifold, respectively. 
Therefore, these degrees of freedom are frozen at the Kondo energy scale $\td\Delta_0$, and the corresponding susceptibilities are not contributed by quasi-particles.
For example, in the one-orbital Anderson impurity model at half-filling, the charge susceptibility $\chi^c$ given by the Bethe ansatz is much smaller than the quasi-particle density of states at Fermi level ($\td\Delta_0^{-1}$) as mentioned in the previous subsubsection.
In fact,  $\chi^c \cdot \td\Delta_0 \ll1$ is a universal behavior of Fermi liquid in the Kondo regime. 
For \cref{eq:chi-c} to reproduce this correct behavior of $\chi^c$, the renormalized interactions in the bracket must cancel the $\td\Delta_0^{-1}$ divergence, implying the constraint
\begin{equation} \label{eq:chic=0-condition}
    2\td U_1 + 2\td U_2 + 2\td U_3 + \td U_4 + \td\mJ = \frac{\pi\td\Delta_0}{\sin^2\delta_f} \ .
\end{equation}
The same argument also applies to valley, orbital, and spin degrees of freedom because they are also frozen at the Kondo energy scale and are not contributed by quasi-particles. 
The constraints $\chi^{c,v,o,s}\cdot\td{\Delta}_0\ll1$ imply 
\begin{equation}\label{eq:Ward-constraint-nu2}
\td{U}_1 = - \frac{\pi\td\Delta_0}{\sin^2\delta_f},\qquad
\td{U}_2 = \td{U}_3 = \frac{\pi\td\Delta_0}{\sin^2\delta_f} - \frac12\td{\mJ},\qquad 
\td{U}_4 = - \frac{\pi\td\Delta_0}{\sin^2\delta_f} + \td{\mJ}\ . 
\end{equation}
The two-electron energies of a generic $H_I$ are given in \cref{sec:flatten-HI}. 
Replacing the bare interaction parameters with the renormalized interaction parameters, we can obtain eigenvalues of $\td{\Gamma}^0$ as 
{\small
\begin{equation}
\td{U}_2 = \frac{\pi\td\Delta_0}{\sin^2\delta_f} - \frac12\td{\mJ} \; \text{(intra-orbital inter-valley)},\qquad 
\td{U}_3 = \frac{\pi\td\Delta_0}{\sin^2\delta_f} - \frac12\td{\mJ} \; \text{(inter-valley intra-orbital)}\ ,
\end{equation}
\begin{equation}
\td{U}_4 = - \frac{\pi\td\Delta_0}{\sin^2\delta_f} + \td{\mJ}\; \text{(intra-orbital intra-valley singlet)}\ ,
\end{equation}
\begin{equation}
\td{U}_1 - \td{\mJ} = - \frac{\pi\td\Delta_0}{\sin^2\delta_f} -  \td{\mJ}\;  \text{(inter-valley inter-orbital singlet)}, \qquad  
\td{U}_1 +\td{\mJ} = - \frac{\pi\td\Delta_0}{\sin^2\delta_f} + \td{\mJ}\; \text{(inter-valley inter-orbital triplet)}\ . 
\end{equation}
}
Here $\delta_f = \frac{\pi}4$. 
One of the last two must be negative. 
Therefore, we prove the statement that the renormalized interaction has at least one negative channel. 

Unlike the charge, valley, orbital, and spin, the angular momentum ($\sigma^z\tau^0\spin^z$) is not quenched in the ground state of $H_I$. 
Thus, the quasi-particle contributed part of $\chi^{a}$ should diverge at the order $\td\Delta_0^{-1}$ in the $ \td\Delta_0 \to 0$ limit. 
$\chi^{a}$ expressed in terms of $\td{\mJ}$ is 
\begin{equation}
    \chi^{a} = 8 \frac{\sin^2\delta_f}{\pi\td{\Delta}_0} \pare{ 8 - 4\frac{\sin^2\delta_f}{\pi \td{\Delta}_0} \td\mJ } \ .  
\end{equation}
Requiring $\chi^{a}$ non-negative leads to the condition $\td{\mJ} \le 2 \frac{\pi\td{\Delta}_0}{\sin^2\delta_f}$.

We can also extract useful conditions from the other quasi-particle susceptibilities in Eqs.~(\ref{eq:chi-0zz}) to (\ref{eq:chi-xxz}) and (\ref{eq:chip-1}) to (\ref{eq:chip-5}). 
Even though they are not related to the exact susceptibilities of the bare particles, they should also be positive for the Fermi liquid theory to be valid. 
Then we have the inequalities 
\begin{equation}
\td{\chi}^{0zz} \approx 8 \frac{\sin^2\delta_f}{\pi\td{\Delta}_0} \pare{ 0 + \frac{\sin^2\delta_f}{\pi\td\Delta_0} 2\td{\mJ} } \gtrsim 0 , \qquad
\td{\chi}^{x00} = \td{\chi}^{xz0} =\td{\chi}^{0x0} = \td{\chi}^{zx0} \approx 8 \frac{\sin^2\delta_f}{\pi\td{\Delta}_0} \cdot  
\pare{ 2 - \frac{\sin^2\delta_f}{\pi\td{\Delta}_0} \frac12 \td{\mJ} } \gtrsim 0\ , 
\end{equation}
\begin{equation}
\td{\chi}^{xx0} \approx 8 \frac{\sin^2\delta_f}{\pi\td{\Delta}_0}
    \pare{0 + \frac{\sin^2\delta_f}{\pi\td{\Delta}_0} 2 \td{\mJ}} \gtrsim 0, \qquad 
\td{\chi}^{xxz} = 8 \frac{\sin^2\delta_f}{\pi\td{\Delta}_0} \cdot 0 \gtrsim 0\ ,
\end{equation}
\begin{equation}
\td{\chi}_{\rm p}^{0xy} = \td{\chi}_{\rm p}^{x0y} 
    \approx \frac{4}{ \td\Delta_0} \pare{ 1 - \frac{1}{4\td\Delta_0} \pare{ 2\pi\td\Delta_0 - \frac12\td\mJ  } } \gtrsim 0,\qquad 
\td{\chi}_{\rm p}^{00y} 
    \approx \frac{4}{ \td\Delta_0}
    \pare{ 1 - \frac{1}{4\td\Delta_0} \pare{ -2\pi\td\Delta_0 + \td\mJ  }  } \gtrsim 0,
\end{equation}
\begin{equation}
\td{\chi}_{\rm p}^{xxy} 
\approx \frac{4}{ \td\Delta_0}
    \pare{ 1 - \frac{1}{4\td\Delta_0} \pare{ -2\pi\td\Delta_0 - \td\mJ  } } \gtrsim 0,\qquad 
\td{\chi}_{\rm p}^{xy0} 
\approx \frac{4}{ \td\Delta_0}
    \pare{ 1 - \frac{1}{4\td\Delta_0} \pare{ -2\pi\td\Delta_0 + \td\mJ  }   } \gtrsim 0 .
\end{equation}
As the susceptibilities are approximate, we use ``$\gtrsim$'' rather than ``$\ge$'' in these inequalities. 
This requirement leads to 
\begin{equation} \label{eq:J-inequality}
    4.6\td\Delta_0 \approx 8\pare{\frac{\pi}2 - 1}   \td\Delta_0
    \lesssim \td \mJ \lesssim  
    4 \pare{1 + \frac{\pi}2 } \td\Delta_0  \approx 10.3\td\Delta_0 \ .
\end{equation}
We have plotted the estimated range for all the five pairing channels ($\td{U}_1 \mp \td{\mJ}$, $\td{U}_{2,3,4}$) correspondingt to \cref{eq:J-inequality} in \cref{fig:renorm_int}. 
One can directly verify that the inter-orbital inter-valley singlet pairing fluctuation ($\chi^{xxy}_{\rm p}$) is the strongest among all the pairings. 
The {\it universality hypothesis} states that $T_{\rm K}$, or $\td\Delta_0$, is the only energy scale in the Kondo regime, implying that the ratio $\td\mJ / \td{\Delta}_0 $ is a universal constant. 
Our analyses above show that this ratio is in the range from 4.6 to 10.3.

For later convenience, here we also express the effective interactions in the pairing channel as functions of $\td\mJ$ (\cref{eq:Ueff1,eq:Ueff2})
\begin{equation}
U^{\rm p}_{2} = U^{\rm p}_{3} \approx \frac{ 2\pi\td\Delta_0 - \frac12 \td\mJ  }{  1 - \frac{1}{4\td\Delta_0} ( 2\pi\td\Delta_0 - \frac12 \td\mJ  ) },\qquad 
U^{\rm p}_4 \approx \frac{ -2\pi\td\Delta_0 + \td\mJ  }{  1 - \frac{1}{4\td\Delta_0} ( -2\pi\td\Delta_0 + \td\mJ  ) }\ ,
\end{equation}
\begin{equation} \label{eq:Ueff}
U^{\rm p}_1 - \mJ^{\rm p} \approx \frac{ -2\pi\td\Delta_0 - \td\mJ    }{  1 - \frac{1}{4\td\Delta_0} ( -2\pi\td\Delta_0 - \td\mJ  ) },\qquad 
U^{\rm p}_1 + \mJ^{\rm p} \approx \frac{ -2\pi\td\Delta_0 + \td\mJ    }{  1 - \frac{1}{4\td\Delta_0} ( -2\pi\td\Delta_0 + \td\mJ  ) } \ .
\end{equation}

\paragraph{The $\nu_f=0$ states.}
The flattened interaction (\cref{eq:HI-def}) is designed for the $\nu_f=\pm2$ states in the actual MATBG in such a way that it reproduces the correct two-electron ground states, as explained at the beginning of \cref{sec:flatten-HI}. 
Thus, it may not apply to the $\nu_f=0$ states in the actual MATBG. 
However, here we still discuss the physics at $\nu_f=0$ for the flattened interaction (\cref{eq:HI-def}) for theoretical interests. 
The four-particle ground state is unique, which can be understood as filling the doubly degenerate $\nu_f=-2$ ground states simultaneously. 
It has 
$N_{\alpha\eta}=1$, $\mathbf{S}_{\alpha\eta}^2=\frac34$ for all $\alpha\eta$ flavors, and $(\mathbf{S}_{1+}+\mathbf{S}_{2-})^2=(\mathbf{S}_{1-}+\mathbf{S}_{2+})^2=0$. 
Its energy can be read from \cref{eq:HI-expand} as $E=6U-2\mJ$. 
Apart from $\chi^{c, s, v, o} \cdot \td{\Delta}_0 \ll 1 $, there is also $\chi^a \cdot \td{\Delta}_0 \ll 1 $ in the $\td\Delta_0\ll \mJ$  limit, which thus fixes $\td{J} = 2\frac{\pi\td{\Delta}}{\sin^2\delta_f} = 2\pi \td{\Delta}_0$, where $\delta_{f}=\frac{\pi}2$ at $\nu_f=0$. 
Substituting the renormalized interactions and $\delta_f\to \frac{\pi}2$ into \cref{eq:chip-1,eq:chip-2,eq:chip-3,eq:chip-4,eq:chip-5}, we have
$\tan\delta_f\cdot (\frac\pi2 - \delta_f) = 1$, 
$\td{U}_2 = \td U_3 = 0$, 
$\td{U}_4  = \pi \td\Delta_0$, 
$\td{U}_1 - \td\mJ  = -3 \pi \td\Delta_0$, 
$\td{U}_1 + \td\mJ  =  \pi \td\Delta_0$
and hence 
\begin{equation}
\td{\chi}_{\rm p}^{0xy} = \td{\chi}_{\rm p}^{x0y} 
\approx \frac{16}{\pi \td\Delta_0} \pare{1 + 0} ,\qquad 
\td{\chi}_{\rm p}^{00y} 
\approx \frac{16}{\pi \td\Delta_0}
    \pare{ 1 -1  } ,
\end{equation}
\begin{equation}
\td{\chi}_{\rm p}^{xxy} 
\approx \frac{16}{\pi \td\Delta_0}
    \pare{ 1 + 3 },\qquad 
\td{\chi}_{\rm p}^{xy0} 
\approx \frac{16}{\pi \td\Delta_0}
    \pare{ 1 - 1    }\ .
\end{equation}
One can see that the inter-orbital inter-valley singlet fluctuation is the only one that is promoted by interaction.  
However, as $\td\Delta_0$ is negligible at $\nu_f=0$ \cite{zhou_kondo_2023,hu_symmetric_2023}, the energy scale of renormalized attractive interactions is also negligible.

\subsubsection{The \texorpdfstring{$U\gg T_{\rm K} \gg \mJ$}{U>>TK>>J} limit}

We only focus the states at $\nu_f=-2$ in this subsubsection. 
On the one hand, only the charge degree of freedom is frozen in the $U\gg T_{\rm K} \gg \td\Delta_0 \sim T_{\rm}$ limit, implying (\cref{eq:chic=0-condition})
\begin{equation}
    2\td U_1 + 2\td U_2 + 2\td U_3 + \td U_4 + \td\mJ = \frac{\pi\td\Delta_0}{\sin^2\delta_f} \ .
\end{equation}
$\td{\mJ}$ is barely renormalized in the Fermi liquid phase because it is much smaller than the Kondo energy scale. 
Thus, there must be $\td\mJ \approx \mJ $. 
On the other hand, the model has an U(8) symmetry in the $\td{\mJ}\to0$ limit. 
This approximate U(8) symmetry implies $\td{U}_1 \approx \td{U}_2 \approx \td{U}_3 \approx \td{U}_4$. 
Therefore, we have 
\begin{equation}
    \td{U}_1 \approx \td{U}_2 \approx \td{U}_3 \approx \td{U}_4 \approx \frac{2\pi}7 \td \Delta_0,\qquad 
    \td{\mJ} = \mJ\ . 
\end{equation}
The ground state of the renormalized interaction is still the $d$-wave $E_2$ states (\cref{eq:2particle-4}). 
It has the energy 
\begin{equation}
    \td{U}_1 - \td\mJ = \frac{2\pi}7 \td \Delta_0 - \mJ\ . 
\end{equation}
This result implies that the pairing potential should start becoming negative when $\mJ \sim T_{\rm K}$. 
Combined with the discussions in the last subsubsection, this shows that the pairing potential is negative when $T_{\rm K}\ll \mJ$ and remains negative to the regime $T_{\rm K}>\mJ$ as $T_{\rm K}$ increases, as shown in \cref{fig:interaction}(e).

\subsection{The counter term *}
\label{sec:perturbation}

In this section we demonstrate how the counter term (\cref{eq:Sc}) is determined perturbatively in terms of $\td{\Gamma}^{0}$.

\begin{figure}
\centering
\includegraphics[width=0.6\linewidth]{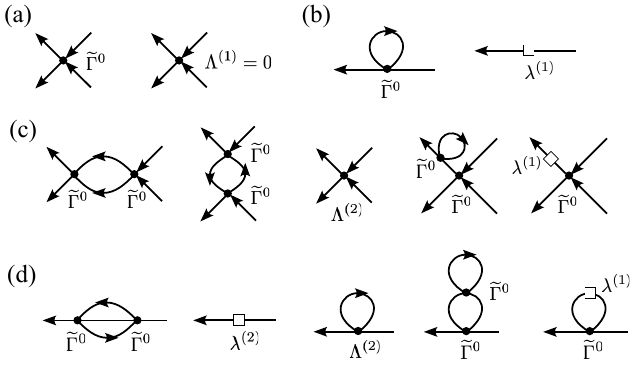}
\caption{Diagrams to determine the counter term. 
(a), (b), (c), (d) are diagrams for the first order vertex, first order self-energy, second order vertex, and second order self-energy, respectively. 
\label{fig:perturbation}
}
\end{figure}

The first order $\td{\Gamma}$ (\cref{fig:perturbation}(a)) is
\begin{equation}
    \td{\Gamma}^{(1)}(00;00) = \td{\Gamma}^{0}(00;00) + \Lambda^{(1)}\ ,
\end{equation}
where $\Lambda^{(1)}$ is the first-order counter term. 
$\td{\Gamma}^{(1)}(00;00)$ equals $\td{\Gamma}^{0}(00;00)$ by definition (\cref{eq:no-further-renormalization}). 
Thus, $\Lambda^{(1)}=0$.

The first order self-energy is given by the diagrams in \cref{fig:perturbation}(b), where the first term is contributed by $\td{\Gamma}^0$ and the second term is contributed by the first-order counter terms $\lambda_{0,1}^{(1)}$.
To derive this self-energy, we consider its correction to the Green's function
{\small
\begin{align}
\delta \td{\mG}^{(1)}(1) =& - \Inn{f(1) \pare{ - \sum_{2} f^\dagger(2) (-\lambda_0^{(1)}\omega_2 +\lambda_1^{(1)}) f(2) - \frac{T}4 \sum_{2345} \delta_{\omega_2+\omega_3,\omega_4+\omega_5}\td{\Gamma}^0(23;45) f^\dagger(2)f^\dagger(3) f(4) f(5) }  f^\dagger(1)}_{0C} \nonumber\\
=& [\td{\mG}^0(1)]^2 \times \pare{ -\lambda_0^{(1)}\omega_1 +\lambda_1^{(1)}  + 4\frac{T}4 \sum_2 \td{\Gamma}^0(12;21) \td{\mG}^0(2) }
\end{align}}
where the factor 4 in the second term of the second row comes from 4 equivalent contractions, $\td{\mG}^0$ is the free propagator of the quasi-particles, and the subscript $0C$ means connected diagrams.
The self-energy can be read as 
\begin{equation}
    \td{\Sigma}^{(1)}(1) = - i\lambda_0^{(1)} \omega_1 + \lambda_1^{(1)} + T \sum_{2} \td{\Gamma}^0(12;21) \td{\mG}^0(2)  
= - i\lambda_0^{(1)} \omega_1 + \lambda_1^{(1)} + \sum_{I_2} \td{\Gamma}^0_{I_1I_2;I_2I_1} \int \frac{d\omega_2}{2\pi} \td{\mG}^0(\omega_2)  e^{i0^+\omega_2}\ ,
\end{equation}
where the factor $e^{i0^+\omega_2}$ factor is due to the time-ordering in the Green's function $-\inn{f(-0^+) f^\dagger(0)}$ such that the creation operator is before the annihilation operator. 
The interaction factor is (\cref{eq:kappa000})
\begin{align}
\sum_{I_2} \td{\Gamma}^0_{I_1I_2;I_2I_1} = \frac18 \sum_{I_1 I_2} \td{\Gamma}^0_{I_1I_2;I_2I_1}
=& 2 \td U_1 + 2 \td U_2 + 2 \td U_3 + \td U_4 + \td \mJ\ . 
\end{align}
The frequency integral is 
\begin{equation}
\int_{-\infty}^{\infty} \frac{d\omega_2}{2\pi} e^{i0^+ \omega_2}\frac{1}{i\omega_2 + i\td{\Delta}_0 \cdot \sgn(\omega_2) - \td{\ee}_f} 
= \int_0^{\infty}  \frac{d\omega_2}{2\pi} e^{i0^+ \omega_2}\frac{1}{i\omega_2 + i\td{\Delta}_0 - \td{\ee}_f} + c.c. 
\end{equation}
Define $z=i\omega_2$, the function $\frac{1}{z+ i\td{\Delta}_0 - \td{\ee}_f}$ is analytical in the half plane $\Im[z]>0$. 
We can continuously deform the integral line from $(0,i\infty)$ to $(0,-\infty)$, then we have 
\begin{equation}
    \int_{0}^{-\infty} \frac{dz}{2\pi i} e^{0^+ z} \frac1{z + i\td{\Delta}_0 - \td{\ee}_f} +  c.c. 
= \frac1{\pi} \int_{-\infty}^0 dz \cdot e^{0^+z} \frac{\td{\Delta}_0}{ (z-\td{\ee}_f)^2 + \td{\Delta}_0^2 } = \pi \delta_f\ . 
\end{equation}
Therefore, there is 
\begin{equation}
    \td{\Sigma}^{(1)}(\omega) = -i \lambda_0^{(1)}\omega + \lambda_1^{(1)} + \pare{2 \td U_1 + 2 \td U_2 + 2 \td U_3 + \td U_4 + \td \mJ} \pi \delta_f
\end{equation}
The condition \cref{eq:no-further-renormalization} implies
\begin{equation}
    \lambda_0^{(1)} = 0,\qquad 
    \lambda_1^{(1)} = -\pare{2 \td U_1 + 2 \td U_2 + 2 \td U_3 + \td U_4 + \td \mJ} \pi \delta_f\ . 
\end{equation}
This also implies $\td{\Sigma}^{(1)}(\omega)=0$.

The second order vertex is contributed by diagrams in \cref{fig:perturbation}(c). 
The last two diagrams cancel each other because $\td{\Sigma}^{(1)}(\omega)=0$. 
Then $\Lambda^{(2)}$ should cancel the first two diagrams at zero frequency. 
We do not attempt to calculate $\Lambda^{(2)}$ in this work. 

We also do not attempt to calculate the full second order self-energy $\td{\Sigma}^{(2)}$. 
Instead, we only show how to calculate its linear frequency-dependence, {\it i.e.}, $\partial_{\omega}\td{\Sigma}^{(2)}$. 
The condition $\partial_{\omega}\td{\Sigma}^{(2)}=0$ (\cref{eq:no-further-renormalization}) will determine the counter term $\lambda_0^{(2)}$. 
$\td{\Sigma}^{(2)}$ is contributed by the diagrams shown in \cref{fig:perturbation}(d).   
The last two diagrams cancel each other because $\td{\Sigma}^{(1)}(\omega)=0$ according to the second paragraph above. 
The third diagram is contributed by $\Lambda^{(2)}$, which by definition is $\omega$-independent (\cref{eq:Sc}) and hence does not contribute to $\partial_{\omega}\td{\Sigma}^{(2)}$. 
Thus, we only need to consider the first two diagrams, dubbed as $\Sigma^{(2,1)}$ and $\Sigma^{(2,2)}$, respectively. 
The first diagram gives the correction 
{\scriptsize
\begin{align}
\delta \td{\mG}^{(2,1)}(1) =& - \Inn{f(1) \pare{ \frac{1}{2} \frac{T^2}{4^2} \sum_{\substack{2345\\5'4'3'2'}} \delta_{\omega_2+\omega_3,\omega_4+\omega_5} \delta_{\omega_5'+\omega_4',\omega_3'+\omega_2'} \td{\Gamma}^0(23;45) \td{\Gamma}^0(5'4';3'2') 
    f^\dagger(2)f^\dagger(3) f(4) f(5) f^\dagger(5')f^\dagger(4') f(3') f(2') }  f^\dagger(1)}_{0C} 
\end{align}}
Applying Wick's theorem, there is 
\begin{align}
\delta \td{\mG}^{(2,1)}(1)  & = - 2^4 \frac{1}{2} \frac{T^2}{4^2} \sum_{\substack{2345\\5'4'3'2'}} \delta_{\omega_2+\omega_3,\omega_4+\omega_5} \delta_{\omega_5'+\omega_4',\omega_3'+\omega_2'} \td{\Gamma}^0(23;45) \td{\Gamma}^0(5'4';3'2')  \nonumber\\
   &\times  \wick{ \c1 f(1) \c1 f^\dagger(2)  \c3 f^\dagger(3) \c2 f(4) \c1 f(5) 
        \c1 f^\dagger(5') \c2 f^\dagger(4') \c3 f(3') \c1 f(2') \c1 f^\dagger(1) }\ .
\end{align}
The factor $2^4$ comes from equivalent contractions. 
It equals 
\begin{align}
\delta \td{\mG}^{(2,1)}(1)  & = - [\mG^0(1)]^2 \cdot \frac{1}{2} \cdot  T^2 \sum_{345} \delta_{\omega_1+\omega_3,\omega_4+\omega_5} 
    \td{\Gamma}^0(13;45) \td{\Gamma}^0(54;31)  
    \td\mG^0(3) \td\mG^0(4) \td\mG^0(5) 
\end{align}
The self-energy can be extracted as 
\begin{align}
\td{\Sigma}^{(2,1)}(\omega_1) = - \frac{1}{2} T^2 \sum_{I_{3,4,5}} |\td{\Gamma}^0_{I_1I_3;I_4I_5}|^2 
    \sum_{\omega_{3,4}} \td\mG^0(\omega_3) \td\mG^0(\omega_4) \td\mG^0(\omega_1+\omega_3-\omega_4) \ ,
\end{align}
where the interaction factor can be calculated as 
\begin{equation}
u^2 = \frac{1}{2} \sum_{I_{3,4,5}} |\td{\Gamma}^0_{I_1I_2;I_3I_4}|^2 = \frac{1}{16} \sum_{I_{1,2,3,4}} |\td{\Gamma}^0_{I_1I_3;I_4I_5}|^2
= \frac{1}{16} \tr[ \td{\Gamma}^0 \cdot \td{\Gamma}^{0\dagger} ]
\end{equation}

Here we regard $I_{1,2}$ and $I_{3,4}$ as the left and right matrix indices respectively. 
$\tr[ \td{\Gamma}^0 \cdot \td{\Gamma}^{0\dagger} ]$ equals the sum of squares of the eigenvalues. 
$\td{\Gamma}^0$ is a 64-dimensional matrix. 
Its eigenspace with nonzero eigenvalues consists of the 56 anti-symmetric states. 
Each of the 28 states discussed in \cref{sec:flatten-HI} appears twice in the eigenspace because states related by permutation have been regarded physically identical. 
Suppose $\Phi_1$ is a two-electron state with the energy $E$.
Define $\Phi_2$ as the permutation of $\Phi_1$, it represents the same state.
The physical energy is given by the generalized eigenvalue problem defined by the Hamiltonian matrix $H_{nm} = \inn{\Phi_n| \hat H | \Phi_m}$ and overlap matrix $S_{nm} = \inn{\Phi_n | \Phi_m}$. 
There must be $H = E (\sigma_0 - \sigma_x) $ and $S=\sigma_0 - \sigma_x$ such that the generalized eigenvalue is $E$. 
However, the Hamiltonian matrix $H$ itself has two eigenvalues: $2E$ and 0. 
Therefore, each of 28 two-electron state contribute to an eigenvalue $2E$ and an eigenvalue 0. 
Thereby, 
\begin{eqnarray}
    u^2 &=& \frac{1}{16} \sum_{E} |2E|^2 
    = \frac14 \pare{  8 \td{U}_2^2 + 8 \td{U}_3^2 + 4\td{U}_4^2 + 2 (\td U_1 - \td\mJ)^2 + 6(\td U_1 + \mJ)^2 } \nonumber \\
    &=& 2 \td{U}_1^2 + 2 \td{U}_2^2 + 2\td{U}_3^2 + \td{U}_4^2 + 2\td{U}_1 \td\mJ +  2 \td{\mJ}^2 \ . 
\end{eqnarray}
Now we have 
{\small
\begin{align}
\td{\Sigma}^{(2,1)}(\omega_1) =& -  u^2 \int \frac{d\omega_3 d\omega_4}{(2\pi)^2} \cdot \frac{1}{i\omega_3+i\td{\Delta}_0\sgn(\omega_3) - \td{\ee}_f} \frac{1}{i\omega_4+i\td{\Delta}_0\sgn(\omega_4) - \td{\ee}_f}  \frac{1}{i(\omega_1+\omega_3-\omega_4) +i\td{\Delta}_0 \sgn(\omega_1+\omega_3-\omega_4) - \td{\ee}_f} \nonumber\\
=& -u^2 \int \frac{d\nu}{2\pi} \cdot F(\nu) \cdot \frac{1}{i(\omega_1+\nu) + i\td{\Delta}_0 \sgn(\omega_1+\nu) - \td{\ee}_f}  
\end{align}}
where $\nu=\omega_3-\omega_4$ and 
\begin{equation}
    F(\nu) = \int  \frac{d\omega}{2\pi} \cdot \frac{1}{i\omega+i\td{\Delta}_0\sgn(\omega) - \td{\ee}_f} \cdot 
    \frac{1}{i(\omega-\nu) +i\td{\Delta}_0 \sgn(\omega-\nu) - \td{\ee}_f} \ .
\end{equation}
By definition $F(\nu)=F(-\nu)$. 
We hence have
\begin{align}
F(\nu) = &   \int_{-\infty}^0 \frac{d\omega}{2\pi}  \frac{1}{i\omega-i\td{\Delta}_0 - \td{\ee}_f}    
    \frac{1}{i(\omega-|\nu|) - i\td{\Delta}_0  - \td{\ee}_f} 
+\int_{0}^{|\nu|} \frac{d\omega}{2\pi}  \frac{1}{i\omega + i\td{\Delta}_0 - \td{\ee}_f}    
        \frac{1}{i(\omega- |\nu|) - i\td{\Delta}_0  - \td{\ee}_f} \nonumber\\
+&   \int_{|\nu|}^{\infty} \frac{d\omega}{2\pi}  \frac{1}{i\omega + i\td{\Delta}_0 - \td{\ee}_f}    
        \frac{1}{i(\omega - |\nu|) + i\td{\Delta}_0  - \td{\ee}_f}     \ ,
\end{align}
Define $z=i\omega$, the above integral along imaginary axes of $z$ can be continuously deformed to integral along real axes:
{\small
\begin{align}
F(\nu) =& \int_{0}^{\infty} \frac{dz}{2\pi i} 
    \pare{  \frac{1}{z + i\td{\Delta}_0 - \td{\ee}_f} -  c.c.  } \frac{1}{z -i |\nu| - i\td{\Delta}_0  - \td{\ee}_f} +
\int_{0}^{\infty} \frac{dz}{2\pi i} \frac{1}{z + i |\nu| + i\td{\Delta}_0 - \td{\ee}_f} 
    \pare{ \frac{1}{z + i\td{\Delta}_0 - \td{\ee}_f} - c.c.  } \nonumber\\
=& \frac{2}{\pi} \int_0^\infty dz \cdot \frac{\td{\Delta}_0}{(z-\td{\ee}_f)^2 + \td{\Delta}_0^2} \cdot 
    \frac{z-\td{\ee}_f}{ (z - \td{\ee}_f)^2 + (|\nu|+\td{\Delta}_0)^2 } 
= \frac{\td{\Delta}_0}{ \pi (\nu^2 + 2\td{\Delta}_0 |\nu|) } 
    \ln \pare{1+\frac{ \nu^2 + 2\td{\Delta}_0 |\nu|  }{ \td{\Delta}_0^2 + \td{\ee}_f^2 }}
\end{align}}
It is direct to verify that $F(0^+)=F(0^-)=\frac{\td{\Delta}_0}{\pi (\td{\Delta}_0^2 + \td{\ee}_f^2)} = \frac{\sin^2\delta_f}{\pi \td{\Delta}_0} $ is the density of states at the Fermi level. 
Due to \cref{eq:G-derivative}, the leading order $\omega_1$-dependence of $\td{\Sigma}^{(2,1)}$ can be obtained as 
\begin{align}
\partial_{\omega}\td{\Sigma}^{(2,1)}(\omega) =& 
 -i \cdot u^2 \int \frac{d\nu}{2\pi} \cdot F(\nu) \pare{ [\td{\mG}^0(\nu)]^2 + 2\pi  \cdot \frac{\sin^2\delta_f}{\pi\td{\Delta}_0} \delta(\nu) } \nonumber\\
=& -i \cdot \pare{\frac{u \cdot \sin^2\delta_f}{\pi\td{\Delta}_0}}^2 
   -i \cdot u^2 \int_0^{\infty} \frac{d\nu}{2\pi} \cdot F(\nu) 
    \frac{2\td{\ee}_f^2 - 2(\nu+\td{\Delta}_0)^2}{(\td{\ee}_f^2+(\nu+\td{\Delta}_0)^2)^2}
\end{align}
We define the dimensionless function 
\begin{equation}
\xi^2(\delta_f) = \sin^4\delta_f + \pi^2 \int_0^{\infty} \frac{dx}{2\pi} \cdot 
    \frac{1}{ \pi (x^2 + 2x) } 
\ln \pare{1+\frac{ x^2 + 2x  }{ 1 + \cot^2\delta_f}} 
    \frac{2\cot^2(\delta_f) - 2(1+x)^2}{[\cot^2(\delta_f)+(1+x)^2]^2}
\end{equation}
then we can write  $\partial_{\omega}\td{\Sigma}^{(2,1)}(\omega) $ as 
\begin{equation}
\partial_{\omega}\td{\Sigma}^{(2,1)}(\omega) = -i \pare{\frac{u\cdot\xi(\delta_f)}{\pi\td{\Delta}_0} }^2 
\end{equation}
We find 
\begin{equation}
\xi\pare{\frac{\pi}2} = \sqrt{3- \frac{\pi^2}4} \approx 0.7298 , \qquad 
\xi\pare{\frac{\pi}4} = \sqrt{  \frac{4-4\mathcal{C} + \pi -\pi \ln2 }{8} } \approx 0.4031\ ,
\end{equation}
where $\mathcal{C}\approx0.915966$ is the Catalan's constant. 
The result $\xi(\pi/2)$ can also be found in Ref.~\cite{hewson_renormalized_1993}. 
The total frequency derivative $\td{\Sigma}^{(2)}$ also contains the contribution from the second diagram of \cref{fig:perturbation}(d), {\it i.e.}, $ - i\lambda_0^{(2)}$. 
Imposing $\partial_{\omega}\td{\Sigma}^{(2)}=0$ gives 
\begin{equation}
    \lambda^{(2)}_0  = -\pare{\frac{u\cdot\xi(\delta_f)}{\pi\td{\Delta}_0} }^2  \ .
\end{equation}

\clearpage
\section{Anderson impurity problem with the original interaction}
\label{sec:original-SIAM}
In this section, we apply a phenomenological susceptibility analysis to the original $\rm U(1)^{\times 3} \times SU(2)$ interaction in the $\td\Delta_0\ll J_{\rm A,H}$ limit. The procedures parallel \cref{sec:flatten}, albeit due to the lower symmetry, less susceptibilities can be exactly expressed through the Ward identities, and approximations using the quasi-particle susceptibilities have to be adopted.  
We find the results can recover our main conclusions drawn from the flattened $\rm U(1)^{\times 4} \times SU(2)^{\times 2}$ symmetry, and in the low-energy end controlled by the same $[E_2,0]$ ground states, the $\rm U(1)^{\times 4} \times SU(2)^{\times 2}$-breaking parameters are restricted to a rather narrow range. 

\begin{table}[h]
    \centering
    \begin{tabular}{c|c|c|c}
    \hline\hline
        Flattened parametrization & $(|N_{o}|, |N_{v}|, |L|)$ & $[\rho, j]$ & Original parametrization \\
    \hline
        $U_1 - \mJ$ & $(0, 0, 2)$ & $[E_2, 0]$ & $U_1 - \mJ$  \\
    \hline
        $U_1 + \mJ$ & $(0, 0, 2)$ & $[E_1, 1]$ & $U_1 + \mJ$ \\
    \hline
        \multirow{4}{*}{$U_2$} & \multirow{4}{*}{$(2, 0, 0)$} & $[A_1, 0]$ & $U_2 - J_2 + (V_2 - K_2)$ \\
    \cline{3-4}
         &  & $[B_2, 0]$ & $U_2 - J_2 - (V_2 - K_2)$ \\
    \cline{3-4}
         &  & $[B_1, 1]$ & $U_2 + J_2 + (V_2 + K_2)$ \\
    \cline{3-4}
         &  & $[A_2, 1]$ & $U_2 + J_2 - (V_2 + K_2)$ \\
    \hline
        \multirow{2}{*}{$U_3$} & \multirow{2}{*}{$(0, 2, 0)$} & $[A_1+B_1, 0]$ & $U_3 - J_3$ \\
    \cline{3-4}
         &  & $[A_2+B_2, 1]$ & $U_3+J_3$ \\
    \hline
        $U_4$ & $(2, 2, 2)$ & $[E_1+E_2, 0]$ & $U_4$ \\
    \hline\hline
    \end{tabular}
    \caption{\label{tab:levelsplit} Level splitting of the two-particle scattering channels by breaking the $\rm U(1)^{\times 4} \times SU(2)^{\times 2}$ symmetry to $\rm U(1)^{\times 3} \times SU(2)$. $N_{o}$, $N_v$, and $L$ denote the orbital ($\sigma^z \tau^0 \spin^0$), valley ($\sigma^0 \tau^z \spin^0$), and angular momentum ($\sigma^z \tau^z \spin^0$) charges, respectively. Since $C_{2x} \in D_2$ anti-commutes with $N_o$ and $L$, and $C_{2z} \in D_2$ anti-commutes with $N_v$, only $|N_o|, |N_v|$ and $|L|$ are preserved. $j$ denotes the total spin of the global SU(2) representation, and $\rho$ denotes the $D_6$ representation. }
\end{table}

The first step is to re-parametrize the zero-frequency vertex function (equivalently, all the two-particle scattering channels) in the most general form according to the continuous and discrete symmetries. 
To make a convenient comparison to the flattened interaction, we start with the general parametrization of the flattened interaction obtained in \cref{sec:flatten-HI} (using $U_{1,2,3,4}$ and $\mJ$), break the $\rm U(1)^{\times 4} \times SU(2)^{\times 2}$ symmetry to $\rm U(1)^{\times 3} \times SU(2)$, and identify the level splittings. 
The results are also necessarily a re-parametrization to the two-particle levels obtained for the bare original interaction, which are listed in \cref{sec:original}. We summarize the results in \cref{tab:levelsplit}.  

For the two spin singlets (triplets) of energy $U_1 - \mJ$ ($U_1 + \mJ$) under the flattened interaction, they already form the $[E_2, 0]$ ($[E_1, 1]$) irrep of the $D_6 \times \mrm{SU}(2)$ group, thus their degeneracy is still protected by the lower symmetry. 
For the four states of energy $U_4$ under the flattened interaction, they form a $[E_1+E_2, 0]$ representation of $D_6\times \mrm{SU}(2)$. Although $E_1+E_2$ is reducible under $D_6$ actions, they are related by a valley U(1) action, hence will be irreducible under the full symmetry group that also includes valley U(1). We keep the corresponding energies parametrized by $U_1 \mp \mJ$ and $U_4$ as before. 

For the eight states of energy $U_2$ under the flattened interaction, due to the breaking of the two independent SU(2) rotations, they will split to two spin singlets and two spin triplets. The two spin singlets (triplets) possess opposite orbital charges, $N_o = \pm2$, which is, however, not related by any symmetry. Therefore, splittings within the two singlets and within the two triplets are both expected. 
We thus need 3 new parameters - $J_2$ describes the overall spin splitting, $V_2 - K_2$ describes the splitting within singlets, and $V_2 + K_2$ describes the splitting within triplets. The two singlet levels $[A_1, 0]$, $[B_2, 0]$ have energies $U_2 - J_2 \pm (V_2 - K_2)$, respectively, and the two triplet levels $[B_1, 1]$, $[A_2, 1]$ have energies $U_2 + J_2 \pm (V_2 + K_2)$, respectively.

The eight states of energy $U_3$ would split as the above eight states if crystalline symmetries were not considered. However, as the two spin singlets (triplets) possess opposite valley charges, $N_v = \pm2$, they necessarily stay degenerate since  $C_{2z}$ anti-commutes with $N_v$. We thus only need 1 new parameter $J_3$ to describe the spin splitting, so that singlets (triplets) have energy $U_3 - J_3$ ($U_3 + J_3$). Written in terms of the representation of $D_6\times\mrm{SU}(2)$, the two spin singlets (triplets) form $[A_1+B_1, 0]$ ($[A_2+B_2, 1]$).

The wave-functions of each two-particle scattering channel, $F_i^\dagger$, can be found in \cref{sec:original} in one-to-one correspondence by tracking the $[\rho, j]$ labels.
With both the energies $E_{i}$, parametrized in terms of the 9 free parameters, and the wave-functions $F^\dagger_i$ known, the effective two-particle interaction can be constructed as $\sum_i E_i F^\dagger_{i} F_{i}$, and the zero-frequency vertex can be obtained accordingly. 
The results, in terms of the renormalized values (with a tilde), read
\begin{align}   \label{eq:Gm0_original}
    \td{\Gamma}^0_{I_1 I_2; I_3 I_4}(0,0;0,0) &= \delta_{\alpha_1, \ovl{\alpha}_2} \delta_{\eta_1, \ovl{\eta}_2} \left[ \td{U}_1 \left( \delta^{\alpha_1\eta_1s_1}_{\alpha_4\eta_4s_4} \delta^{\alpha_2\eta_2s_2}_{\alpha_3\eta_3s_3} - (3\leftrightarrow4) \right) + \td{\mJ} \left( \delta^{s_1}_{s_3} \delta^{s_2}_{s_4} \delta^{\alpha_1\eta_1}_{\alpha_4\eta_4} \delta^{\alpha_2\eta_2}_{\alpha_3\eta_3} - (3\leftrightarrow4) \right) \right] \\\nonumber 
    &+ \delta_{\alpha_1, \alpha_2} \delta_{\eta_1, \ovl{\eta}_2} \left[ \td{U}_2 \left(\delta^{\alpha_1 \eta_1 s_1 }_{\alpha_4 \eta_4 s_4} \delta^{\alpha_2 \eta_2 s_2}_{\alpha_3 \eta_3 s_3} - (3\leftrightarrow4) \right) + \td{J}_2 \left( \delta^{s_1}_{s_3} \delta^{s_2}_{s_4} \delta^{\alpha_1\eta_1}_{\alpha_4\eta_4} \delta^{\alpha_2\eta_2}_{\alpha_3\eta_3} - (3\leftrightarrow4) \right) \right. \\\nonumber
    & ~~~~~~~~ \left. + \td{V}_2  \left( \delta^{\ovl{\alpha}_1}_{\alpha_4} \delta^{\ovl{\alpha}_2}_{\alpha_3} \delta^{\eta_1 s_1}_{\eta_4s_4} \delta^{\eta_2s_2}_{\eta_3s_3} - (3\leftrightarrow4) \right) + \td{K}_2 \left( \delta^{\ovl{\alpha}_1}_{\alpha_4} \delta^{\ovl{\alpha}_2}_{\alpha_3} \delta^{s_1}_{s_3} \delta^{s_2}_{s_4} \delta^{\eta_1}_{\eta_4} \delta^{\eta_2}_{\eta_3}  - (3\leftrightarrow4) \right)  \right] \\\nonumber
    &+ \delta_{\alpha_1, \ovl{\alpha}_2} \delta_{\eta_1, {\eta}_2} \left[ \td{U}_3 \left( \delta^{\alpha_1\eta_1s_1}_{\alpha_4\eta_4s_4} \delta^{\alpha_2\eta_2s_2}_{\alpha_3\eta_3s_3} - (3\leftrightarrow4) \right) + \td{J}_3 \left( \delta^{s_1}_{s_3} \delta^{s_2}_{s_4} \delta^{\alpha_1\eta_1}_{\alpha_4\eta_4} \delta^{\alpha_2\eta_2}_{\alpha_3\eta_3} - (3\leftrightarrow4) \right) \right] \\\nonumber
    &+ \delta_{\alpha_1, \alpha_2} \delta_{\eta_1, \eta_2} \left[ \td{U}_4 \left( \delta^{\alpha_1\eta_1s_1}_{\alpha_4\eta_4s_4} \delta^{\alpha_2\eta_2s_2}_{\alpha_3\eta_3s_3} - (3\leftrightarrow4)   \right) \right]\ . 
\end{align}
It can be seen that, $\td{J}_2$, and $\td{J}_3$ are flavor-dependent Hund's, similar to $\td{\mJ}$, which merely exchange spins but do not violate any flavor charge conservation, while $\td{V}_2$ and $\td{K}_2$ are pair-scatterings that violate the orbital charge conservation. $\td{V}_2$ is spin-independent, and $\td{K}_2$ exchanges spins in the meanwhile. 
In particular, the bare interaction obeys parametrization of \cref{eq:Gm0_original} as well. 
The corresponding values (without tildes) can be easily read off by comparing \cref{tab:levelsplit} to the bare levels solved in \cref{sec:original}, 
\begin{align}
    & U_1 = U + J'_H , \quad \mJ = J_A - J'_H,  \\
    & U_2 = U + J_H , \quad J_2 = J_A - J_H, \quad V_2 = J_H' , \quad K_2 = J_A - J_H'  \\
    & U_3 = U + J'_H , \quad J_3 = - J'_H \\
    & U_4 = U + 2J_H 
\end{align}
It can also be seen that, $V_2$ and $K_2$ correspond to the last and second-last terms in \cref{eq:HI123}. 

Next, we need to express the susceptibilities of all the charge-0 operators through the zero-frequency vertex function $\td{\Gamma}^0$, using \cref{eq:chi-qp}. If any susceptibility is asymptotically frozen in the Kondo regime, then the asymptotic behavior of $\td{\Gamma}^0$ will be restricted. 
The charge $O^{000}$, valley charge $O^{0z0}$, and the total spin $O^{00z}$ are still conserved, hence by virtue of the Ward identities, \cref{eq:chi-qp} provides an exact relation between the exact charge $\chi^c$, valley $\chi^v$ and spin $\chi^s$ susceptibilities and the exact vertex $\td{\Gamma}^0$.  
When the ground states are dominated by the $[E_2, 0]$ states, $\chi^{c,v,s}$ still freeze as in the flattened case, giving rise to 3 exact equations constraining the 9 unknown parameters,  
\begin{align}  \label{eq:freeze_chi_c_ori}
    0 = \frac{\pi\td{\Delta}_0}{\sin^2\delta_f} \chi^c &= 8 \left( 1 - \frac{\sin^2\delta_f}{\pi \td{\Delta}_0}  \kappa[O^{000}, \td{\Gamma}^0]  \right) \\
    0 = \frac{\pi\td{\Delta}_0}{\sin^2\delta_f} \chi^v &= 8 \left( 1 - \frac{\sin^2\delta_f}{\pi \td{\Delta}_0}  \kappa[O^{0z0}, \td{\Gamma}^0]  \right) \\
    0 = \frac{\pi\td{\Delta}_0}{\sin^2\delta_f} \chi^s &= 8 \left( 1 - \frac{\sin^2\delta_f}{\pi \td{\Delta}_0}  \kappa[O^{00z}, \td{\Gamma}^0]  \right) \ .
\end{align}
The $\kappa$ functions are defined by \cref{eq:kappa}, and will be evaluated in \cref{eq:kapp_ori}. 
The `orbital charge' $O^{z00}$, on the other hand, no longer remains conserved, thus \cref{eq:chi-qp} only estimates the corresponding quasi-particle susceptibility $\td{\chi}^o$. As the exact susceptibility $\chi^o$ will exactly equal to $\td{\chi}^o$ when the orbital U(1) is recovered, we argue that \cref{eq:chi-qp} still approximates $\chi^o$ if the orbital U(1) is weakly broken in the effective theory. In the following, we show the self-consistency of this approximation by checking that the renormalized interaction indeed approximately respects the higher $\rm U(1)^{\times 4} \!\times\! SU(2)^{\times2} $ symmetry. Since the $[E_2, 0]$ states still hold as  eigenstates of zero eigenvalue for the $O^{z00}$ operator, $\chi^o$ should be asymptotically frozen as well. 
This contributes the fourth constraint, albeit with approximation,  
\begin{align}
    0 = \frac{\pi\td{\Delta}_0}{\sin^2\delta_f} \chi^o &\approx 8 \left( 1 - \frac{\sin^2\delta_f}{\pi \td{\Delta}_0}  \kappa[O^{z00}, \td{\Gamma}^0] \right) 
\end{align}

Crucially, we note that, in the flattened case, the operators $O^{zzz}$ and $O^{00z}$ have been shown as linked by a symmetry action in the large $\rm U(1)^{\times 4} \times SU(2)^{\times 2}$ group (\cref{sec:chi}), therefore, for the $[E_2, 0]$ ground states, the two susceptibilities, $\chi^{zzz}$ and $\chi^s$, are frozen simultaneously. Physically, this can be understood as follows: $\frac{O^{00z} \pm O^{zzz}}{2}$ probes the spin susceptibility of each angular momentum flavor, while for both states in the $[E_2, 0]$ irrep, each angular momentum flavor either is empty, or owns a spin singlet, hence neither flavor can exhibit a net spin. 
With the symmetry broken to $\rm U(1)^{\times 3} \times SU(2)$, 1) $O^{zzz}$ no longer remains conserved, hence we can only approximate $\chi^{zzz} \approx \td{\chi}^{zzz}$ with $\td{\chi}^{zzz}$ expressed through \cref{eq:chi-qp}, and 2) $\chi^{zzz}$ should rely on the new 4 parameters of $\td{\Gamma}^0$ in a different way from $\chi^{s}$. 
However, as long as the same $[E_2, 0]$ ground states dominate in the heavy Fermi liquid, $\chi^{zzz}$ will be frozen as well as $\chi^s$, and by its different reliance on $\td{\Gamma}^0$, this imposes a fifth constraint, 
\begin{align}   \label{eq:freeze_chi_zzz_ori}
    0 = \frac{\pi\td{\Delta}_0}{\sin^2\delta_f} \chi^{zzz} &\approx 8 \left( 1 - \frac{\sin^2\delta_f}{\pi \td{\Delta}_0}  \kappa[O^{zzz}, \td{\Gamma}^0]  \right)\ .
\end{align}

The five $\kappa$ functions are evaluated according to \cref{eq:kappa} as follows, 
\begin{align}   \label{eq:kapp_ori}
    \begin{pmatrix}
        \kappa[O^{000}, \td{\Gamma}^0] \\
        \kappa[O^{00z}, \td{\Gamma}^0] \\
        \kappa[O^{0z0}, \td{\Gamma}^0] \\
        \kappa[O^{z00}, \td{\Gamma}^0] \\
        \kappa[O^{zzz}, \td{\Gamma}^0] \\
    \end{pmatrix} &= \begin{pmatrix}
        2 & 2 & 2 & 1 & 1 & 1 & 1 & 0 & 0 \\
        0 & 0 & 0 & -1 & 1 & 1 & 1 & 0 & 0 \\
        -2 & -2 & 2 & 1 & -1 & -1 & 1 & 0 & 0 \\
        -2 & 2 & -2 & 1 & -1 & 1 & -1 & 0 & 0 \\
        0 & 0 & 0 & -1 & 1 & -1 & -1 & 0 & 0 \\
    \end{pmatrix} \begin{pmatrix}
        \td{U}_1 \\
        \td{U}_2 \\
        \td{U}_3 \\
        \td{U}_4 \\
        \td{\mJ} \\
        \td{J}_2 \\
        \td{J}_3 \\
        \td{V}_2 \\ 
        \td{K}_2 \\
    \end{pmatrix}\ .
\end{align}
Eqs.~(\ref{eq:freeze_chi_c_ori}) to (\ref{eq:freeze_chi_zzz_ori}) thus imply
\begin{align}  \label{eq:chi_c_ori}
    2\pi \td{\Delta}_0 &= 2 \td{U}_1 + 2\td{U}_{2} + 2\td{U}_3 + \td{U}_4 + \td{\mJ} + \td{J}_2 + \td{J}_3 \\ \label{eq:chi_s_ori}
    2\pi \td{\Delta}_0 &= -\td{U}_4 + \td{\mJ} + \td{J}_{2} + \td{J}_3 \\ \label{eq:chi_v_ori}
    2\pi \td{\Delta}_0 &= -2 \td{U}_1 - 2 \td{U}_2 + 2\td{U}_3 + \td{U}_4 - \td{\mJ} -\td{J}_2 + \td{J}_3 \\ \label{eq:chi_o_ori}
    2\pi \td{\Delta}_0 &\approx  -2 \td{U}_1 + 2 \td{U}_2 - 2\td{U}_3 + \td{U}_4 - \td{\mJ} + \td{J}_2 - \td{J}_3 \\ \label{eq:chi_zzz_ori}
    2\pi \td{\Delta}_0 &\approx  -\td{U}_4 + \td{\mJ} - \td{J}_2 - \td{J}_3 
\end{align}
where $\sin^2\delta_f = \frac{1}{2}$ is employed. By summing up \cref{eq:chi_s_ori} to \cref{eq:chi_zzz_ori}, importantly, we directly arrive at
\begin{align}   \label{eq:U1_ori}
    \td{U}_1 \approx - 2\pi \td{\Delta}_0 
\end{align}
which immediately recovers our main results obtained using the flattened approximation - one of the two pairing channels with energies $\td{U}_1 \mp \td{\mJ}$ must be attractive, albeit now with an approximate sign due to the lack of Ward identities. 

Besides, by comparing \cref{eq:chi_s_ori} and \cref{eq:chi_zzz_ori} only, we obtain that
\begin{align}
    \td{J}_2 &\approx - \td{J}_3  \\ \label{eq:U4_ori}
    \td{U}_4 &\approx \td{\mJ} - 2\pi\td{\Delta}_0
\end{align}
where the first equation reduces the dimension of the `flattened-symmetry-breaking' parameter space by one, and the second equation again recovers the flattened result. Moreover, we obtain
\begin{align}
    \td{U}_2 &= 2\pi\td{\Delta}_0 - \frac{\td{\mJ}}{2} + \frac{\td{J}_3}{2} \\
    \td{U}_3 &= 2\pi\td{\Delta}_0 - \frac{\td{\mJ}}{2} - \frac{\td{J}_3}{2}
\end{align}
which differs from the flattened result with only one new parameter $\td{J}_3$. 

As is also exploited in the flattened case, the not-yet-restricted parameters cannot vary unboundedly, otherwise the non-frozen susceptibilities will become negative, leading to an unphysical interactions. This imposes inequality constraints to the parameters in $\td{\Gamma}^0$, 
\begin{align}
    0 \leq \frac{\pi\td{\Delta}_0}{\sin^2\delta_f} \chi^{\mu\nu\rho} \approx 8 \left( 1 - \frac{\sin^2\delta_f}{\pi\td{\Delta}_0} \kappa[O^{\mu\nu\rho}, \td{\Gamma}^0]  \right)
\end{align}
where $\chi^{\mu\nu\rho}$ is still approximated through \cref{eq:chi-qp}, if $O^{\mu\nu\rho}$ is not conserved. In total, we find thirteen independent inequalities, six out of which arise from spin-0 operators, 
\begin{align}
    \begin{pmatrix}
        \kappa[O^{0x0}, \td{\Gamma}^0] \\
        \kappa[O^{x00}, \td{\Gamma}^0] \\
        \kappa[O^{y00}, \td{\Gamma}^0] \\
        \kappa[O^{xx0}, \td{\Gamma}^0] \\
        \kappa[O^{yx0}, \td{\Gamma}^0] \\
        \kappa[O^{zz0}, \td{\Gamma}^0] \\
    \end{pmatrix} = \begin{pmatrix}
        0 & -1 &  0 &  0 &  0 &  -2 & 0 & 0 & 0  \\
        0 &  0 & -1 &  0 &    0 & 0 & -2 & 2 & 1 \\
        0 &  0 & -1 &  0 &    0 & 0 & -2 & -2 & -1 \\
        -1 & 0 &  0 &  0 &   -2 & 0 & 0 & -1 & -2 \\
        -1 & 0 &  0 &  0 &   -2 & 0 & 0 & 1 & 2 \\
        2 & -2 & -2 &  1 &    1 & -1 & -1 & 0 & 0 \\
    \end{pmatrix} \begin{pmatrix}
        \td{U}_1 \\
        \td{U}_2 \\
        \td{U}_3 \\
        \td{U}_4 \\
        \td{\mJ} \\
        \td{J}_2 \\
        \td{J}_3 \\
        \td{V}_2 \\ 
        \td{K}_2 \\
    \end{pmatrix}
\end{align}
and seven out of which arise from spin-1 operators, 
\begin{align}
    \begin{pmatrix}
        \kappa[O^{0zz}, \td{\Gamma}^{0}] \\
        \kappa[O^{z0z}, \td{\Gamma}^{0}] \\
        \kappa[O^{0xz}, \td{\Gamma}^{0}] \\
        \kappa[O^{x0z}, \td{\Gamma}^{0}] \\
        \kappa[O^{y0z}, \td{\Gamma}^{0}] \\
        \kappa[O^{xxz}, \td{\Gamma}^{0}] \\
        \kappa[O^{yxz}, \td{\Gamma}^{0}] \\
    \end{pmatrix} = \begin{pmatrix}
        0 &    0 &  0 & -1 &   -1 & -1 & 1 & 0 & 0 \\
        0 &    0 &  0 & -1 &   -1 & 1 & -1 & 0 & 0 \\ 
        0 & -1 &  0 &  0 &    0 &    0 & 0 &  0 & 0 \\
        0 &    0 & -1 &  0 &    0 &  0 & 0 & 0 & 1 \\
        0 &    0 & -1 &  0 &    0 &  0 & 0 & 0 & -1 \\ 
        -1 & 0 &  0 &  0 &    0 &    0 &    0 & -1 & 0 \\
        -1 & 0 &  0 &  0 &    0 &    0 &    0 &  1 & 0 \\
    \end{pmatrix}
    \begin{pmatrix}
        \td{U}_1 \\
        \td{U}_2 \\
        \td{U}_3 \\
        \td{U}_4 \\
        \td{\mJ} \\
        \td{J}_2 \\
        \td{J}_3 \\
        \td{V}_2 \\ 
        \td{K}_2 \\
    \end{pmatrix}
\end{align}
The constraints from other channels are not independent, as they are related to the above ones by \textit{e.g.} valley U(1), spin SU(2), and discrete symmetries. 

Specially, let us observe the last two expressions associated with $O^{xxz}$ and $O^{yxz}$. By inserting $\td{U}_1 \approx -2\pi \td{\Delta}_0$, $\chi^{xxz} \gtrsim 0$ and $\chi^{yxz} \gtrsim 0$ actually reduce to $\td{V}_2 \gtrsim 0$ and $\td{V}_2 \lesssim 0$, respectively, which essentially requires $\td{V}_2$ to vanish. 
Once again, the dimension of the `flattened-symmetry-breaking' parameters is further reduced by one. 

In addition, the susceptibilities of charge-2 operators (pairing susceptibilities) also ought to be non-negative, which are calculated through \cref{eq:chi-pairing} and are equivalent to finding independent two-electron states. The $\kappa_{\rm p}$ functions are given by two-electron eigen-energies expressed by the renormalized parameters. They contribute nine more independent inequalities. 
As a result, we are left with five equations and the twenty two inequalities. Altogether, they restrict $\td{\mJ}$ to the following range, 
\begin{align} \label{eq:mJ_ori}
    \td{\mJ} \in [4.6, 10.3] \td{\Delta}_0
\end{align}
which again recovers the results of the flattened interaction. 

Inspired by the well recovery of the flattened results, it is natural to ask, whether the $\rm U(1)^{\times 4} \times SU(2)^{\times 2}$ symmetry is a good approximation in the Kondo regime when the low energy (local) Hilbert space is nearly restricted to the $[E_2,0]$ ground states. 
For this sake, we investigate the allowed range for the 4 new parameters, $\td{J}_{2,3}$, $\td{V}_2$ and $\td{K}_2$. If they are all forced to vanish, then the above statement holds true. 
It has already been shown that, by the susceptibility requirements, they are only allowed to vary in a 2-dimensional subspace satisfying $\td{J}_2 + \td{J}_3 \approx 0$ and $\td{V}_2 \approx 0$, hence it suffices to check the allowed range for $\frac{\td{J}_2 - \td{J}_3}{2}$ and $\td{K}_2$. We find, 
\begin{align}
    \frac{\td{J}_2 - \td{J}_3}{2} & \in [-1.91, 1.91 ] \td{\Delta}_0 \\ 
    \td{K}_2 &\in [-2.86, 2.86]   \td{\Delta}_0\ .
\end{align}
Their deviation from 0 is considerably smaller than the estimated size for $\td{U}_1$ or $\td{U}_1 - \td{\mJ}$. 
This indeed suggests that the $\rm U(1)^{\times 4} \times SU(2)^{\times 2}$ symmetry can be deemed as a good approximation for the low-energy physics when the low energy (local) Hilbert space is nearly restricted to the $[E_2,0]$ ground states.

\begin{figure}[tb]
    \centering
    \includegraphics[width = 0.5\linewidth]{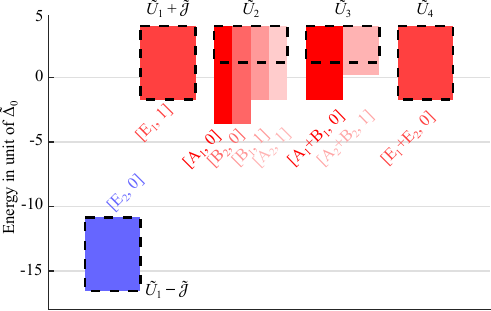}
    \caption{\label{fig:renorm_int} Comparison of the renormalized two-particle scattering channels that are obtained for the flattened interaction (with $\rm U(1)^{\times 4} \times SU(2)^{\times2}$ symmetry) and the original interaction (with $\rm U(1)^{\times 3} \times SU(2)$ symmetry). The limit $\td\Delta_0 \ll J_{\rm A,H}$ is assumed. The five dashed rectangles indicate the estimated energy ranges for the five independent channels of the flattened interaction, namely $\td{U}_1 - \td{\mJ}$, $\td{U}_{1} + \td{\mJ}$, and $\td{U}_{2,3,4}$, respectively, and the nine colored boxes indicate the estimated energy ranges for the nine independent channels of the original interaction. }
\end{figure}

Finally, in \cref{fig:renorm_int}, we compare the estimated energy range for all the renormalized two-particle scattering channels (pairing channels) that are obtained for the flattened interaction (\cref{sec:mott-limit}) in the $\td\Delta_0 \ll \mJ$ limit, and for the original interaction in the $\td\Delta_0 \ll J_{\rm A,H}$ limit. 
As has been obtained in \cref{eq:U1_ori}, \cref{eq:U4_ori} and \cref{eq:mJ_ori}, the allowed energy ranges for the $\td{U}_1 \mp \td{\mJ}$ and $\td{U}_4$ channels are the same for the original interaction and the flattened interaction. 
For the other channels, breaking the flattened $\rm U(1)^{\times 4} \times SU(2)^{\times2}$ symmetry to $\rm U(1)^{\times 3} \times SU(2)$ only slightly broadens the allowed energy ranges. 
For either the flattened or the original interaction, in the renormalized vertex, the phenomenological susceptibility analysis finds the $d$-wave singlets as the lowest pairing channel with attractive strength.

\clearpage
\section{Symmetry-breaking insulator favored by the flattened interaction at \texorpdfstring{$\nu=-2$}{v=-2}}

\begin{figure}[h]
    \centering
    \includegraphics[width=0.5\linewidth]{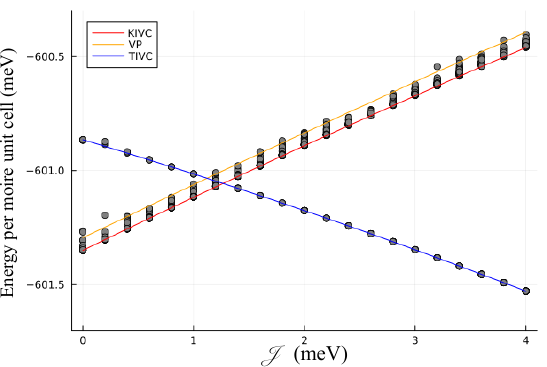}
    \caption{\label{fig:sbci} Symmetry-breaking states at $\nu=-2$ in Hartree-Fock self-consistent calculations using the flattened interaction. 
    }
\end{figure}

We carry out self-consistent Hartree-Fock calculation to study the symmetry-breaking orders stabilized by the flattened interaction at $\nu=-2$. 
We find the results are consistent with the results obtained using the realistic interaction Hamiltonian detailed \cref{sec:interaction}. 
Throughout the calculation, we fix $U=58$meV, and vary $\mJ$ between 0-4meV. Other parameters are set as the same in Ref. \cite{song_magic-angle_2022}. The results are shown in \cref{fig:sbci}. 

At $\mJ=0$, there is only Coulomb interaction but no multiplet splitting interactions. 
The Hartree-Fock ground state is thus found as Kramers inter-valley coherence (KIVC) state, in accordance with Refs.~\cite{bultinck_ground_2020,song_magic-angle_2022}. 
Over a threshold of $\mJ \approx 1.2$meV, time-reversal inter-valley coherent (TIVC) state becomes the ground state, which is consistent with previous mean-field calculations using phonon-mediated interaction \cite{kwan_electron-phonon_2023,shi_moire_2024,wang_tbg_epc_2024}, where TIVC is always found beyond a threshold of the $K$-phonon-mediated interaction. 
The TIVC order also exhibits the charge Kekul\'e pattern as observed in experiments \cite{nuckolls_quantum_2023}. 

\clearpage
\section{Quasi-particle mean-field theory of the superconductivity} \label{sec:mfsc}

\subsection{Effective BdG Hamiltonian}
\label{sec:H-lattice-symmetric}

The free part of topological heavy fermion model is given by \cite{song_magic-angle_2022}
{\small
\begin{equation} \label{eq:H0-THF}
{H}_0 = -\mu\sum_{\RR \alpha\eta s} f_{\RR\alpha\eta s}^\dagger f_{\RR\alpha\eta s} + \sum_{\eta s}\sum_{aa'}\sum_{|\kk|<\Lambda_c} (H^{(c,\eta)}_{aa'}(\kk) - \mu \delta_{aa'})  c^\dagger_{\kk a\eta s}c_{\kk a\eta s}
 + \sum_{\eta s \alpha a} \sum_{|\kk|<\Lambda_c} \pare{e^{-\frac{|\kk|^2\lambda^2}2} H^{(cf,\eta)}_{a\alpha}(\kk)c^\dagger_{\kk a\eta s}f_{\kk\alpha\eta s}+h.c.}
\end{equation}}%
Here $\mu$ is the chemical potential, $c_{\kk a\eta s}$ is the fermion operator for the $c$-electron of the momentum $\kk$, orbital $a$ ($=1,2,3,4$), valley $\eta$ ($=\pm$), and spin $s$ ($=\uparrow,\downarrow$), $f_{\RR \alpha \eta s}$ is the fermion operator for the $f$-electron of the site $\RR$ orbital $\alpha$ (=1,2),  valley $\eta$, and spin.
$\RR$'s form the triangular lattice generated by the Bravais lattice basis $\mbf{a}_{\rm M1} = \frac{2\pi}{3k_\theta} (\sqrt3,1)$, $\mbf{a}_{\rm M2} = \frac{2\pi}{3k_\theta} (-\sqrt3,1)$, where $k_\theta=2|\mbf{K}|\sin \frac{\theta}2$, $\theta=1.08^\circ$ is the first magic angle, $|\mbf{K}|=1.703\mathring{A}^{-1}$. 
The triangular lattice corresponds to the AA stacking regions of MATBG (\cref{fig:model}(a)). 
$f_{\kk\alpha\eta s} = \frac1{N_M} \sum_{\RR} e^{-i\RR\cdot\kk} f_{\RR\alpha\eta s}$ is the $f$-electron operator at the momentum $\kk$, with $N_{\rm M}$ being the number of moir\'e cells and  $\kk$ taking values in the moir\'e Brillouin zone spanned by $\mbf{b}_{\rm M1} = k_\theta(\frac{\sqrt2}2, \frac32)$, $\mbf{b}_{\rm M2} = k_\theta(-\frac{\sqrt2}2, \frac32)$.
However, $c$-electron is described by a continuous field that cannot be regularized to a lattice due to the symmetry anomaly \cite{TBG2}. 
Thus, momentum $\kk$ of $c$-electrons is within a  cutoff $\Lambda_c$ that will be sent to infinity in the end \cite{song_magic-angle_2022}. 
The matrices $H^{(c,\eta)}$ and $H^{(cf,\eta)}$ are defined by 
\begin{equation}
H^{(c,\eta)}(\kk) = 
\begin{pmatrix}
0_{2\times2}   &  v_\star (\eta k_x \sigma^0 + i k_y \sigma^z) \\
v_\star (\eta k_x \sigma^0 - i k_y \sigma^z)  & M\sigma^x 
\end{pmatrix},\qquad 
H^{(cf,\eta)}(\kk) = 
\begin{pmatrix}
\gamma \sigma^0 + v_\star' (\eta k_x \sigma^x + k_y \sigma^y) \\
0_{2\times2} 
\end{pmatrix}\ . 
\end{equation}
The parameter $\lambda$  is the spread of the Wannier functions of $f$-electrons, and it suppresses the hybridization at large momentum. 
In this work we adopt the parameters: $\gamma=-24.75$meV, $v_\star=-4.303\mathrm{eV\cdot\mathring{A}}$, $v_\star'= 1.623 \mathrm{eV\cdot\mathring{A}}$, $\lambda = 1.413 1/k_\theta$, $M=3.697$meV. 
The resulting band structure is shown in the inset of \cref{fig:model}(a).

The free Hamiltonian has time-reversal $T$ and $D_6$ point group, generated by $C_{2z}$-rotation, $C_{3z}$-rotation, and $C_{2x}$-rotation symmetries. 
Their representations on $f$-electrons and $c$-electrons are given by (\cref{eq:discrete-sym})
\begin{equation} \label{eq:discrete-sym-f}
T = \sigma^0\tau^x\spin^0 K,\qquad 
C_{2z} = \sigma^x\tau^x\spin^0,\qquad 
C_{3z} = e^{i\frac{2\pi}3\sigma^z\tau^z\spin^0},\qquad 
C_{2x} = \sigma^x\tau^0\spin^0\ , 
\end{equation}
and 
\begin{equation}
T=(\sigma^0\tau^x\spin^0 \oplus \sigma^0\tau^x \spin^0) K, \qquad 
C_{2z} = \sigma^x\tau^x\spin^0 \oplus \sigma^x\tau^x \spin^0,\qquad 
C_{3z} = e^{i\frac{2\pi}3\sigma^z\tau^z\spin^0} \oplus \sigma^0\tau^0\spin^0,\qquad 
C_{2x} = \sigma^x\tau^0\spin^0 \oplus \sigma^x\tau^0\spin^0\ ,
\end{equation}
respectively.
Detailed derivations of these symmetries are given sections S2A and S2B of the supplementary material Ref.~\cite{song_magic-angle_2022}.
Readers may directly verify that \cref{eq:H0-THF} is invariant under these symmetries. 
In this work we {\it do not} distinguish the single-particle and the second-quantized representations of symmetry operators. 
The correspondence between the two representations are given after \cref{eq:discrete-sym}.

\cref{eq:H0-THF} with Coulomb interaction is solved within the framework of dynamical mean-field theory (DMFT) in Refs.~\cite{zhou_kondo_2023,datta_heavy_2023,rai_dynamical_2023}. 
Here we are only interested in the doped correlated insulator at the filling $\nu=-2-\delta\nu$, where the $\td\Delta_0\ll \mJ$ is assumed. 
In the mean-field calculations, we will regard $z$ as an additional input to $\delta\nu$. 
In the $\td\Delta_0\ll \mJ$  limit, the $f$-filling is frozen to $\nu_f=-2$, indicating that the doping is fully undertaken by $c$-electrons, {\it i.e.}, $\nu_c=\nu-\nu_f = -\delta\nu$. 

To describe the quasi-particles at $\nu=-2-\delta\nu$, we express $f$-electron operators in terms of the quasi-particles operators as $f_{\RR \alpha\eta s} = z^{\frac12} \td{f}_{\RR\alpha\eta s} $.
We also introduce the DMFT self-consistent on-site energies $\ee_f$, $\ee_{c1}=\ee_{c2}$, $\ee_{c3}=\ee_{c4}$ for $\td f$-electrons, $a=1,2$ $c-$electrons, $a=3,4$ $c$-electrons, respectively \cite{zhou_kondo_2023}. 
Then the effective free Hamiltonian for the quasi-particle can be written as
\begin{align} \label{eq:H0-lattice-quasi-symmetric}
{H}_0 =& \ee_f \sum_{\RR \alpha\eta s} \td f_{\RR\alpha\eta s}^\dagger \td f_{\RR\alpha\eta s} + \sum_{\eta s}\sum_{aa'}\sum_{|\kk|<\Lambda_c} (H^{(c,\eta)}_{aa'}(\kk) +  \delta_{aa'}\ee_{c,a} )  c^\dagger_{\kk a\eta s}c_{\kk a\eta s} \nonumber\\
& +  z^{\frac12} \sum_{\eta s \alpha a} \sum_{|\kk|<\Lambda_c} \pare{e^{-\frac{|\kk|^2\lambda^2}2} H^{(cf,\eta)}_{a\alpha}(\kk)c^\dagger_{\kk a\eta s} \td f_{\kk\alpha\eta s}+h.c.}\ .
\end{align}
One can see that the $cf$ hybridization is suppressed by $z^{\frac12}$. 
$\ee_f$, $\ee_{c,1}$, $\ee_{c,3}$ should be determined self-consistently in the DMFT plus Hartree-Fock calculations. 
Here we provide a quick approach to estimate $\ee_f$, $\ee_{c,1}$, $\ee_{c,3}$ for given $z$ and $\nu=-2-\delta\nu$. 
First, it is found that $G = \epsilon_{c,3} - \epsilon_{c,1}$ is insensitive to doping and can be approximated as a fixed quantity (taken as 8meV here) \cite{zhou_kondo_2023}.  
We then search for $\epsilon_f$ and $\epsilon_{c,1}$ that give the correct fillings $\nu_f=-2$ and $\nu_c=-\delta\nu$.

In DMFT, the two-particle irreducible vertex in the lattice is purely local and equals the two-particle irreducible vertex of the impurity model, for which the susceptibility in the lattice model can be evaluated as $\chi_q^{-1} = \left(\chi_q^{0}\right)^{-1}+\Gamma_{loc},\Gamma_{loc}=\chi_{loc}^{-1} -\left(\chi_{loc}^0\right)^{-1}$, where $\chi_q$ ($\chi_q^{0}$) is the momentum dependent interacting (free) susceptibility in the lattice model, $\chi_{loc}(\chi_{loc}^0)$ is the susceptibility in the interacting (free) single impurity model, and $\Gamma_{loc}$ is the two-particle irreducible vertex in the single impurity model \cite{georges_infinite_d_1996}. We emphasize that $\chi_{loc}$ can not diverge as there is no symmetry breaking in zero dimension, while $\chi_{q}$ from the above expression can diverge and lead to symmetry breaking, \textit{e.g.}, the pairing instability. A pairing instability will yield a mean field ground state with non zero pairing order parameter.
We take the irreducible vertex $z^2\Gamma^{\rm p}$ - multiplied by $z^2$ - in the pairing channel as the effective interaction in mean-field calculation, where $z^2\Gamma^{\rm p}$ are calculated in Appendix \ref{sec:irreducible}.
(One would over-count Feynman diagrams if $z^2\Gamma$ is used instead \cite{georges_infinite_d_1996}.)
For the sake of investigating superconducting orders, it suffices to keep the strongest channel in $\Gamma^{\rm p}$, {\it i.e.}, the $d$-wave $E_2$ states (\cref{eq:2particle-4}).
The effect of other channels will be discussed in the end of this subsection. 
At $\nu_f =-2$, the pairing potential is determined by the renormalized interaction $\td\mJ$ via (\cref{eq:Ueff})
\begin{equation}
    U^{\rm p}_1 - \mJ^{\rm p} \approx \frac{-2\pi \td\Delta_0 - \td\mJ}{ 1 + \frac{\pi}2  + \td\mJ/(4\td\Delta_0) },\qquad 
    4.6\td\Delta_0 \lesssim \td \mJ \lesssim 10.3 \td \Delta_0 \ . 
\end{equation}
The explicit form of the corresponding interaction can be constructed as projectors to  \cref{eq:2particle-4}
\begin{align}
H_{\rm p} = & (U_1^{\rm p}- \mJ^{\rm p} ) \sum_{\RR} \sum_{\eta }
\frac{( \td f_{\RR 1\eta  \up}^\dagger \td f_{\RR 2 \ovl\eta \down}^\dagger - \td f_{\RR 1 \eta \down}^\dagger \td f_{ \RR 2 \ovl \eta \up }^\dagger) 
    ( \td f_{\RR 2 \ovl\eta \down} \td f_{\RR 1\eta  \up} - \td f_{ \RR 2 \ovl \eta \up } \td f_{\RR 1 \eta \down} ) }{2} \nonumber\\
=& \frac{U_1^{\rm p}- \mJ^{\rm p}}2 \sum_{\RR} \sum_{\eta }
    (  \td f_{\RR 2 \ovl\eta \down}^\dagger \td f_{\RR 1\eta  \up}^\dagger -\td f_{ \RR 2 \ovl \eta \up }^\dagger  \td f_{\RR 1 \eta \down}^\dagger ) 
    (  \td f_{\RR 1\eta  \up} \td f_{\RR 2 \ovl\eta \down} -  \td f_{\RR 1 \eta \down} \td f_{ \RR 2 \ovl \eta \up } )\ . 
\end{align}

We only consider the order parameters
\begin{equation} \label{eq:order-parameter-symmetric}
\Psi_{\eta} = \inn{ \Phi_{\rm G}|  \td f_{\RR 1 \eta \uparrow} \td f_{\RR 2 \ovl\eta \downarrow} | \Phi_{\rm G} }\ . 
\end{equation}
where $\Phi_{\rm G}$ is the ground state. 
Due to the spin-SU(2) symmetry ($i\spin^y)$, there must be 
\begin{equation}
\Psi_{\eta} = -\inn{ \Phi_G|  \td f_{\RR 1 \eta \downarrow} \td f_{\RR 2 \ovl\eta \uparrow} | \Phi_G }\ . 
\end{equation}
The mean-field decomposition of $H_{\rm p}$ reads
\begin{align}
H_{\rm p} \approx & \frac{U^{\rm p}_1- \mJ^{\rm p}}{2} \sum_{\RR \eta}
    \Big[ 2\Psi_{\eta} ( \td f^\dagger_{\RR 2 \ovl\eta \down} \td f^\dagger_{\RR 1 \eta \up} - \td f^\dagger_{\RR 2 \ovl \eta \up} \td f^\dagger_{\RR 1 \eta \down} )
    + 2\Psi_{\eta}^* (  \td f_{\RR 1\eta  \up} \td f_{\RR 2 \ovl\eta \down} -  \td f_{\RR 1 \eta \down} \td f_{ \RR 2 \ovl \eta \up } )
    - 2|\Psi_{\eta}|^2 \Big]  
\end{align} 
We can see that $(\eta,s)=(+,\up)$ electrons only couples to $(-,\down)$ electrons, whereas $(-,\up)$ electrons only couples to $(+,\down)$ electrons.
Thus, the BdG Hamiltonian is block diagonal. 
We can write the total Hamiltonian as 
\begin{align}
H \approx & - N_{\rm M}  (U^{\rm p}_1 - \mJ^{\rm p}) (|\Psi_+|^2 + |\Psi_-|^2)  + \ovl H_+ + \ovl H_-
\end{align}
$\ovl H_+$ is the BdG Hamiltonian for the $(\eta,s)=(+,\up)$ and $(-,\down)$ electrons
{\small
\begin{align} \label{eq:BdG-symmetric}
\ovl H_+ =& \sum_{\kk} \begin{pmatrix}
    \td f_{\kk+\up}^\dagger  &  c_{\kk +\up}^\dagger  & \td f_{-\kk-\down} & c_{-\kk -\down} 
    \end{pmatrix}
    \begin{pmatrix}
        \td\ee_f & z^\frac12 \mcl{H}^{(cf,+)\dagger}(\kk) &  \mV  & 0 \\
        z^\frac12 \mcl{H}^{(cf,+)}(\kk) &  \mcl{H}^{(c,+)}(\kk) & 0 & 0 \\
        \mV^\dagger & 0 & - \td\ee_f & -z^{\frac12}\mcl{H}^{(cf,-)T}(-\kk) \\
        0 & 0 & -z^{\frac12}\mcl{H}^{(cf,-)*}(-\kk) & - \mcl{H}^{(c,-)T}(-\kk) 
    \end{pmatrix}
    \begin{pmatrix}
    \td f_{\kk+\up} \\  c_{\kk +\up} \\ \td f_{-\kk-\down}^\dagger \\ c_{-\kk -\down}^\dagger
    \end{pmatrix}
\end{align}}
where 
\begin{equation}
    \mV = -(U^{\rm p}_1 - \mJ^{\rm p})
    \begin{pmatrix}
    0 & \Psi_+ \\ \Psi_- & 0 
    \end{pmatrix}\ ,
\end{equation}
is a two-by-two matrix. 
Here $\td f_{\kk + \up}$ and $\td f_{-\kk - \down}^\dagger$ are column vectors consisting of $\td f_{\kk \alpha + \up}$ and $\td f_{-\kk \alpha- \down}^\dagger$, respectively. 
$c_{\kk + \up}$ and $c_{-\kk - \down}$ are column vectors consisting of $c_{\kk+\mathbf{G},a + \up}$ and $c_{-\kk-\mathbf{G}, a - \down}$, respectively, where $\kk$ is in the moir\'e Brillouin zone and $\mathbf G$, regarded as a matrix index, is moir\'e reciprocal lattice. 
The $\mcl{H}^{(c,\eta)}$ and $\mcl{H}^{(cf,\eta)}$ matrices are defined as 
\begin{equation}
\mcl{H}^{(c,\eta)}_{\mathbf{G} a, \mathbf{G}'a'}(\kk) = \delta_{\bf G,G'} \pare{ \ee_{c,a} \delta_{aa'} + H^{(c,\eta)}_{aa'}(\kk+\mbf{G})}, \qquad
\mcl{H}^{(cf,\eta)}_{\mathbf{G}a, \alpha}(\kk) = e^{-\frac{|\kk + \mbf{G}|^2\lambda^2}2} H^{(cf,\eta)}_{a\alpha} (\kk+\mathbf{G})\ . 
\end{equation}
$\ovl H_-$ is the BdG Hamiltonian for $(\eta,s)=(+,\down), (-,\up)$ electrons. 
It can be obtained by applying the $i\spin^y$ rotation to $\ovl H_+$, and hence it has identical spectrum as $\ovl H_+$. 
The total energy (per moir\'e unit cell) can be calculated as 
\begin{equation}
    \frac{E_{\rm tot}}{N_{\rm M}} = - N_{\rm M}  (U^{\rm p}_1 - \mJ^{\rm p}) (|\Psi_+|^2 + |\Psi_-|^2)
    + \frac{2}{N_{\rm M}} \sum_{n\in{\rm occ}}\sum_{\kk} \mathcal{E}_{n\kk}\ ,
\end{equation}
where $\mathcal{E}_{n\kk}$ is the $n$-th BdG band of $\ovl H_+$, and ``$\sum_{n\in{\rm occ}}$'' means summation over the negative branches.


There are two relevant saddle points of the pairings in the numerical calculation: 
\begin{equation}
\mV = e^{i\theta_1} \mV_1 (\cos\varphi \cdot \sigma^x + \sin\varphi \cdot \sigma^y), \quad e^{i\theta_2} \mV_2  \frac{\sigma^x \pm i\sigma^y}2 \ .
\end{equation}
Here $\mV_1$ ($>0$), $\mV_2$ ($>0$) represent the pairing mean-field Hamiltonian terms
\begin{equation} \label{eq:V1pairing-symmetric}
    e^{i\theta_1} \mV_1 \sum_{\kk} \brak{ e^{-i\varphi} \td{f}_{\kk 1 + \up}^\dagger \td{f}_{-\kk 2 - \down}^\dagger +  e^{i\varphi} \td{f}_{\kk 2 + \up}^\dagger \td{f}_{-\kk 1 - \down}^\dagger }  - (\up \leftrightarrow \down) \ . 
\end{equation}
\begin{equation} \label{eq:V2pairing-symmetric}
     e^{i\theta_2} \mV_2 \sum_{\kk} \td{f}_{\kk 1 + \up}^\dagger \td{f}_{-\kk 2 - \down}^\dagger- (\up \leftrightarrow \down)  \quad \text{or}\quad   
    e^{i\theta_2} \mV_2 \sum_{\kk} \td{f}_{\kk 2 + \up}^\dagger \td{f}_{-\kk 1 - \down}^\dagger  - (\up \leftrightarrow \down) \ ,
\end{equation}
respectively. 
$\theta_{1,2}$ are the U(1) phases spontaneously chosen in the symmetry breaking. 
$\sum_{\kk} \td{f}_{\kk 1 + \up}^\dagger \td{f}_{-\kk 2 - \down}^\dagger - \td{f}_{\kk 1 + \down}^\dagger \td{f}_{-\kk 2 - \up}^\dagger$ 
and $\sum_{\kk} \td{f}_{\kk 2 + \up}^\dagger \td{f}_{-\kk 1 - \down}^\dagger - \td{f}_{\kk 2 + \down}^\dagger \td{f}_{-\kk 1 - \up}^\dagger$ 
transform as $d_{x^2-y^2} \pm i d_{xy}$ orbitals under the symmetry operations and form the two-dimensional representation $E_2$ of $D_6$ (\cref{eq:discrete-sym-f}). 
The linear combination in \cref{eq:V1pairing-symmetric} breaks the $C_{3z}$ symmetry and respects $C_{2z}$ and time-reversal ($T=\tau^x K$) symmetries. 
The linear combinations in \cref{eq:V2pairing-symmetric} break the time-reversal symmetry and preserve $C_{6z}$. 
Thus, $\mV_1$ and $\mV_2$ correspond to the nematic $d$-wave pairing and chiral $d$-wave pairing, respectively. 
As we will show in next subsubsection, $\mV_1$ leads to a $p$-wave-like nodal structure in the BdG spectrum due to the Berry's phase on Fermi surface.

Now we discuss the omitted higher on-site pairing channels. By referring to \cref{fig:renorm_int}, it can be seen that, for the allowed parameter range we obtained, the $\td{U}_{2,3}$ (${U}^{\rm p}_{2,3}$ take the same sign) channels always remain repulsive, hence can be directly dropped in the SC mean-fields. Even if they are incorporated, they necessarily lead to vanishing expectation values on these channels. 
On the other hand, the $\td{U}_1+\td{\mJ}$ and $\td{U}_4$ channels (${U}_1^{\rm p}+{\mJ}^{\rm p}$ and ${U}_4^{\rm p}$ take the same sign) can be weakly attractive in a small range. We therefore check with an enlarged the BdG basis, where inter-valley/intra-valley singlets/triplets can be investigated simultaneously, whether the mean-field solution can have an inter-valley triplet or intra-valley singlet component, aside from the leading inter-valley singlet order. 
For the parameter range $\td{\mJ} \in [4.6, 10.3] \td{\Delta}_0 $ and $\td{\Delta}_0 \in [1.7, 7.1]\mrm{meV}$ adopted in the main-text, we do \textit{not} observe their formation. The solution is purely the inter-valley $d$-wave singlet SC. 

We also discuss the possible effect of the omitted non-local interactions on the SC order. 
A non-local interaction either 1) involves the $c$-electrons, hence involving a mean-field decomposition like $\langle \td{f}_{\RR\alpha\eta s} c_{\kk a'\eta's'} \rangle$ or $\langle c_{\kk a\eta s} c_{-\kk a'\eta's'} \rangle$, or 2) acts inter-$f$-site, hence involving a mean-field decomposition $\langle \td{f}_{\RR\alpha\eta s} \td{f}_{\RR' a'\eta's'} \rangle$ with $\RR\not=\RR'$. 
As the $c$-electrons scarcely participate in the heavy flat bands, type 1) should mainly act on remote bands away from the Fermi level, hence can be neglected. 
Moreover, by discussion in \cref{sec:inter-site}, the strength of inter-$f$-site interactions for the quasi-particles should be reduced to $\sim z^2 U^{\rm inter-site}$, as a weak repulsive interaction. By raising the energy of inter-site pairings, such interactions cannot blockade the on-site pairings. Therefore, they can be neglected as well.

\subsection{The \texorpdfstring{$p$-wave-like}{p-wave-like} nodal structure in the nematic \texorpdfstring{$d$-wave}{d-wave} phase}

We now show that the $\mV_1$ pairing leads to a $p$-wave-like nodal structure in the BdG spectrum. 
We consider the weak pairing limit and project \cref{eq:BdG-symmetric} onto the Fermi surface basis. 
As shown in \cref{fig:mf}(b), there are two Fermi surfaces around the $\Gamma_M$ point, and both of them are dominated by the $f$-orbitals. 
For either Fermi surface, we can approximate the Bloch band state around the Fermi surface as
\begin{equation}
    \psi_{\kk \eta s}^\dagger = \sum_{\alpha} U^{(\eta)}_{\alpha}(\kk) \td{f}^\dagger_{\kk \alpha\eta s} \ .
\end{equation}
The time-reversal $T=\tau^xK$ and $C_{2z}T=\sigma^xK$ symmetries (\cref{eq:discrete-sym-f}) indicate 
\begin{equation}
  T  \psi_{\kk + s}^\dagger T^{-1} = \psi_{-\kk - s}^\dagger \quad \Rightarrow \quad    U_{\alpha}^{(-)} (-\kk) = U_{\alpha}^{(+)*}(\kk)
\end{equation}
\begin{equation}
  C_{2z}T \psi_{\kk + s}^\dagger (C_{2z}T)^{-1} = \psi_{\kk + s}^\dagger e^{i\phi_\kk} \quad \Rightarrow \quad    U_{2}^{(+)} (\kk) = U_{1}^{(+)*}(\kk) e^{-i\phi_\kk}\ .
\end{equation}
Without loss of generality, we can choose the gauge 
\begin{equation}
    U^{(+)}(\kk) = \frac1{\sqrt2} \begin{pmatrix}
        1 \\ e^{-i\phi_\kk}
    \end{pmatrix},\qquad 
    U^{(-)}(-\kk) = \frac1{\sqrt2} \begin{pmatrix}
        1 \\ e^{i\phi_{\kk}}
    \end{pmatrix}\ .
\end{equation}
According to Ref.~\cite{ahn_failure_2019}, $\phi_\kk$ must have an odd winding number along the Fermi surface, {\it i.e.},
\begin{equation}
    \mathcal{W} = \frac1{2\pi} \oint_{\rm FS} d\kk \cdot \partial_\kk \phi_\kk = 1 \mod 2\ ,
\end{equation}
if the Fermi surface encloses an odd number of Dirac points. 
In the symmetric Fermi liquid,  $\phi_\kk$ winding number is $\pm1$, as shown in \cref{fig:mf}(b). 
Now we project the $\mV_1$ pairing onto the Fermi surface
\begin{align} \label{eq:V1-pairing-symmetric}
  &  \sum_{\kk} e^{i\theta_1} \mV_1  \pare{   e^{-i\varphi} \td{f}_{\kk 1 + \up}^\dagger   \td{f}_{-\kk 2 - \down}^\dagger +  e^{i\varphi} \td{f}_{\kk 2 + \up}^\dagger \td{f}_{-\kk 1 - \down}^\dagger  }  - (\up \leftrightarrow \down) \nonumber\\
=&\sum_{\kk} 2   e^{i\theta_1} \mV_1 \cdot \cos(\phi_\kk + \varphi) \cdot
     \psi_{\kk + \up}^\dagger \psi_{-\kk-\down}^\dagger 
     - (\up \leftrightarrow \down) \ . 
\end{align}
As $\phi_\kk$ winds $\pm2\pi$ along the Fermi surface, there will be two nodes.

The nodes are robust against perturbations that preserve the $C_{2z}T$ symmetry. 
For example, we consider a mixture of $s$-wave pairing and the nematic $d$-wave pairing. 
Due to the $D_6$ point group symmetry, the $s$-wave pairing must have the form $e^{i\theta_0} \cdot \mV_0 \sigma^0$ ($\mV_0>0$).  
The projected pairing on Fermi surface becomes 
\begin{align} \label{eq:V0V1-pairing-symmetric}
  &  \sum_{\kk} e^{i\theta_0} \mV_0  \pare{   \td{f}_{\kk 1 + \up}^\dagger   \td{f}_{-\kk 1 - \down}^\dagger +  \td{f}_{\kk 2 + \up}^\dagger \td{f}_{-\kk 2 - \down}^\dagger  }
  + e^{i\theta_1} \mV_1  \pare{   e^{-i\varphi} \td{f}_{\kk 1 + \up}^\dagger   \td{f}_{-\kk 2 - \down}^\dagger +  e^{i\varphi} \td{f}_{\kk 2 + \up}^\dagger \td{f}_{-\kk 1 - \down}^\dagger  }   - (\up \leftrightarrow \down)  \nonumber\\
=&\sum_{\kk} 2 \pare{  e^{i\theta_0} \mV_0  + 
     e^{i\theta_1} \mV_1 \cdot \cos(\phi_\kk + \varphi) 
     } 
     \psi_{\kk + \up}^\dagger \psi_{-\kk-\down}^\dagger 
     - (\up \leftrightarrow \down) \ . 
\end{align}
The $C_{2z}T$ symmetry requires $\theta_0=\theta_1$ mod $\pi$ such that the pairing term gains a uniform phase $e^{-i2\theta_0}$ under $C_{2z}T$. 
There are two pairing nodes on the Fermi surface as long as $\mV_0<\mV_1$.

The above discussions are based on the weak pairing presumption. 
If the pairing energy is much larger than the band separation, then the nodes on inner and outer Fermi surfaces will merge each other, leading to a gapped phase.

Refs.~\cite{wu_theory_2018,liu_electron-k-phonon_2023} reported that the $d$-wave pairing has four nodes on the Fermi surface. 
The difference is because Refs.~\cite{wu_theory_2018,liu_electron-k-phonon_2023} used the bare band structure of MATBG, where $\phi_\kk$ has a winding number 2.

We also discuss the relation of our results to the Euler obstruction \cite{yu_euler_2022}, which states that a $C_{2z}T$-symmetric pairing diagonal in the Chern basis must have zeros in the Brillouin zone if the Euler class \cite{ahn_failure_2019} of the normal state bands is nontrivial, as it is MATBG. Since $f_{\alpha\eta s}$ has a large overlap with the $C\!=\!(-1)^{\alpha}\eta$ Chern basis \cite{song_magic-angle_2022}, the nematic $d$-wave pairing has a large component (more than $95\%$) in the obstructed channel.

\end{document}